\documentclass[%
 reprint,
preprintnumbers,
nofootinbib,
 amsmath,amssymb,
 aps,
]{revtex4-2}

\usepackage{graphicx}
\usepackage{dcolumn}
\usepackage{bm}
\usepackage[marginpar]{todo}
\usepackage{colordvi}
\usepackage{color}
\usepackage[dvipsnames]{xcolor}
\usepackage{tikz}
\usepackage{tikz-feynman}
\usetikzlibrary{patterns}
\usepackage{graphics}
\usepackage{hyperref}
\usepackage[mathlines]{lineno}
\usepackage{dcolumn}
\usepackage{tabularx}
\usepackage{orcidlink}

\begin{document}

\newcommand{\upsilonfours}{\ensuremath{\Upsilon}(4S)}
\newcommand{\bbbar}{\ensuremath{B\bar{B}}}
\newcommand{\epem}{\ensuremath{e^+e^-}}

\newcommand{\btou}{\ensuremath{b \rightarrow u}}
\newcommand{\btoc}{\ensuremath{b \rightarrow c}}
\newcommand{\btoxlnu}{\ensuremath{B \rightarrow X \ell \nu}}
\newcommand{\btoxulnu}{\ensuremath{B \rightarrow X_u \ell \nu}}
\newcommand{\btoxclnu}{\ensuremath{B \rightarrow X_c \ell \nu}}
\newcommand{\btodlnu}{\ensuremath{B \rightarrow D \ell \nu}}
\newcommand{\btoddstlnu}{\ensuremath{B \rightarrow D^{(*)} \ell \nu}}
\newcommand{\btodstlnu}{\ensuremath{B \rightarrow D^* \ell \nu}}
\newcommand{\btodststlnu}{\ensuremath{B \rightarrow D^{**} \ell \nu}}
\newcommand{\btodgaplnu}{\ensuremath{B \rightarrow D^{**}_{\textrm{Gap}} \ell \nu}}
\newcommand{\vub}{\ensuremath{V_{ub}}}
\newcommand{\vcb}{\ensuremath{V_{cb}}}

\newcommand{\btopilnu}{\ensuremath{B \rightarrow \pi \ell \nu}}
\newcommand{\btorholnu}{\ensuremath{B \rightarrow \rho \ell \nu}}
\newcommand{\btoetalnu}{\ensuremath{B \rightarrow \eta \ell \nu}}
\newcommand{\btoetaplnu}{\ensuremath{B \rightarrow \eta' \ell \nu}}
\newcommand{\btoetaetaplnu}{\ensuremath{B \rightarrow \eta/\eta' \ell \nu}}
\newcommand{\btoomegalnu}{\ensuremath{B \rightarrow \omega \ell \nu}}
\newcommand{\btorhoomegalnu}{\ensuremath{B \rightarrow \rho/\omega \ell \nu}}
\newcommand{\xlnu}{\ensuremath{X \ell \nu}}
\newcommand{\xulnu}{\ensuremath{X_u \ell \nu}}
\newcommand{\xclnu}{\ensuremath{X_c \ell \nu}}

\newcommand{\dst}{\ensuremath{D^{*}}}
\newcommand{\dstst}{\ensuremath{D^{**}}}
\newcommand{\kshort}{\ensuremath{K^0_S}}
\newcommand{\klong}{\ensuremath{K^0_L}}

\newcommand{\qtwo}{\ensuremath{q^2}}
\newcommand{\mx}{\ensuremath{M_X}}
\newcommand{\elb}{\ensuremath{E_{\ell}^B}}
\newcommand{\plb}{\ensuremath{p_{\ell}^B}}

\newcommand{\mmsqu}{\ensuremath{M_{\rm{miss}}^2}}
\newcommand{\qtot}{\ensuremath{Q_{\rm tot}}}
\newcommand{\nk}{\ensuremath{N_{K^{\pm}}}}
\newcommand{\nks}{\ensuremath{N_{K_S}}}
\newcommand{\mmsqupich}{\ensuremath{M^2_{\rm{miss}} (\pi^{\pm}_{\rm{slow}})}}
\newcommand{\mmsqupin}{\ensuremath{M^2_{\rm{miss}} (\pi^{0}_{\rm{slow}})}}
\newcommand{\mmsqupi}{\ensuremath{M^2_{\rm{miss}} (\pi_{\rm{slow}})}}
\newcommand{\vertexfit}{\ensuremath{\log_{10} (\chi^2_{\rm{vtx}} / n_{\rm{dgf}})}}
\newcommand{\costhetacpich}{\ensuremath{\cos \theta_c (\pi^{\pm})}}
\newcommand{\costhetacpin}{\ensuremath{\cos \theta_c (\pi^0)}}
\newcommand{\costhetacpi}{\ensuremath{\cos \theta_c (\dst \to D \pi)}}
\newcommand{\cosbypich}{\ensuremath{\cos \theta_{BY} (\pi^{\pm})}}
\newcommand{\cosbypin}{\ensuremath{\cos \theta_{BY} (\pi^0)}}
\newcommand{\cosbypi}{\ensuremath{\cos \theta_{BY} (\dst \to D \pi)}}

\newcommand{\pion}{\ensuremath{\pi}}
\newcommand{\pionp}{\ensuremath{\pi^{\pm}}}
\newcommand{\pionz}{\ensuremath{\pi^0}}

\newcommand{\pBR}{\ensuremath{\Delta\mathcal{B}}}
\newcommand{\Btag}{\ensuremath{B_{\mathrm{tag}}}}

\newcommand{\BRBToXuLNu}{\ensuremath{\mathcal{B} (B \to X_u \ell \nu) }}
\newcommand{\pBRBToXuLNu}{\ensuremath{\Delta\mathcal{B} (B \to X_u \ell \nu) }}
\newcommand{\pBRBzToXuLNu}{\ensuremath{\Delta\mathcal{B} (B^0 \to X_u \ell \nu) }}
\newcommand{\pBRBpToXuLNu}{\ensuremath{\Delta\mathcal{B} (B^{\pm} \to X_u \ell \nu) }}
\newcommand{\pBRBToXuENu}{\ensuremath{\Delta\mathcal{B} (B \to X_u e \nu) }}
\newcommand{\pBRBToXuMuNu}{\ensuremath{\Delta\mathcal{B} (B \to X_u \mu \nu) }}
\newcommand{\jpsi}{\ensuremath{J/\psi}}

\newcommand{\nub}{\ensuremath{\bar{\nu}}}

\newcommand{\deltaE}{\ensuremath{\Delta E}}
\newcommand{\mbc}{\ensuremath{M_{\rm bc}}}
\newcommand{\pfei}{\ensuremath{\mathcal{P}_{\rm FEI}}}
\newcommand{\costbto}{\ensuremath{\cos_{\rm TBTO}}}

\newcommand{\klow}{\ensuremath{_{K, \mathrm{low}}}}
\newcommand{\zlow}{\ensuremath{_{0, \mathrm{low}}}}
\newcommand{\khigh}{\ensuremath{_{K, \mathrm{high}}}}
\newcommand{\zmid}{\ensuremath{_{0, \mathrm{mid}}}}
\newcommand{\kmid}{\ensuremath{_{K, \mathrm{mid}}}}


\preprint{\large Belle II Preprint 2025-030}
\preprint{\large KEK Preprint 2025-37}

\title{Measurement of inclusive \btoxulnu\ partial branching fractions and $|\vub|$ at Belle~II}


  \author{M.~Abumusabh\,\orcidlink{0009-0004-1031-5425}} 
  \author{I.~Adachi\,\orcidlink{0000-0003-2287-0173}} 
  \author{K.~Adamczyk\,\orcidlink{0000-0001-6208-0876}} 
  \author{L.~Aggarwal\,\orcidlink{0000-0002-0909-7537}} 
  \author{H.~Ahmed\,\orcidlink{0000-0003-3976-7498}} 
  \author{Y.~Ahn\,\orcidlink{0000-0001-6820-0576}} 
  \author{H.~Aihara\,\orcidlink{0000-0002-1907-5964}} 
  \author{N.~Akopov\,\orcidlink{0000-0002-4425-2096}} 
  \author{S.~Alghamdi\,\orcidlink{0000-0001-7609-112X}} 
  \author{M.~Alhakami\,\orcidlink{0000-0002-2234-8628}} 
% ...
\collaboration{Belle II Collaboration}



\begin{abstract}
A sample of 365 fb$^{-1}$ of $\epem \to \upsilonfours \to \bbbar$ data collected by the Belle~II experiment is used to measure the partial branching fractions of charmless semileptonic $B$ meson decays and determine the magnitude of the Cabibbo-Kobayashi-Maskawa matrix element $\vub$. Events containing a signal electron or muon $\ell$ and a fully reconstructed hadronic $B$ decay that constrains the signal kinematics are selected, while the rest of the event defines the hadronic system $X_u$ associated with the signal. To discriminate the signal from the 50-times larger background originating from CKM-favored semileptonic $B$ decays, a template fit is performed in both signal and control regions after applying an optimized selection. The partial branching fraction measured for lepton energies greater than 1~GeV in the signal $B$ meson rest frame is $\pBR(\btoxulnu) = (1.54 \pm 0.08 \, {\rm (stat.)} \pm 0.12 \, {\rm (syst.)}) \times 10^{-3}$. From this measurement, using the Gambino, Giordano, Ossola, Uraltsev theoretical framework, $|\vub| = (4.01 \pm 0.19 ^{+0.07} _{-0.08}) \times 10^{-3}$ is determined, where the uncertainties are experimental and theoretical, respectively. This value is consistent with the world average obtained from previous inclusive measurements. Different theoretical predictions and partial branching fractions measured in other phase-space regions, defined by additional selections on the $X_u$ and leptonic system masses, are also used to determine $|\vub|$. This allows for a comparison of the resulting values across theoretical frameworks and phase-space regions.

\end{abstract}

\maketitle

\section{Introduction}\label{sec:intro}
Precise measurements of the Cabibbo-Kobayashi-Maskawa (CKM) matrix elements~\cite{Cabibbo:1963yz, Kobayashi:1973fv} are crucial to test the quark weak interactions described by the standard model of particle physics. For three fermion generations the CKM matrix is unitary, which imposes strict relations between its elements. The precision of experimental tests of several of these relations is limited by the current uncertainty in the determination of the strength of the $b$-to-$u$ quark coupling, $|\vub|$. $B$-meson decays to either an electron or a muon, $\ell$, a neutrino, $\nu$, and a charmless hadronic system, $X_u$, depend on $|\vub|$ as illustrated in Fig.~\ref{fig:feynman_vub}. Such decays are denoted as \btoxulnu. Despite the experimental challenges due to the neutrino in the final state,\footnote{Semitauonic decays are not considered, being even more challenging due to multiple neutrinos in the final state.} charmless semileptonic decays offer the best sensitivity to $|\vub|$. Determinations from fully hadronic $B$ meson decays suffer from significant QCD uncertainties~\cite{Aaij2021}, while fully leptonic decays of charged $B$ mesons are extremely challenging to study with the available datasets as the light-lepton channels are helicity suppressed and the tau channel suffers from large backgrounds~\cite{Snowmass2022}.
\begin{figure}[h!]
    \centering
\begin{center}
        
    \begin{tikzpicture}[scale = 1] 
    \begin{feynman}

    \vertex (base) ;
    \vertex [below = 0.6cm of base] (lowerbase);
    \vertex [above left = 0.05cm and 0.05cm of base] (labelbase) {\textcolor{purple}{\vub}};

    \vertex [left  = 1.3cm of base] (left) {$b$};
    \vertex [below = 0.6cm of left] (lowerleft) {$\bar{q}$};
    \vertex [below right = 1cm and 2cm of base] (right);
    \vertex [below = 0.6cm of right] (lowerright);
    \vertex [below = 0.3cm of left]   (leftmiddle);
    \vertex [below right = 0.8cm and 0.3 of right]  (rightmiddle);

    \draw[thick] (leftmiddle) ellipse (0.35cm and 0.75cm) node [left=0.3cm] {$\bar{B}$};
    
    \vertex [left  = 1.5cm of base] (left);
    \vertex [below = 0.6cm of left] (lowerleft);
    \vertex [below right = 1cm and 2cm of base] (right);
    \vertex [below = 0.6cm of right] (lowerright);
    \vertex [above right = 0.8cm and 0.8cm of base] (Wvertex);
    \vertex [below right = 0.5cm and 1.3cm of Wvertex] (neutrino) {$\bar{\nu}$}; 
    \vertex [above right = 0.5cm and 1.3cm of Wvertex] (lepton) {$\ell^-$}; 
    \vertex [below = 0.3cm of left]   (leftmiddle);
    \vertex [below right = 0.3cm and 0.3 of right]  (rightmiddle);
    
    \vertex[below = 0.5cm of lowerright] (lowerlowerright);
    \vertex[below = 0.5cm of lowerlowerright] (lowerlowerlowerright);

    \vertex [right = 0.25cm of left] (gu1);
    \vertex [right = 0.5cm of left] (gu2);
    \vertex [right = 0.75cm of left] (gu3);
    \vertex [right = 1cm of left] (gu4);
    \vertex [right = 1.25cm of left] (gu5);
    \vertex [right = 1.5cm of left] (gu6);
    \vertex [right = 0.25cm of lowerleft] (gd1);
    \vertex [right = 0.5cm of lowerleft] (gd2);
    \vertex [right = 0.75cm of lowerleft] (gd3);
    \vertex [right = 1cm of lowerleft] (gd4);
    \vertex [right = 1.25cm of lowerleft] (gd5);
    \vertex [right = 1.5cm of lowerleft] (gd6);
    
    \vertex [below right = 0.112cm and 0.224cm of base] (gu7);
    \vertex [below right = 0.224cm and 0.447cm of base] (gu8);
    \vertex [below right = 0.335cm and 0.671cm of base] (gu9);
    \vertex [below right = 0.447cm and 0.894cm of base] (gu10);
    \vertex [below right = 0.559cm and 1.118cm of base] (gu11);
    \vertex [below right = 0.671cm and 1.342cm of base] (gu12);
    \vertex [below right = 0.783cm and 1.565cm of base] (gu13);
    \vertex [below right = 0.9cm and 1.8cm of base] (gu14);
    \vertex [below right = 0.712cm and 0.224cm of base] (gd7);
    \vertex [below right = 0.824cm and 0.447cm of base] (gd8);
    \vertex [below right = 0.95cm and 0.671cm of base] (gd9);
    \vertex [below right = 1.047cm and 0.894cm of base] (gd10);
    \vertex [below right = 1.75cm and 1.118cm of base] (gd11);
    \vertex [below right = 1.18cm and 1.19cm of base] (gd12);
    \vertex [below right = 1.383cm and 1.565cm of base] (gd13);
    \vertex [below right = 1.495cm and 1.789cm of base] (gd14);
    
    \vertex [below = 0.3cm of gu8] (gm8);
    \vertex [below right = 0.35cm and 0.05cm of gu9] (gm9);
    \vertex [below = 0.3cm of gu10] (gm10);
    \vertex [below = 0.3cm of gu11] (gm11);
    
    \vertex[below = 0.1cm of gd11] (gdd11);
    \diagram* {
        (left) -- [shorten <=0.32cm, black, thick,fermion] (base),
        (lowerleft) -- [shorten <=0.32cm, black, thick,anti fermion] (lowerbase),
        
        (base) -- [black, thick,boson,] (Wvertex),
        (Wvertex) -- [black, thick,fermion,] (lepton),
        (Wvertex) -- [black, thick,anti fermion] (neutrino),
        
        (base) -- [shorten >=0.10cm, black, thick,fermion, edge label=$u$] (right),
        (lowerbase) -- [shorten >=0.15cm, black, thick,anti fermion] (lowerright),
        
         (gu2) -- [black, gluon] (gd4) ,
         (gd5) -- [black, gluon] (gu6) ,
         (gu7) -- [black, gluon] (gm8),
         (gm8) -- [black, half left] (gm9),
         (gm8) -- [black, half right] (gm9),
         (gm9) -- [black, gluon] (gm11),
         (gm11) -- [black, gluon] (gu14),
         (gm11) -- [black, gluon] (gd12),
         
    };
    
    \draw[thick] (rightmiddle) ellipse (0.45cm and 0.7cm) node [circle, minimum size=0, inner sep=0] {$X_u$};
    \end{feynman}
    \end{tikzpicture}
    \end{center}
    \caption{Feynman diagram of the leading \btoxulnu\ transition. The $X_u$ system can either represent a specific meson or a nonresonant system comprising several hadrons produced by the fragmentation of the $u\bar{q}$ system.}
    \label{fig:feynman_vub}
\end{figure}
\\
\\
There are two complementary strategies to measure $|\vub|$ via semileptonic decays: exclusive and inclusive measurements. Exclusive measurements select a specific decay mode, such as \btopilnu, while inclusive measurements place little or no restriction on the content of the $X_u$ system. Both strategies have been applied extensively at $B$ factories over the past two decades (see Ref.~\cite{hflav_2023} for a summary of all measurements).
\\
\\
The two most recent averages of $|\vub|$ extracted from exclusive and inclusive decays read~\cite{hflav_2023}
\begin{align}
    |\vub|_{\rm excl} &= (3.43 \pm 0.12) \times 10^{-3}, \nonumber \\
    |\vub|_{\rm incl} &= (4.06 \pm 0.16) \times 10^{-3},
\end{align}
where the total (experimental and theoretical) uncertainty is given. The two values disagree by about three standard deviations, which limits the precision of the CKM unitarity test. This discrepancy calls for a deeper investigation of the theoretical and experimental factors that may contribute to it. The measurement presented here uses the inclusive method.
\\
\\
The decay width of inclusive charmless semileptonic decays can be described by the \textit{heavy quark expansion} (HQE)~\cite{MANNEL1994396, CHAY1990399}, an expansion in powers of the strong coupling constant $\alpha_s$ and inverse powers of the $b$ quark mass $m_b$. Decay rate predictions in the HQE framework rely on quark-hadron duality which states that predictions made at the parton level, when integrated over sufficiently large portions of the phase space, should yield reliable predictions of hadronic observables~\cite{PhysRevD.57.2691}. However, selections on kinematic variables such as the lepton energy or the $X_u$ mass are necessary in experimental analyses to suppress backgrounds. The most important source of background emerges from CKM-favored semileptonic decays to a hadronic system containing a charm quark. Such decays, denoted as \btoxclnu, are enhanced by about a factor 50 compared to \btoxulnu\ decays. These kinematic selections break the inclusivity requirement of the HQE~\cite{Gambino:2007rp} and, instead, predictions of differential decay rates are necessary. These predictions in restricted kinematical regions rely on an effective description of the $b$-quark motion inside the meson through a nonperturbative \textit{shape function}, whose development has been a major theoretical challenge.\footnote{Beyond the leading order in the HQE, additional shape functions arise.} Treatment of such nonperturbative effects has led to the development of several theoretical frameworks that yield different predictions of the partial decay rate $\Delta\tilde{\Gamma}(\btoxulnu)$ ($\Delta\tilde{\Gamma}(\btoxulnu) = \Delta\Gamma(\btoxulnu) / |\vub|^2$) over a given region of phase space~\cite{Lange:2005yw, Andersen:2005mj, Gambino:2007rp}. A value of $|\vub|$ can be extracted from the measurement of the inclusive partial branching fraction $\pBR(\btoxulnu)$ via
\begin{equation}\label{equ:vub_extraction}
    |\vub| = \sqrt{\frac{\pBR(\btoxulnu)}{\tau_B\Delta\tilde{\Gamma}(\btoxulnu)}},
\end{equation}
where $\tau_B$ is the $B$ meson lifetime.
\\
\\
In this paper, an inclusive measurement of partial branching fractions of \btoxulnu\ ($\ell= e, \mu$) decays using $\epem \to \Upsilon(4S) \to \bbbar$ data collected by the Belle~II experiment is presented. Three variables are used to describe the kinematics of inclusive semileptonic decays: the lepton energy $E_\ell$ -- with the superscript $B$ added when the energy is calculated in the signal $B$ meson rest frame; the squared momentum transfer defined as $\qtwo = (p_\ell + p_\nu)^2$ where $p_\ell$ and $p_\nu$ are the charged lepton and neutrino four momenta; and the hadronic system mass \mx. The measurement is performed in three phase-space regions covering about 87\%, 57\% and 32\% of the full \btoxulnu\ phase space and defined by three sets of selections on \elb, \qtwo\ and \mx\ designed to suppress the \btoxclnu\ background. For each measured partial branching fraction, $|\vub|$ is extracted from Eq.~\ref{equ:vub_extraction} using different decay rate predictions.
\\
\\
The partner $B$ meson produced in the $\epem \to \bbbar$ process is reconstructed in specific hadronic decay channels. This method, referred to as \textit{hadronic tagging}, constrains the kinematics of the inclusively reconstructed hadronic system $X_u$ despite the neutrino being undetected. The partner $B$ is referred to as the tag $B$ or \Btag.
\\
\\
The remainder of this manuscript is organized as follows: the collected and simulated datasets used for the measurement are presented in Sec.~\ref{sec:modelling}; the event reconstruction and signal selections are outlined in Sec.~\ref{sec:event_reconstruction}; checks on the modeling of the \btoxclnu\ background and associated corrections are detailed in Sec.~\ref{sec:mismodelling}; the signal extraction procedure is discussed in Sec.~\ref{sec:signal_extraction}; finally, the results of the measurement are presented in Sec.~\ref{sec:results} and discussed in Sec.~\ref{sec:conclusions}. Throughout this article, charge conjugation is implied and natural units $\hbar = c = 1$ are used.

\section{Dataset and simulation}\label{sec:modelling}
The measurement is performed using a 365 fb$^{-1}$ sample of electron-positron collision data recorded at a center-of-mass energy of $\sqrt{s} = 10.58$~GeV corresponding to the \upsilonfours\ resonance mass (\textit{on-resonance} data). This sample contains $387 \times 10^6$ \upsilonfours\ events. In addition, 43 fb$^{-1}$ of data collected at a center-of-mass energy 60 MeV below the \upsilonfours\ resonance (\textit{off-resonance} data) are used to study background from $\epem \to q\bar{q}$ ($q \in \{u, d, s, c\}$) production (\textit{continuum} background).
\\
\\
The Belle~II detector is situated at the SuperKEKB asymmetric-energy electron–positron collider, based at KEK in Tsukuba, Japan~\cite{AKAI2018188}. It is a large solid-angle cylindrical detector consisting of several layers of subdetectors around the \epem\ interaction point (IP). The $z$ axis is defined as the symmetry axis of the detector in the direction of the electron beam, the $y$ axis corresponds to the vertical axis, pointing upward and the $x$ axis points towards the outside of the accelerator ring. At the innermost part of Belle~II is the pixel vertex detector (PXD), which consists of two layers of silicon pixel sensors located at 1.4 cm and 2.2 cm from the IP, which provides precise vertexing capabilities. In the dataset used in this measurement, the second pixel layer is installed in only one-sixth of the solid angle. Surrounding the PXD is the silicon vertex detector (SVD), which comprises four layers of silicon strip sensors. The SVD covers polar angles ranging from $17^{\circ}$ to $150^{\circ}$, which further enhances the vertex reconstruction and improves the overall tracking performance. The central drift chamber (CDC) surrounds the SVD and is specifically designed for charged particle tracking. It contains 14,336 sense wires operating in a helium-ethane gas mixture and also covers polar angles ranging from $17^{\circ}$ to $150^{\circ}$. The CDC provides subpercent accuracy in measuring charged particle momenta, essential for precise kinematic reconstruction. A time-of-propagation (TOP) detector positioned outside the CDC in the barrel region uses Cherenkov light to identify charged particles. In the forward end-cap, this role is fulfilled by a ring-imaging Cherenkov detector equipped with an aerogel radiator (ARICH). Beyond the TOP and ARICH lies the electromagnetic calorimeter (ECL), composed of 8,736 homogeneous thallium-doped caesium iodide crystals. The ECL is divided into three regions: the barrel and two end-caps, and covers polar angles from $12.4^\circ$ to $155.1^\circ$. This subdetector excels in detecting electromagnetic showers from both charged and neutral particles and provides percent-level energy resolution for electrons and photons. Additionally, the ECL provides three-momentum information for neutral particles through energy deposition (cluster) analysis. The ECL is surrounded by a superconducting solenoid which generates a magnetic field of 1.5 T. Charged tracks are bent in the $x$-$y$ plane by the magnetic field and the curvature of the trajectory can be used to determine the transverse momentum of the particle. The outermost subdetector of Belle~II is the \klong\ and muon detector (KLM). The KLM consists of 14 iron layers, each 4.7 cm thick, interspersed with 15 active layers in the barrel region and 14 (12) in the forward (backward) endcap region. The KLM covers polar angles from $20^\circ$ to $155^\circ$.
\\
\\
Simulated samples are used to optimize the signal selections, train multivariate algorithms to suppress various sources of background, study signal and background components and validate the signal extraction. The events are simulated using dedicated generators and they are processed using a full detector simulation implemented with GEANT4~\cite{AGOSTINELLI2003250}. Experimental and simulated data events are reconstructed and analyzed with the open-source BASF2 framework~\cite{Kuhr:2018lps}.
\\
\\
In the simulation, the assumed branching fractions for \btoxulnu\ and \btoxclnu\ decays are summarized in Table~\ref{tab:Xlnu_bf}. The branching fractions used for the modeling of resonant \btoxulnu\ decays are taken from the most up-to-date Heavy Flavor Averaging Group (HFLAV) report~\cite{hflav_2023}. The inclusive \btoxulnu\ branching fractions are taken separately for $B^0$ and $B^+$ as quoted in the latest Particle Data Group (PDG) Review of Particle Physics~\cite{PhysRevD.110.030001}. The \btoxclnu\ branching fractions are computed from the HFLAV averages assuming isospin symmetry. Dedicated samples of signal events are produced with the EVTGEN generator~\cite{LANGE2001152}. For background contributions, $\upsilonfours \to \bbbar$ samples, which contain known semileptonic and hadronic $B$ decays, are produced with EVTGEN and PYTHIA~8~\cite{SJOSTRAND2008852}.
\renewcommand{\arraystretch}{1.3} 
\begin{table}[h!]
    \centering
    \begin{tabular}{lcc}
    \hline
    \hline
         & \multicolumn{2}{c}{$\mathcal{B}$ (\%)} \\
         \hline
        Decay mode & $B^+$ & $B^0$ \\
        \hline
        Incl. \btoxulnu & $0.192 \pm 0.024$ & $0.176 \pm 0.022$  \\
        \hline
        \btopilnu & $0.0078 \pm 0.0003$ & $0.0150 \pm 0.0006$ \\
        \btorholnu & $0.0158 \pm 0.0011$ & $0.0294 \pm 0.0021$ \\
        \btoomegalnu & $0.0119 \pm 0.0009$ & - \\
        \btoetalnu & $0.0035 \pm 0.0004$ & - \\
        \btoetaplnu & $0.0024 \pm 0.0007$ & - \\
        \hline
        \hline
        Incl. \btoxclnu & $11.05 \pm 0.16$ & $10.27 \pm 0.15$ \\
        \hline
        \btodlnu & $2.27 \pm 0.06$ & $2.11 \pm 0.05$ \\
        \btodstlnu & $5.27 \pm 0.12$ & $4.90 \pm 0.11$ \\
        $B \to D_1 \ell \nu$ & $0.64\pm 0.10$ & $0.59\pm 0.10$ \\
        $B \to D^*_0 \ell \nu$ & $0.13 \pm 0.19$ & $0.12 \pm 0.18$ \\
        $B \to D^\prime_1 \ell \nu$ & $0.28 \pm 0.04$ & $0.26 \pm 0.04$ \\
        $B \to D^*_2 \ell \nu$ & $0.32 \pm 0.03$ & $0.30 \pm 0.03$ \\
        $B \to D_s K \ell \nu$ & $0.03 \pm 0.01$ & - \\
        $B \to D^*_s K \ell \nu$ & $0.03 \pm 0.02$ & - \\
        \hline
        $B \to D  \eta \ell \nu$ & $0.90 \pm 0.90$ & $0.86 \pm 0.86$ \\
        $B \to D^* \eta \ell \nu$ & $0.90 \pm 0.90$ & $0.86 \pm 0.86$ \\
        $B \to D \pi \pi \ell \nu$ & $0.07 \pm 0.09$ & $0.07 \pm 0.08$ \\
        $B \to D^* \pi \pi \ell \nu$ & $0.22 \pm 0.10$ & $0.20 \pm 0.10$ \\
    \hline
    \hline
    \end{tabular}
    \caption{Assumed branching fractions used for simulated semileptonic $B$ decays.}
    \label{tab:Xlnu_bf}
\end{table}
\\
\\
The sample of \btoxulnu\ events is a mixture of nonresonant modes (such as $B \to \pi\pi\ell\nu$, $B \to \pi\rho\ell\nu$...), simulated via inclusive \btoxulnu\ frameworks, and resonant modes (\btopilnu, \btoomegalnu...). The theoretical description of resonant modes is based on functions known as \textit{form factors}, which encapsulate the nonperturbative dynamics of the transition. Nonresonant and resonant decays are combined using the \textit{hybrid model} initially proposed in Ref.~\cite{Ramirez:1989yk}. After presenting the samples of \btoxulnu\ decays, the hybrid model is described.
\\
\\
The simulation of nonresonant events is based on the De Fazio, Neubert (DFN) model~\cite{DeFazio:1999ptt} implemented in the EVTGEN generator. This model describes the differential \btoxulnu\ rate at order $\mathcal{O}(\alpha_s)$ and at leading order in the HQE and requires two input parameters for the modeling of events, $m_b$ and a nonperturbative parameter $a$ used in the modeling of the shape function. In the Kagan, Neubert renormalization scheme~\cite{Kagan:1998ym} their values read: $m_b^{\mathrm{KN}} = 4.66 \pm 0.04$~GeV and $a^{\mathrm{KN}} = 1.3 \pm 0.5$~\cite{PhysRevD.73.073008}. They are related to the difference between the $B$ meson mass and the $b$ quark mass and the average momentum squared of the $b$ quark inside the $B$ meson. The fragmentation of the hadronic system produced in nonresonant events is simulated by PYTHIA~8.
\\
\\
In addition, five resonant decays are included in the \btoxulnu\ modeling: $B \to (\pi/ \rho/ \omega/ \eta/ \eta') \ell \nu$. Their theoretical descriptions rely on specific form factor parametrizations which are also implemented in EVTGEN. The Bourrely, Caprini, Lellouch~\cite{Bourrely:2008za} parametrization is used to model \btopilnu\ decays. The form factor parameters are taken from Table 57 in Ref.~\cite{FlavourLatticeAveragingGroupFLAG:2021npn}; \btorhoomegalnu\ decays are modeled with the Bharucha, Straub, Zwicky parametrization~\cite{Bharucha2016} whose values are taken from Table 4 in Ref.~\cite{Bernlochner:2021rel}; \btoetaetaplnu\ decays are modeled via the Isgur, Scora, Grinstein, Wise approach~\cite{PhysRevD.52.2783} with input parameters taken from the light-cone sum rules framework described in Ref.~\cite{Duplancic:2015zna}.
\\
\\
Resonant and nonresonant \btoxulnu\ contributions are combined using the hybrid approach. The resonant and nonresonant contributions are summed in three-dimensional bins of \elb, \mx\ and \qtwo. In each bin, the inclusive contribution ($\Delta \mathcal{B}^{\textrm{inc}}$) is obtained from the nonresonant component and the exclusive contribution ($\Delta \mathcal{B}^{\textrm{exc}}$) is obtained from the sum of exclusive components. The nonresonant part is scaled down such that the sum of nonresonant and resonant contributions matches the inclusive partial branching fraction. The nonresonant contribution is therefore scaled by weights defined as:
\begin{equation}\label{equ:hybrid}
    w_{ijk} = \frac{\Delta \mathcal{B}^{\textrm{inc}}_{ijk} - \Delta \mathcal{B}^{\textrm{exc}}_{ijk}}{\Delta \mathcal{B}^{\textrm{inc}}_{ijk}},
\end{equation}
where the indices $i, j, k$ denote a three-dimensional bin. These three-dimensional bins are defined by the combinations of the following one-dimensional bins:
\begin{align*}
    E_\ell^B &: [0, 0.50, 1.00, 1.25, 1.50, 1.75, 2.00, 2.25, 3.00] \rm~GeV, {\rm~and} \\
    q^2 &: [0, 2.50, 5.00, 7.50, 10.0, 12.5, 15.0, 20.0, 25.0] \rm~GeV ^2, \\
    M_X &: [0, 1.40, 1.60, 1.80, 2.00, 2.50, 3.00, 3.50] \rm~GeV.
\end{align*}
\\
The \btoxclnu\ background is modeled as a sum of different types of resonant decays. Each resonance is modeled with a specific form factor parametrization implemented in EVTGEN. The \btodlnu\ and \btodstlnu\ decays are modeled using the Bernlochner, Ligeti, Papucci, Robinson, Xiong, Prim form factor model~\cite{Bernlochner:2022ywh} with eight common parameters. The four heavy $1P$ states $D_0^*$, $D_1'$, $D_1$ and $D_2^*$ are collectively referred to as \dstst~\cite{PhysRevD.85.094033}. The \btodststlnu\ decays are modeled using the Bernlochner, Ligeti, Robinson (BLR) model from Ref.~\cite{Bernlochner:2017jxt} with three parameters for the broad $D_0^*$ and $D_1'$ states and four parameters for the narrower $D_1$ and $D_2^*$ states. The \btoxclnu\ modeling also includes the decays $B \to D^{(*)}\pi\pi\ell\nu$ and $B \to D_s^{(*)}K\ell\nu$. The $B \to D^{(*)}\eta\ell\nu$ modes are simulated to fill the gap between the inclusively measured \btoxclnu\ rate and the sum of exclusive decay rates~\cite{Belle:2021eni}. The $D^{(*)}\pi\pi$ and $D^{(*)}\eta$ modes are collectively referred to as \textit{gap modes} and are denoted \btodgaplnu. As there exists no theoretical description of these decays, they are simulated via intermediate decays to \dstst\ resonances using the BLR form factor parametrization. The $D\eta$ decays are simulated via $B \to D_0^* (\to D \eta) \ell \nu$ and $D^*\eta$ decays via $B \to D_1^\prime (\to D^* \eta) \ell \nu$. The $D^{(*)}\pi\pi$ modes are equally split between intermediate decays to $D_0^*$ and $D_1^\prime$.
\\
\\
Another source of background stems from continuum events which are simulated by the KKMCee~\cite{JADACH2023108556} generator (the resulting particle decays are simulated by EVTGEN).

\section{Event reconstruction and selection}\label{sec:event_reconstruction}
The inclusive \btoxulnu\ analysis relies on the reconstruction of the \Btag\ meson in hadronic decay channels. The rest frame of the signal $B$ meson is determined from momentum conservation in $\Upsilon(4S) \to B\bar{B}$ decays. The lepton in the semileptonic decay is reconstructed as a single track. All detected tracks and clusters not associated with the \Btag\ or the lepton are assumed to come from the $X$ system and are grouped into an object called the rest-of-event (ROE). The hadronic system four-momentum $p_X$, from which the hadronic mass \mx\ is derived, is computed from the four-momenta of all tracks and clusters in the ROE. All ROE clusters are assumed to be produced by photons and, since events with kaons and leptons in the ROE are rejected at a later stage, all ROE tracks are assigned a pion mass hypothesis. Particles originating from beam background can pollute the ROE and they are therefore suppressed by the selections given in Sec.~\ref{sec:roe_selections}. The neutrino cannot be detected and its momentum is therefore identified with the missing momentum of the event. Using the four momenta of the initial \epem\ state, the \Btag, the lepton and $X$, the neutrino four-momentum can be inferred from the missing four-momentum as:
\begin{equation}
    p_\nu = p_{\rm miss} = p_{\epem} - p_{\Btag} - p_\ell - p_X.
\end{equation}
The momentum transfer squared \qtwo\ is therefore computed as $(p_\ell + p_\nu)^2$. Finally, the missing mass squared \mmsqu\ is calculated as $\mmsqu = p_{\rm miss}^2$.
\\
\\
In the following, the requirements applied on \Btag\ candidates, ROE particles and signal lepton candidates are presented. The requirements described in Secs~\ref{sec:tag_side_reco} and~\ref{sec:sig_side_reco} are also referred to as \textit{preselections}.

\subsection{Tag side}\label{sec:tag_side_reco}
The \Btag\ is reconstructed using the Full Event Interpretation algorithm (FEI)~\cite{Keck2019} which reconstructs $B$ mesons in $\mathcal{O}(10^4)$ decay chains using a series of boosted decision tree (BDT) classifiers. They are required to have an FEI output score greater than 0.01 and a beam constrained mass
\begin{equation}
    \mbc = \sqrt{E_{\rm beam}^{*2} - |p^*_{\rm tag}|^2} > 5.27 \; \mathrm{GeV},
\end{equation}
where $p^*_{\rm tag}$ is the momentum of the \Btag\ candidate in the \epem\ center-of-mass frame and $E_{\rm beam}^* = \sqrt{s}/2$ is half of the center-of-mass energy of the \epem\ collision. Furthermore, the \Btag\ candidates are required to have an energy difference
\begin{equation}
    \deltaE = E_{\rm tag}^* - E_{\rm beam}^*
\end{equation}
in the range $-0.15$~GeV $< \deltaE < 0.10$~GeV, where $E_{\rm tag}^*$ is the energy of the \Btag\ candidate in the \epem\ center-of-mass frame.
\\
\\
The thrust axis is defined as the axis $\hat{t}$ along which the summed projections of the momenta of a group of particles, $\sum_{i=1}^{n} | \hat{t} \cdot \vec{p}_i |$, is maximized. To suppress continuum background, the absolute value of the cosine of the angle between the thrust axes of the \Btag\ and the ROE is required to be lower than 0.9. Finally, if more than one \Btag\ candidate has been selected, only the candidate with the highest FEI score is retained.

\subsection{Signal side}\label{sec:sig_side_reco}
\subsubsection{Signal lepton}
The selections applied on signal lepton candidates aim to suppress three main sources of background. The first source consists of events with secondary leptons which mostly come from $D$ meson decays and are referred to as \textit{secondary lepton} events. The second consists of, events in which a hadron has been misidentified as the signal lepton. These mostly correspond to pions misidentified as muons and are referred to as \textit{fake lepton} events. The third consists of, events where the lepton originates from a $J/\psi$ decay or from photon conversion.
\\
\\
To ensure that signal lepton candidates originate near the interaction point, the distances between the interaction point and the track point-of-closest-approach along the $z$ axis ($dz$) and in the $xy$ plane ($dr$) are required to satisfy $dr < 1$ cm and $|dz| < 3$ cm. Tracks are also required to fall in the CDC acceptance by having a polar angle between $17^{\circ}$ and $150^{\circ}$. In addition, they are required to have a laboratory-frame momentum greater than 300 MeV. Electrons and muons are identified using a BDT-based and a likelihood-ratio-based identification score, respectively~\cite{Milesi:2020esq}. The identification scores are calculated from information collected by all subdetectors except the PXD. The lepton identification requirements result in a sample where 45\% of the events contain electron candidates and the remaining fraction contain muon candidates. About 10\% of electron candidates and 35\% of muon candidates are misidentified particles (mostly pions). Furthermore, the four-momentum of an electron candidate is corrected for energy loss from bremsstrahlung photons by identifying photons with an energy below 1~GeV within a 0.05~rad cone around the electron initial momentum direction. Finally, to suppress $J/\psi$ and photon conversion leptons, each candidate is combined with oppositely charged tracks found in the ROE which are assigned the same mass hypothesis as the signal lepton candidate. Of all possible pairs, the one with an invariant mass closest to zero or the nominal \jpsi\ mass is selected. Signal lepton candidates are then rejected if the invariant mass is found to be below 50 MeV or between $3.043$ and $3.129$~GeV for electron pairs and between $3.072$ and $3.122$~GeV for muon pairs. These selections are optimized to reject about 95\% of photon conversion tracks and \jpsi, except for electrons below the \jpsi\ mass where a broad tail caused by bremsstrahlung can be observed. In this case, the threshold rejects 68\% of true \jpsi. The signal lepton candidates are required to have a charge consistent with the flavor of the signal $B$ meson, taken as the opposite of the \Btag\ charge. Even though this requirement vetoes about a fifth of neutral $B$ signal events, it also rejects events in which a secondary lepton has been reconstructed as the signal lepton. Finally, events with multiple lepton candidates are rejected as their presence is indicative of secondary semileptonic decays of $D$ mesons.

\subsubsection{Rest-Of-Event}\label{sec:roe_selections}
ROE selections are optimized to select charged particles originating near the interaction point and suppress particles that hit the edges of the detector where the response is poorly modeled. Selections applied on neutral clusters are optimized to suppress beam background and misidentified photons. A fit is performed using all tracks found in the ROE to match them to a single decay vertex. Tracks in the ROE are required to pass similar $dr$, $dz$ and polar angle selections to signal lepton tracks. Duplicated tracks due to particles with low transverse momentum are suppressed by a BDT. The classifier uses the track helix parameters, the product of the track charges, the angle between the tracks and the track transverse and longitudinal momenta.
\\
\\
ROE neutral clusters are required to have a minimum energy of 70, 60 and 70 MeV in the forward, barrel, and backward regions of the ECL respectively and a polar angle between $17^{\circ}$ and $150^{\circ}$. The barrel covers a region between $32^{\circ}$ and $129^{\circ}$ and there are $1^{\circ}$ gaps between the barrel and each end cap. Photon candidates must have a reconstructed time within 70 ns of the time at which the collision occurred. Misidentified photons are suppressed by a BDT classifier trained on numerous cluster properties, such as energy, polar angle or distance from the nearest track~\cite{Cheema:2024iek}.

\subsubsection{Charged pions}
The multiplicity of charged pions in an event is used to derive an uncertainty on the $X_u$ hadronization modeling as described in Sec.~\ref{sec:xu_fragmentation_modelling}. These pions are identified using a likelihood ratio score calculated from information collected by all subdetectors except the PXD and are required to have a momentum and polar angle within the phase-space region considered in hadron identification calibration studies~\cite{HadronIDBelleII}. The likelihood requirement selects about 47\% of the charged pion candidates and about 2\% of pion candidates are misidentified particles.

\subsection{Continuum}\label{sec:continuum_suppression}
In $\epem \to q\bar{q}$ events, light mesons are boosted and therefore result in two back-to-back jetlike structures, whereas $B$ mesons are produced nearly at rest in the center-of-mass frame resulting in isotropically distributed decay products. Variables such as the two $B$ meson thrust angles and magnitudes, the modified Fox-Wolfram moments~\cite{Fox:1978vu, leeEvidencePiPi2003} and the CLEO cones~\cite{CLEO:1995rok} can be used to distinguish continuum events from $B$ meson decays (30 variables are considered). In order to suppress the continuum background, a multi layer perceptron (MLP) is trained. The MLP is configured with two hidden layers with 256 and 128 neurons, respectively. The network uses the Swish activation function~\cite{ramachandranSearchingActivationFunctions2017} and a binary cross-entropy loss function. The AdamW optimizer~\cite{loshchilov2019} is used to train the network, with a learning rate of $9.3 \times 10^{-3}$. The classifier is trained using a total of 14 variables: the charged track multiplicity, and a subset of 13 variables among the 30 describing the event shape as mentioned above.\footnote{The other 17 variables are dropped because they display poor agreement between experimental and simulated data in the region used to train the classifier.} All chosen variables exhibit good agreement between simulated and observed data. A subset consisting of 5\% of simulated \btoxulnu\ events and 20\% of simulated continuum events is used for training. The sample of continuum events is corrected following the procedure described in Sec.~\ref{sec:continuum_calibration}. The neural network is optimized using the OPTUNA package~\cite{optuna}. The optimal selection on the classifier output score is chosen according to the maximal significance defined as $S/\sqrt{S+B}$, with $S$ the number of \btoxulnu\ events and $B$ the number of continuum events that pass the selection. The yields are counted in the region where $\elb > 1$~GeV (see Sec.~\ref{sec:kinematic_selections} for a discussion on kinematic selections). The continuum suppression selection rejects 95\% of continuum events while retaining 68\% of \btoxulnu\ events.

\subsection{\dst\ reconstruction from low-momentum pions}
Most decays of \dst\ mesons produce a $D$ meson and low-momentum (or slow) pion, denoted as $\pi_s$, whose properties can be used to suppress \btodstlnu\ events. Using $\pi_s$, an inclusive reconstruction of \dst\ mesons in the ROE is performed. Charged $\pi_s$ are identified as single tracks with a charge opposite to that of the signal lepton, and neutral $\pi_s$ are reconstructed from pairs of photons. The $\pi_s$ are required to have a laboratory-frame momentum between 50 and 200 MeV. Duplicated tracks left by charged $\pi_s$ are suppressed by a BDT as mentioned in the previous section. The \dst\ candidate is assumed to have a momentum vector along the $\pi_s$ flight direction in the laboratory-frame and its energy is calculated from
\begin{equation}
    E_{D^*} = \frac{M_{\dst}}{M_{\dst} - M_{D}} E_{\pi_s},
\end{equation}
where $M_{D/\dst}$ is the known mass of the $D/\dst$ meson~\cite{PhysRevD.110.030001} and $E_{\pi_s}$ the energy of the $\pi_s$. Having computed the \dst\ four-momentum and knowing the \Btag\ and signal lepton candidate four-momenta, the missing mass of each \btodstlnu\ candidate can be computed as:
\begin{equation}\label{equ:mmiss_slow_pi}
   \mmsqu(\dst \to D \pi) = \left(p_{\epem} - p_{\Btag} - p_\ell - p_{\dst}\right)^2.
\end{equation}
When multiple $\pi_s$ candidates are reconstructed, the one with $\mmsqu(\dst \to D \pi)$ closest to zero is selected.

\subsection{\btoxclnu\ suppression}\label{sec:btoxclnu_suppression}
\begin{figure*}
    \centering
    \includegraphics[scale=0.7]{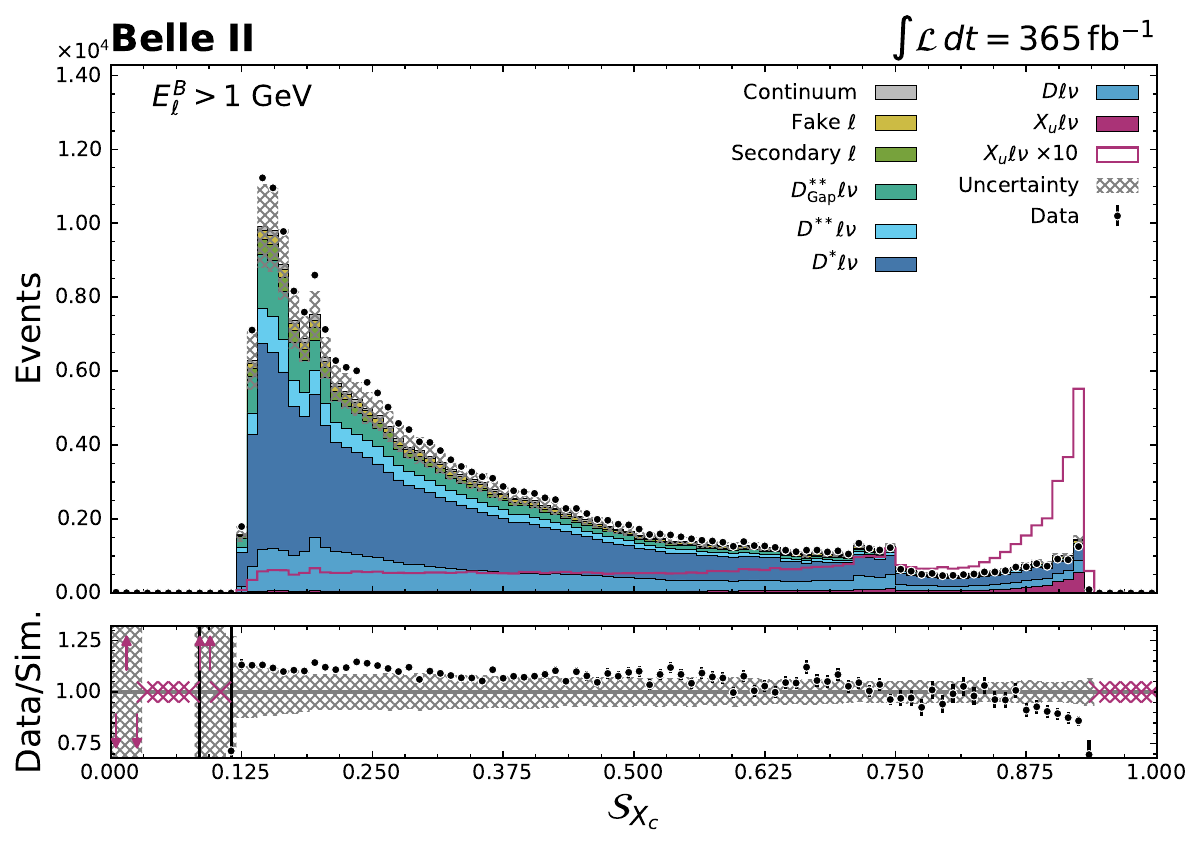}
    \caption{\btoxclnu\ suppression classifier output score. Simulated templates are shown as stacked histograms and experimental data are shown as black points. The signal component is shown as a purple line with its yield multiplied by 10 to enhance visibility. The bottom panel shows the ratio between data and simulation yields in each bin. The purple arrows indicate that the point is outside of the plotting range and the purple crosses indicate the absence of simulated data in a particular bin. The step in the distribution in the range $0.70-0.75$ is caused by the large number of events with a $p_{\rm vtx}^{\rm ROE}$ close to zero (see Fig.~\ref{fig:xclnu_mva_input} in Appendix~\ref{app:xclnu_suppression}).}
    \label{fig:b2xclnu_lambda_0_score}
\end{figure*}

Since most charm meson weak decays produce kaons, events where a $K^\pm$ or \kshort\ candidate is found in the ROE are rejected. No attempt is made to reconstruct \klong. Similarly to charged pions, charged kaons are selected by combining into a likelihood ratio the particle identification information from all subdetectors except the PXD. The likelihood requirement selects about 18\% of all charged kaon candidates and about 9\% of the selected kaon candidates are misidentified particles. They are also required to have a momentum and polar angle within the phase-space region considered in hadron identification calibration studies~\cite{HadronIDBelleII}.
\\
\\
\kshort\ candidates are reconstructed from two oppositely charged tracks that are fit to a common production vertex. As \kshort\ are relatively long-lived particles, the ROE track selections are not applied in this case. To ensure that the momentum vector of the candidates is consistent with the vector connecting the interaction point and the fitted decay vertex, the cosine of the angle between these two vectors is required to be greater than 0.998. Finally, \kshort\ candidates are required to have a mass between 470 and 530 MeV, consistent with the known $K^0$ mass~\cite{PhysRevD.110.030001}. The \kshort\ mass resolution is about 10~MeV.
\\
\\
To suppress the dominant \btoxclnu\ background, an MLP is trained on nine kinematic variables. The trained network is similar to that used to suppress continuum background but is configured with four hidden layers with 64, 32, 128 and 128 neurons, respectively, and a learning rate of $3.3 \times 10^{-3}$. The input features exploit various differences between semileptonic decays to charmed and charmless mesons:
\begin{itemize}
    \item the missing mass squared, \mmsqu;
    \item the total event charge, $Q_{\rm{tot}}$;
    \item the $p$-value of the ROE vertex fit, $p_{\rm vtx}^{\rm ROE}$;
    \item when a $D^{*+}$ or $D^{*0}$ has been inclusively reconstructed from a $\pi_s$ (either neutral or charged):
    \begin{itemize}
        \item the missing mass squared $\mmsqu(D^* \to D \pi_s)$ as defined in equation~\ref{equ:mmiss_slow_pi};
        \item the cosine of the angle between the flight directions of the reconstructed \dst\ and the signal lepton in the center-of-mass frame, $\cos \theta_c (D^* \to D \pi_s)$;
        \item the $\cos \theta_{BY} (D^* \to D \pi_s)$ variable defined as
        \[
        \cos \theta_{BY} = \frac{2E^*_{\rm beam}E^*_Y - m_B^2-m_Y^2}{2|p^*_B||p^*_Y|},
        \]
        where $Y = \dst\ell$; $E^*_Y$, $|p^*_Y|$ and $m_Y$ are the energy, three-momentum magnitude and mass of the $Y$ system, respectively; $|p^*_B|$ and $m_B$ are the three-momentum magnitude and mass of the $B$ meson, respectively. Energies and three-momenta are defined in the center-of-mass frame. This variable is expected to be between $-1$ and 1 if only a massless particle such as a neutrino is missing in the reconstructed event.
    \end{itemize}
\end{itemize}
Since the average particle multiplicity of \btoxclnu\ events is expected to be larger than for \btoxulnu\ events, the event reconstruction quality is expected to be poorer because of missed particles or because of the limited detector resolution. The first three variables therefore help to reject poorly reconstructed \btoxclnu\ events which tend to populate the high \mmsqu\ tail, have a low $p_{\rm vtx}^{\rm ROE}$ and a $Q_{\rm{tot}}$ different than zero. The $\pi_s$ variables allow the identification of events with a \dst\ decay. The distributions of the nine input features are shown in Appendix~\ref{app:xclnu_suppression}. The optimal selection on the classifier output score is chosen as the one which minimizes the $\pBR(\btoxulnu)$ total uncertainty as obtained from the nominal fit described in Sec.~\ref{sec:signal_extraction}. The output score is shown in Fig.~\ref{fig:b2xclnu_lambda_0_score} and the optimal selection value for the score is found to be 0.87. The \btoxclnu\ suppression requirement rejects 98\% of \btoxclnu\ events while retaining 25\% of \btoxulnu\ events.

\subsection{Kinematic selections}\label{sec:kinematic_selections}
In addition to the selections discussed so far, requirements on the three kinematic variables \elb, \mx\ and \qtwo\ can be used to further suppress background. The three sets of kinematic selections listed in Table~\ref{tab:fit_selections} are considered. In the following, a selection of $\elb > 1$~GeV is applied to all samples, as it rejects a portion of phase space mainly populated by continuum and fake and secondary lepton events. Furthermore, a region where an $\mx < 1.7$~GeV selection is applied is also explored. The mass of the lightest charm state -- the $D$ meson -- being around 1.86~GeV, this selection efficiently rejects \btoxclnu\ decays. Finally, since \btoxulnu\ decays are expected to dominate at higher leptonic energies, a region with a $\qtwo > 8$~GeV$^2$ selection, in addition to the \elb\ and \mx\ selections, is also considered.
\\
\\
The three signal regions used to measure $\pBR(\btoxulnu)$ are defined by applying all preselections, the continuum suppression, the \btoxclnu\ suppression, and one of the three sets of kinematic selections. The signal phase-space acceptance ($i.e.$ the covered fraction of the accessible \btoxulnu\ phase-space) for the three sets of kinematic selections is given in Table~\ref{tab:fit_selections}. Inclusive \btoxulnu\ models are most reliable in the broadest phase-space region but the additional selections on \mx\ and \qtwo\ increase the signal purity and provide additional tests of the theoretical predictions.
\begin{table}[h!]
    \caption{\btoxulnu\ phase-space acceptance for the three sets of kinematic selections used to define the signal regions.}
    \label{tab:fit_selections}
    \centering
    \begin{tabular}{ccc}
    \hline
    \hline
        Region label & Selections & Acceptance \\[2.0ex]
        1 & $\elb > 1.0$~GeV & 87\% \\[2.0ex]
         & $\elb > 1.0$~GeV  & \\
        2 & $\mx < 1.7$~GeV & 57\% \\[2.0ex]
          & $\elb > 1.0$~GeV &  \\
        3 & $\mx < 1.7$~GeV & 31\% \\
          & $\qtwo > 8$~GeV$^2$ & \\
    \hline
    \hline
    \end{tabular}
\end{table}

\section{Modeling corrections}\label{sec:mismodelling}
\subsection{Continuum modeling corrections}\label{sec:continuum_calibration}
\begin{figure*}
    \centering
    \includegraphics[scale=0.5]{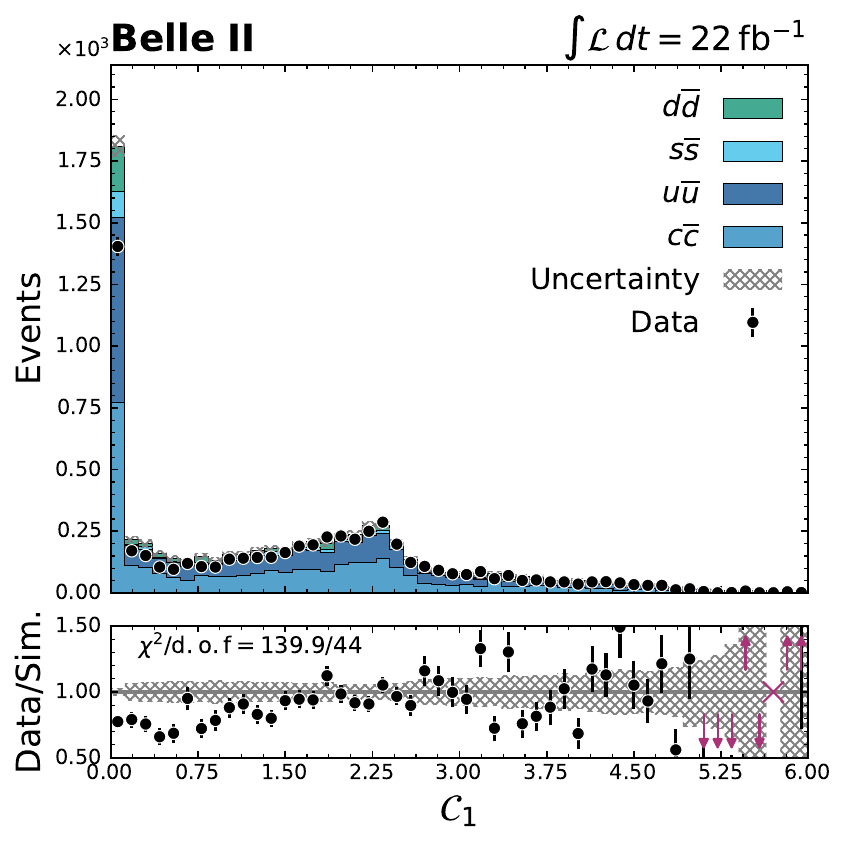}
    \includegraphics[scale=0.5]{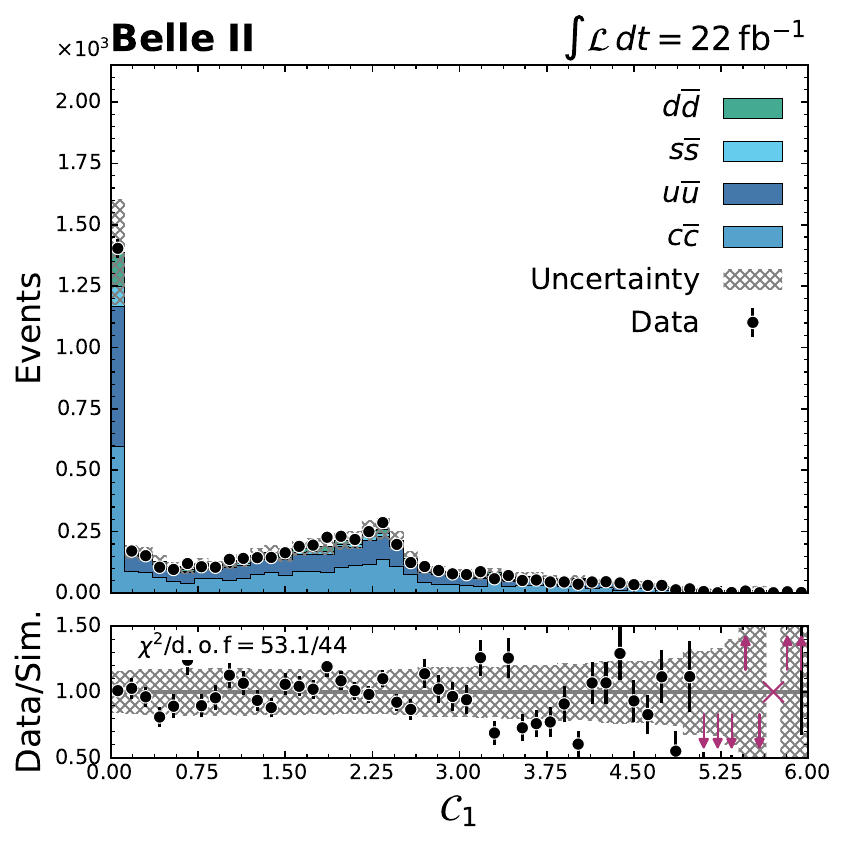}
    \caption{Distribution of $C_1$ (first CLEO cone~\cite{CLEO:1995rok}) in an off-resonance sample before (left) and after (right) the continuum modeling correction is applied. The modeling uncertainties, described in Sec.~\ref{sec:cont_calibration_error}, are included in the error band after applying the continuum calibration. The bottom panel shows the ratio between data and simulation yields in each bin. The purple arrows indicate that the point is outside of the plotting range and the purple crosses indicate the absence of simulated data in a particular bin.}
    \label{fig:continuum_calibration}
\end{figure*}

The modeling of continuum events is known to be poorly understood and is therefore corrected using a data-driven approach. First, the total number of expected events is corrected by comparing the number of events in 40\% of the off-resonance experimental and simulated data samples. The remaining 60\% are used to test and validate the BDT described below. The correction factor is extracted independently for events in which a charged or neutral $B$ meson candidate has been reconstructed and is found to be $0.87 \pm 0.02$ and $0.94 \pm 0.03$, respectively, where the uncertainties are statistical (uncertainties on these factors are derived as described in Sec.~\ref{sec:cont_calibration_error}).
\\
\\
In a second step, a BDT with a maximum tree depth of five is trained to distinguish between experimental and simulated off-resonance data using 30 variables that describe the event shape~\cite{Martschei:2012pr} (see Sec.~\ref{sec:continuum_suppression}). In addition, the missing mass squared and the scalar sum of the transverse momenta of the signal and tag-side $B$ final-state particles are included in the BDT input features. The BDT is trained on the off-resonance sample from which the normalization correction factors are extracted. The remainder of the sample is split into a validation sample (10\% of the full sample) and a test sample (50\%). The classifier output score $\mathcal{S}_{\rm cont}$ is well calibrated and takes values between 0 and 1. The ratio
\begin{equation}
    w_{\rm cont} = \frac{\mathcal{S}_{\rm cont}}{1 - \mathcal{S}_{\rm cont}}
\end{equation}
is an approximation to the ratio of experimental to simulated data likelihoods: it is applied as a weight to continuum events in the on-resonance simulated data sample. The continuum correction includes both the overall normalization factors and the BDT-extracted shape factors. The modeling correction is applied on the continuum sample after applying the preselections and before training the continuum suppression classifier. Its impact is illustrated in Fig.~\ref{fig:continuum_calibration} for the most discriminating variable in the BDT training.

\subsection{\btoxclnu\ modeling corrections}\label{sec:xclnu_modelling_corrections}
\begin{figure*}
    \centering
    \includegraphics[scale=0.4]{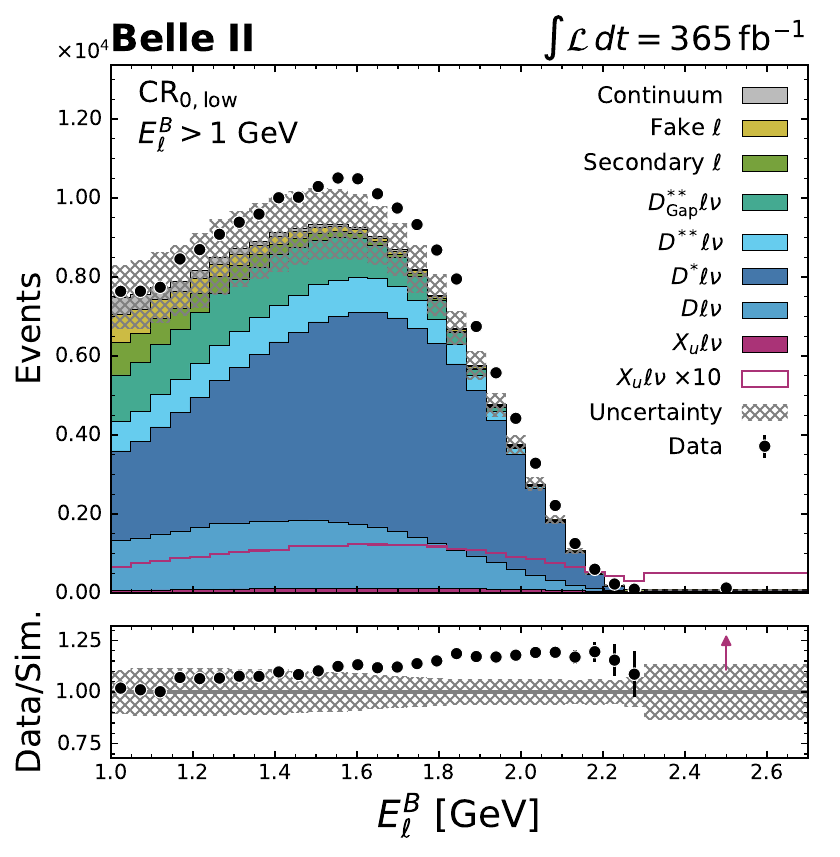}
    \includegraphics[scale=0.4]{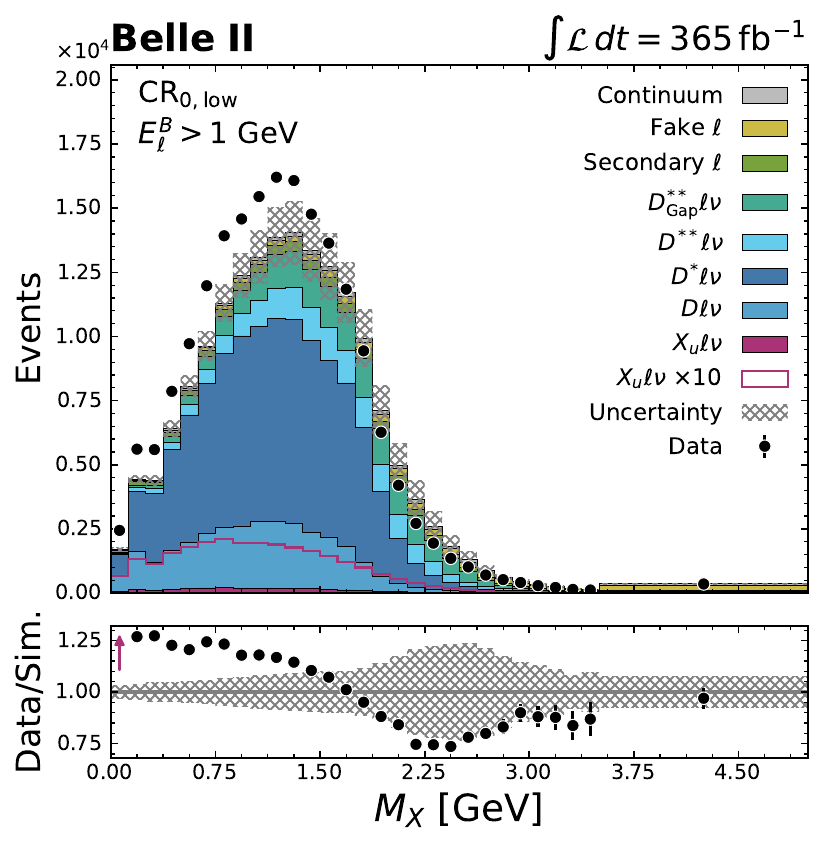}
    \includegraphics[scale=0.4]{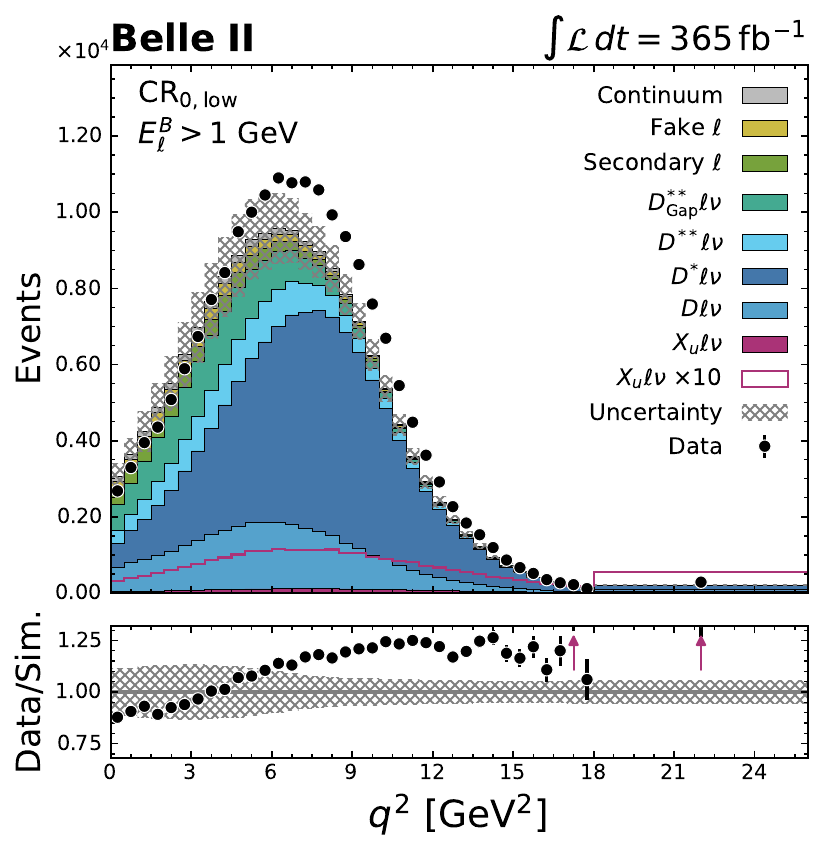}
    \caption{\elb, \mx\ and \qtwo\ distributions in data and simulation in the CR\zlow\ region defined in the text. Only the $\elb > 1$~GeV kinematic selection is applied. All systematic uncertainties as described in Sec.~\ref{sec:syst_uncert} are included. The bottom panel shows the ratio between data and simulation yields in each bin. The purple arrows indicate that the point is outside of the plotting range.}
    \label{fig:cr0l_data_mc}
\end{figure*}

The agreement between data and simulation is evaluated in control regions. Three control regions are defined for each kinematic selection defined in Table~\ref{tab:fit_selections} by inverting the kaon veto and/or requiring a low \btoxclnu\ suppression classifier score (see Fig.~\ref{fig:b2xclnu_lambda_0_score}):
\begin{itemize}
    \item CR\zlow: events with kaons are rejected and the \btoxclnu\ suppression classifier score is required to be lower than 0.60;
    \item CR\khigh: the presence of at least one kaon and a \btoxclnu\ suppression classifier score higher than 0.87 are required;
    \item CR\klow: the presence of at least one kaon and a \btoxclnu\ suppression classifier score lower than 0.60 are required.
\end{itemize}
A comparison of the \btoxclnu\ composition in the three control regions and the signal region is shown in Appendix~\ref{app:xc_compo}. In addition, two validation regions, VR\kmid\ and VR\zmid, with classifier scores between 0.60 and 0.87 are defined. In the former region, a kaon is required whereas in the latter, kaons are vetoed. The subdivided selection plane is illustrated in Fig.~\ref{fig:cr_illustration}. In Fig.~\ref{fig:cr0l_data_mc}, clear disagreements between data and simulation distributions of \elb, \mx\ and \qtwo\ can be seen.
\begin{figure}[h!]
    \centering
    \begin{tikzpicture}
  \draw (0,2) rectangle (2,4); 
  \draw (2,2) rectangle (4,4); 
  \draw (4,2) rectangle (6,4); 
  \draw (0,0) rectangle (2,2); 
  \draw (2,0) rectangle (4,2); 
  \draw (4,0) rectangle (6,2); 

  \node at (1,3) {\small CR\klow};
  \node at (3,3) {\small VR\kmid};
  \node at (5,3) {\small CR\khigh};
  \node[Purple] at (1,1) {\small CR\zlow};
  \node at (3,1) {\small VR\zmid};
  \node[Green] at (5,1) {SR};

  \node at (-0.5,3) {\Large $>0$};
  \node at (-0.5,1) {\Large $0$};
  \node[rotate=90] at (-1.2,2) {$K$ multiplicity};

  \node at (0, -0.5) {0};
  \node at (2, -0.5) {0.60};
  \node at (4, -0.5) {0.87};
  \node at (6, -0.5) {1};
  \node at (3, -1.0) {\btoxclnu\ suppression classifier score};

  \draw[Purple, very thick] (0,0) rectangle (2,2); 
  \draw[Green, very thick] (4,0) rectangle (6,2); 

\end{tikzpicture}
    \caption{The data samples after reconstruction, preselection and continuum suppression are subdivided into six regions illustrated here. The selection plane is subdivided based on the kaon multiplicity (vertical axis) and the \btoxclnu\ suppression classifier score (horizontal axis). The signal and control regions used for the signal extraction are highlighted in green and purple, respectively.}
    \label{fig:cr_illustration}
\end{figure}
\\
\\
Background originating from continuum as well as fake and secondary leptons is known to be poorly modeled in simulation. However, continuum events are corrected following the procedure described in Sec.~\ref{sec:continuum_calibration} and these background sources represent in total less than 10\% of all events in the control regions. The mismodeling is therefore expected to stem from \btoxclnu\ events. There are several known shortcomings in the simulation of \btoxclnu\ decays. The $B \to D^{(*)}\eta\ell\nu$ modes have never been observed and are used as a hypothesis for the decays that fill the \btoxclnu\ gap. Therefore, a relative uncertainty of 100\% is assigned to their assumed branching fractions. This uncertainty doesn't cover the full difference between data and simulation distributions shown in Fig.~\ref{fig:cr0l_data_mc}. Furthermore, most $D$ meson decays with three or more particles are not simulated according to measured decay distributions but use simple phase-space models, which can bias the kinematics of charmed $B$ decays. Moreover, it was observed in Belle~II studies~\cite{PhysRevD.109.112006} that measurements can be affected by the poor modeling of charm decays involving \klong\ mesons. The impact of the modeling of \btoxclnu\ events with a $D$ meson decaying to a \klong\ is studied in detail by scaling their contribution by arbitrary factors. Based on studies performed in the measurement of the $B^+ \to K^+ \nu \bar{\nu}$ branching fraction~\cite{PhysRevD.109.112006}, this component is increased by 30\% which only marginally improves the agreement between data and simulation. This indicates that it is unlikely that the mismodeling of $D \to \klong X$ decays can explain the observed mismodeling. As no single source of mismodeling could be identified, the \btoxclnu\ component is corrected as a whole using the control regions defined above. The normalization and shape from simulation are corrected separately.

\subsection{\btoxclnu\ normalization correction}\label{sec:xclnu_norm_correction}
The signal-region \btoxclnu\ normalization is corrected before the signal extraction via a method relying on the observed normalizations in the three control regions. Since the kaon multiplicity and classifier score are only weakly correlated, it is assumed that the ratio of experimental over simulated \btoxclnu\ events in the signal region, $r_{\rm SR}$, can be determined from the same ratio in the control regions as
\begin{equation}\label{equ:abcd}
    r_{\rm SR} = \frac{r\zlow \cdot r\khigh}{r\klow}.
\end{equation}
The \btoxclnu\ normalization correction factors $r\zlow$, $r\klow$ and $r\khigh$ are calculated by subtracting from data and simulation all \btoxulnu, continuum, as well as fake and secondary lepton simulated events. These components represent between 2\% and 10\% of all events in CR\zlow\ and CR\klow\ and between 8\% and 25\% of events in CR\khigh, depending on kinematic selections. Since this computation relies on various sources of background and signal being well modeled, an uncertainty is derived on the three ratios by varying the total yields of \btoxulnu, continuum and fake and secondary lepton events by 50\% and computing the normalization correction factors for each variation. The calculation defined in Eq.~\ref{equ:abcd} is tested by replacing CR\khigh\ and the signal region by VR\kmid\ and VR\zmid\, respectively. Although the signal purity is larger in VR\zmid\ (about 10--20\% depending on kinematic selections) than in all other control regions, it remains lower than in the signal region (30--70\%). The \btoxclnu\ normalization correction obtained for VR\zmid\ via the control region method (1.07, 1.15, 1.21, from the broadest set of kinematic selections to the tightest) and the factor obtained by directly taking the \btoxclnu\ data over simulated data yield ratio in that region (1.03, 1.16, 1.22) are compared. The relative difference between these two factors is considered as an additional uncertainty on $r_{\rm SR}$. In the signal extraction, two nuisance parameters are included to float the normalization of the \btoxclnu\ component in CR\zlow\ and in the signal region separately. The normalization parameters are constrained by Gaussian functions with widths equal to the uncertainty on the normalization correction of each region. The values of $r\zlow$ and $r_{\rm SR}$ in the three phase-space regions considered in the signal extraction are given in Table~\ref{tab:xc_norms}.
\renewcommand{\arraystretch}{1.2} 
\begin{table}[h!]
    \caption{Values of the normalization correction factors $r\zlow$ and $r_{\rm SR}$ in the three phase-space regions considered for the signal extraction. The \elb, \mx\ and \qtwo\ kinematic selections are: $\elb > 1.0$~GeV, $\mx < 1.7$~GeV and $\qtwo > 8$~GeV$^2$. The total uncertainties computed as described in the text are reported.}
    \label{tab:xc_norms}
    \centering
    \begin{tabular}{ccc}
        \hline
        \hline
        Kinematic selection & CR\zlow & SR \\
        \elb & $1.11 \pm 0.05$ & $1.01 \pm 0.08$ \\
        \elb, \mx & $1.18 \pm 0.04$ & $1.07 \pm 0.12$ \\
        \elb, \mx, \qtwo & $1.24 \pm 0.03$ & $1.09 \pm 0.21$ \\
        \hline
        \hline
    \end{tabular}
\end{table}
\\
\\
The \btoxclnu\ shape correction is implemented directly in the signal extraction procedure by simultaneously fitting the signal region and CR\zlow\ as described in Sec.~\ref{sec:fitted_regions}.

\section{Signal extraction}\label{sec:signal_extraction}
The \btoxulnu\ partial branching fraction is extracted from a binned template fit with the PYHF package~\cite{Heinrich2021}.

\subsection{Statistical model}
The statistical model relies on the simultaneous measurement of event counts $n$ in two independent regions\footnote{Referred to as \textit{channels} in the PYHF documentation.} -- a signal region and a control region -- using binned distributions of a given variable. These distributions are modeled as the sum of three different components (\textit{i.e.} samples or templates, specified in Sec.~\ref{sec:fit_templates}). Within the statistical model, a set of free and constrained parameters, $\boldsymbol{\eta}$ and $\boldsymbol{\chi}$ respectively, parametrize the variations of each template yield. The impact of each nuisance parameter $\chi$ is limited by a Gaussian constraint term $c_{\chi}$. The constraints are defined using a set of auxiliary measurements $\boldsymbol{a}$ (one measurement $a_{\chi}$ for each parameter), which correspond to component yield variations caused by various systematic effects. Thus, the statistical model $f$, in its most general form, can be written as
\begin{equation}
    f(\boldsymbol{n}, \boldsymbol{a} | \boldsymbol{\eta}, \boldsymbol{\chi}) = \prod_c \prod_b \mathrm{Pois}(n_{cb} | \nu_{cb}(\boldsymbol{\eta}, \boldsymbol{\chi})) \prod_{\chi \in \boldsymbol{\chi}} c_{\chi}(a_{\chi}|\chi),
\end{equation}
where $\boldsymbol{n} \equiv \{n_{cb}\}$ are the observed event yields. The products run over all regions $c$, bins $b$ and constrained parameters $\chi$; and $\nu_{cb}$ are the expected yields.
\begin{eqnarray}
    \nu_{cb} & = & \sum_{s} \nu_{scb}(\boldsymbol{\eta}, \boldsymbol{\chi}) \nonumber\\
     & = & \sum_{s} \bigg(\prod_{\kappa \in \boldsymbol{\kappa}} \kappa_{scb}(\boldsymbol{\eta}, \boldsymbol{\chi}) \bigg) \nonumber\\
     & \times & \bigg( \nu_{scb}^{0}(\boldsymbol{\eta}, \boldsymbol{\chi}) + \sum_{\Delta \in \boldsymbol{\Delta}} \Delta_{scb}(\boldsymbol{\eta}, \boldsymbol{\chi}) \bigg),
\end{eqnarray}
where the sum runs over all templates $s$, $\kappa_{scb}$ and $\Delta_{scb}$ are multiplicative and additive modifiers respectively and $\nu_{scb}^{0}$ is the nominal yield in a given bin. The different types of modifiers implemented in the fit are discussed in the following subsections.

\subsection{Fitted regions and \btoxclnu\ shape correction}\label{sec:fitted_regions}
\begin{figure*}
    \centering
    \includegraphics[scale=0.55]{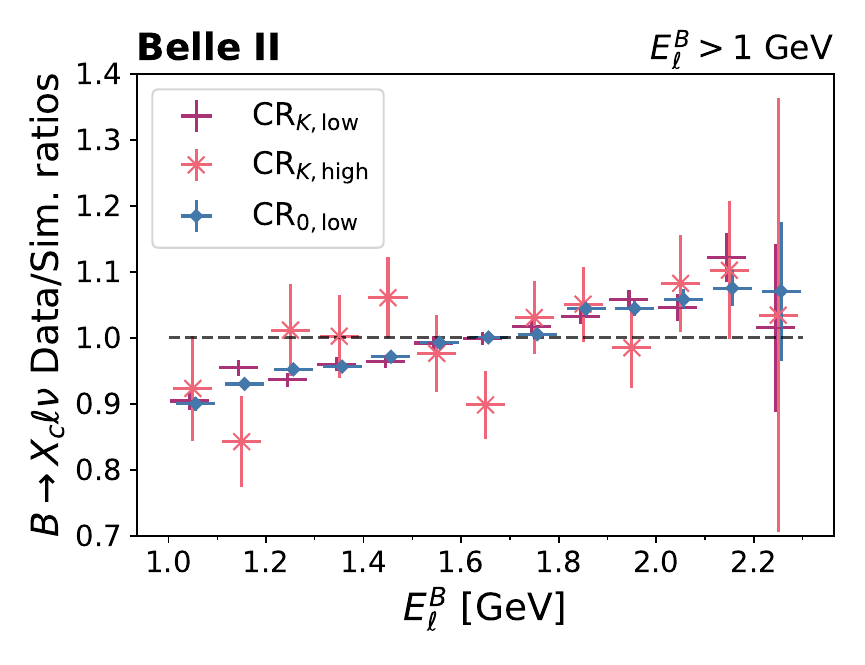}
    \includegraphics[scale=0.55]{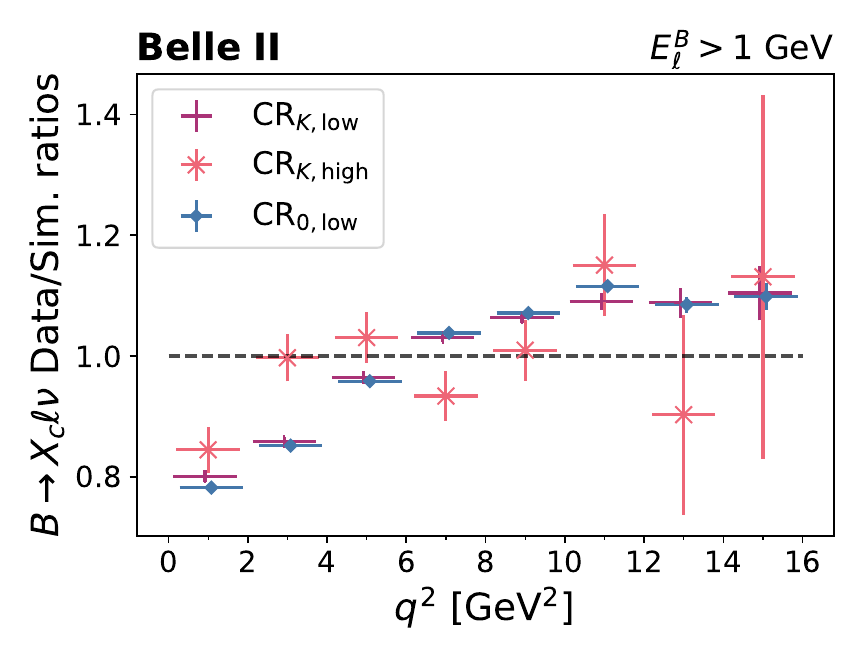}
    \caption{The normalized ratios of \btoxclnu\ yields in data to simulation are shown for the three control regions CR\klow\ (purple), CR\khigh\ (pink) and CR\zlow\ (blue) in \elb\ (top) and \qtwo\ (bottom). Only the $\elb > 1$~GeV kinematic selection is applied. The error bars include the statistical uncertainty of the experimental and simulated data samples.}
    \label{fig:data_mc_ratio_comp}
\end{figure*}

As detailed in Sec.~\ref{sec:xclnu_modelling_corrections}, \btoxclnu\ decays appear to be poorly modeled. Normalization correction factors are calculated and applied before the fit whereas the correction of the \btoxclnu\ component shape is implemented directly into the fit. In order to constrain the shape of the \btoxclnu\ background in the signal region, distributions are fitted simultaneously in the signal region and CR\zlow, the latter being orthogonal to the signal region and enriched in \btoxclnu\ events. This method not only corrects the \btoxclnu\ shape in the signal region, it also properly accounts for all correlations between correction factors and other fit parameters to determine a \btoxulnu\ branching fraction value that incorporates uncertainties related to the \btoxclnu\ background corrections. Using any control region to correct the shape of the \btoxclnu\ component relies on the assumption that the disagreement between data and simulation is similar in the signal region and the chosen control region. This assumption is supported by the fact that the ratio of \btoxclnu\ data over simulated data yields follows a similar trend in relevant kinematic variables in all three control regions as illustrated in Fig.~\ref{fig:data_mc_ratio_comp}. To choose the control region used for correcting the signal region mismodeling, two considerations must be taken into account: the similarity in composition and shape of the \btoxclnu\ component in the control and signal region and the sample size relative to the signal region. Based on these two considerations, CR\zlow\ is chosen for the fit.
\\
\\
The partial branching fraction of \btoxulnu\ is measured in three different phase-space regions defined by applying the  kinematic selections of Table~\ref{tab:fit_selections} in the signal region. Each set of selections is also applied in the control region and therefore characterizes a separate fit setup. The differences between these setups are highlighted in the following sections. The setups are referred to as fits 1, 2 and 3 from the broadest phase-space region to the tightest (see Table~\ref{tab:fit_selections}).

\subsection{Fit templates}\label{sec:fit_templates}
The fit is performed using the following templates:
\begin{itemize}
    \item Signal (\btoxulnu): this template includes all signal events whose generated kinematics fall in the chosen phase-space region (\textit{signal-in}). Signal decays are modeled via the DFN framework~\cite{DeFazio:1999ptt}. The template is assigned an unconstrained normalization factor in the statistical model whose postfit value can be directly translated into a measurement of the branching fraction. This factor is fixed to its value from simulation in CR\zlow\ where its contribution is negligible. An additional template (\textit{signal-out}) is defined to cover the small fraction of signal events which pass the reconstruction-level selections given in Table~\ref{tab:fit_selections} but whose generated kinematic parameters fall outside the target signal region (based on the generator-level selections). The resolution of the \elb\ distribution is extremely good, allowing the \btoxulnu-out yield to be neglected in fit 1. In fits 2 and 3, this component is kept as it represents 2\% and 5\% of all signal events in their respective signal regions. The template is assigned the same unconstrained normalization as the signal-in template but nuisance parameters can vary its normalization and shape independently.
    \item \btoxclnu\ background: this template represents the largest source of background. It is composed of \btodlnu, \btodstlnu, \btodststlnu\ and \btodgaplnu\ decays. To capture the shape of this template in CR\zlow, one free-floating normalization factor per bin, common to the signal and control-region distributions, is used. The normalization of this template is corrected prior to fitting, following the procedure described in Sec.~\ref{sec:xclnu_norm_correction}. A constrained normalization parameter is assigned to the \btoxclnu\ template in each region to float the overall normalization within the uncertainties given in Table~\ref{tab:xc_norms}.
    \item \textit{Other backgrounds}: all other backgrounds are summed up in a single template. They include continuum events as well as fake and secondary lepton events. After applying the continuum suppression classifier, the yields of these components are relatively small compared to semileptonic event yields. In fit 1, this template is assigned a normalization modifier which is constrained by a Gaussian function with a width of 50\% of the total yields to conservatively vary this component. In fits 2 and 3, the contribution from other backgrounds is negligible and this modifier is therefore not added.
\end{itemize}

\subsection{Fitted variables}
Different variables are tested for the fit. To combine information from different shapes, two and three-dimensional fits can be performed. In this analysis, the two-dimensional \elb:\qtwo\ variable is fitted in fits 1 and 2 by splitting \elb\ into four bins and \qtwo\ into three bins. When applying both the tight \mx\ and \qtwo\ selections, only \elb\ spans a range large enough to extract the signal branching fraction, and thus only \elb, split into eight bins, is used in fit 3. The binning for each variable is chosen to mitigate model dependence. Since inclusive \btoxulnu\ models are known to poorly describe the shape of kinematic distributions in the high \elb\ and \qtwo\ region, a single bin is defined for this region. The variable and binning for each fit are given in Table~\ref{tab:fit_vars_bins}.
\renewcommand{\arraystretch}{1.3} 
\begin{table}[h!]
    \caption{Variable and binning for the three fit configurations described in the text.}
    \label{tab:fit_vars_bins}
    \centering
    \begin{tabular}{ccc}
    \hline
    \hline
         & Variable & Binning \\
    \hline
        Fit 1 & $\elb:\qtwo$ & $[1.0, 1.3, 1.6, 1.9, 2.7]$~GeV $\times [0, 4, 8, 26]$~GeV$^2$ \\
        Fit 2 & $\elb:\qtwo$ & $[1.0, 1.3, 1.6, 1.9, 2.7]$~GeV $\times [0, 4, 8, 26]$~GeV$^2$ \\
        Fit 3 & \elb & $[1.0, 1.4, 1.6, 1.7, 1.8, 1.9, 2.0, 2.1, 2.7]$~GeV \\
    \hline
    \hline
    \end{tabular}
\end{table}

\subsection{Systematic uncertainties}\label{sec:syst_uncert}
Sources of systematic uncertainty are directly implemented in the fit as nuisance parameters. Each nuisance parameter can affect either the shape or the normalization of the template(s) to which it is assigned, or both. An overview of systematic uncertainties considered is given in the following. Two sources of uncertainty (number of $\Upsilon(4S)$ decays and $X_u$ hadronization modeling) which are not considered in the fit are also described.
\\
\\
For the uncertainties related to the $D$ decay branching fractions, the continuum calibration, the charged particle identification and the hadronic tagging, the number of parameters is truncated to avoid implementing a large number of nuisance parameters with a negligible impact. To do so, the principal components of each covariance matrix are taken such that at least 99\% of the total variance is covered. This method reduces the total number of parameters in each fit from about $1,000$ to about $150$.

\subsubsection{\btoxulnu\ modeling}
The resonant and nonresonant branching fractions (given in Table~\ref{tab:Xlnu_bf}) are varied within their respective uncertainties therefore yielding one nuisance parameter per decay type. The modeling of \btoxulnu\ resonant modes relies on assumed form factor models which depend on a set of parameters extracted from lattice QCD and dedicated measurements. The form factor parameter variations are calculated using the eFFORT~\cite{EFFORT2022} package. The parameter values, uncertainties and correlation matrices are taken from the references given for each model in Sec.~\ref{sec:modelling}. Two nuisance parameters are added to represent the uncertainties in $m_b^{\mathrm{KN}} = 4.66 \pm 0.04$~GeV and $a^{\mathrm{KN}} = 1.3 \pm 0.5$, including their covariance. In addition, to factor in differences between inclusive \btoxulnu\ models, a set of events is simulated via the Bosch, Lange, Neubert, Paz (BLNP) framework~\cite{Lange:2005yw, BOSCH2004335}. The difference between the nominal fitted distribution and the BLNP distribution is added as a symmetric uncertainty. Moreover, applying a kaon veto to suppress charm background in the signal region also rejects \btoxulnu\ events where the $X_u$ system hadronizes as a pair of kaons. The branching fraction of $B \to K\bar{K}(X)\ell\nu$ decays (where $X$ is a charmless hadronic system) is not known and an uncertainty related to this contribution is therefore extracted by varying the production of $s$ quarks relative to $u$ or $d$ production, $\gamma_s$. The two values measured by the TASSO~\cite{TASSO:1984nda} ($\gamma_s = 0.35 \pm 0.05$) and JADE~\cite{JADE:1983KaonProd} ($\gamma_s = 0.27 \pm 0.06$) collaborations are averaged and the uncertainty is chosen to cover the full spread of $1\sigma$ variations from the two values, giving $\gamma_s = 0.30 \pm 0.09$ (which corresponds to a branching fraction of roughly 0.02\%). Three sets of events with $\gamma_s$ equal to 0.21, 0.30 and 0.39 are produced. The sample simulated with $\gamma_s$ equal to its central value is used as nominal and the variations with respect to the other two distributions are added as systematic uncertainties. For each variation related to the \btoxulnu\ modeling, the hybrid weights are recalculated (see Eq.~\ref{equ:hybrid}).

\subsubsection{\btoxclnu\ modeling}
All \btoxclnu\ and all $D$ meson decay branching fractions are varied within their respective uncertainties (see Table~\ref{tab:Xlnu_bf}). The uncertainties of the unmeasured $B \to D^{(*)}\eta\ell\nu$ branching fractions are chosen to be equal to 100\%. The branching fraction uncertainty of $B \to D_0^* \ell \nu$ decays is increased by about a factor of six to partially cover the large difference between the two measurements entering the HFLAV average calculation~\cite{PhysRevLett.101.261802, PhysRevD.107.092003}. The number of parameters related to the branching fractions of $D$ meson decays is truncated from 227 to one. For \btoxclnu\ decays, the form factor parameter variations are computed using the Hammer~\cite{Bernlochner:2020tfi} package. The parameter values, uncertainties and correlation matrices are taken from the references given for each model in Sec.~\ref{sec:modelling}.

\subsubsection{$f^{+-}/f^{00}$}
The ratio of production fractions of charged and neutral pairs of $B$ mesons in $\Upsilon(4S)$ decays is taken from Ref.~\cite{hflav_2023}: $f^{+-}/f^{00} = 1.052 \pm 0.031$. A nuisance parameter is added in the fit to vary the relative fraction of $B^+$ and $B^0$ decays accordingly.

\subsubsection{$\pi_s$ efficiency}
Efficiency correction factors for $\pi_s$ originating from \dst\ decays are extracted separately for charged and neutral pions from studies of $B^0 \to D^{*-} (\to \bar{D}^0 \pi_s^-) \pi^+$ and $B^+ \to \bar{D}^{*0} (\to \bar{D}^0 \pi_s^0) \pi^+$ events. The correction factors are extracted in three laboratory-frame momentum bins between 50 and 200 MeV. Separately for charged and neutral $\pi_s$, covariance matrices are derived to encode the uncertainties versus momentum. This results in a total of six nuisance parameters.

\subsubsection{Tracking efficiency}
Track finding efficiency differences between data and simulation for tracks with a momentum higher than 200~MeV are studied in $\epem \to \tau^+\tau^-$ events with one $\tau$ decaying to three charged tracks. A correction factor consistent with unity with an associated uncertainty of 0.24\% per track is found. Therefore, the efficiency of each reconstructed track is varied up and down by 0.24\% in simulated events and the difference between the resulting up and down variations of each template is added as an uncertainty.

\subsubsection{Charged particle identification}
The efficiencies and misidentification rates of $e^\pm$, $\mu^\pm$ and $K^\pm$ are extracted from independent studies in bins of the track charge, laboratory-frame momentum and polar angle and applied to simulated events~\cite{Milesi:2020esq, HadronIDBelleII}. The correction factors, which combine efficiency and misidentification corrections, are typically in the range 0.95 to 1.02 for kaons and 0.90 to 1.03 for leptons. Each correction factor has associated statistical and systematic uncertainties. A total of 200 variations built from random samples are drawn from the probability density function for lepton and hadron identification factors separately. The number of parameters implemented in the fit is truncated from 400 to six in each fit.

\subsubsection{\kshort\ reconstruction efficiency}
The reconstruction efficiency of \kshort\ as a function of the flight distance from the interaction point is studied in $D^{*+} \to D^0 (\to \kshort \pi^+ \pi^-) \pi^+$ decays. An efficiency correction factor of 0.55\% per cm of flight distance is extracted. The correction factor is applied to each \kshort\ candidate to reweight simulated events and the full difference between the nominal and the corrected template is added as a symmetric uncertainty.

\subsubsection{Hadronic tagging}\label{sec:fei_correction}
Hadronic tagging efficiency correction factors are extracted from independent studies where the signal side $B$ meson is reconstructed in \btoxlnu\ and $B \to D^{(*)}\pi$ decays. The correction factors are extracted in 12 (11) separate tag-side $B^+$ ($B^0$) channels covering all decay modes used by the FEI algorithm and applied to simulated events. The number of parameters included in the fit is truncated from 23 to two.

\subsubsection{Continuum calibration}\label{sec:cont_calibration_error}
Simulated continuum events are corrected using the data-driven approach described in Sec.~\ref{sec:continuum_suppression}. The full difference between the nominal distribution and the distribution corrected by the normalization factors is added as an uncertainty. Uncertainties on the continuum component shape also arise because of the limited size of the samples used to extract the correction factors. A total of 200 bootstrapped variations of the off-resonance experimental and simulated data samples are trained separately yielding as many sets of correction factors. The number of parameters implemented in the fit is truncated from 200 to five.

\subsubsection{Finite size of simulated data samples}
To account for the limited size of simulated samples, one constrained nuisance parameter per fit bin is added.

\subsubsection{Fit template normalizations}
In fit 1, a normalization parameter is assigned to the \textit{other backgrounds} template, which contributes to the total signal branching fraction uncertainty. The nuisance parameter is constrained by a Gaussian function with a width of 50\% of the total component yields to vary the component normalization within a conservative range. In fits 2 and 3, the yields of nonsemileptonic events being negligible, this parameter is not added but other nuisance parameters related to this template can vary both its normalization and shape. Furthermore, the control region and signal region \btoxclnu\ template normalizations are corrected before the fit and they are allowed to float separately within the correction factor uncertainties given in Table~\ref{tab:xc_norms}.

\subsubsection{Number of $\Upsilon(4S)$ decays}
To extract a \btoxulnu\ branching fraction from the fit result, the output signal normalization is multiplied by the assumed input branching fraction. Since charged and neutral \btoxulnu\ decays are simulated separately but the measurement is performed simultaneously on all decays, the combined inclusive \btoxulnu\ input branching fraction is computed from the number of simulated signal events $N_{X_u}$ and the number of collected $\Upsilon(4S)$ events such that $\mathcal{B}(\btoxulnu) = N_{X_u} / (2N_{\Upsilon(4S)})$. The small number of $\Upsilon(4S)$ decays to final states without open-$b$ mesons has been neglected. The number of $\Upsilon(4S)$ events collected by Belle~II has been measured to be $N_{\Upsilon(4S)} = (387 \pm 6) \times 10^6$ and its uncertainty is therefore added to the total uncertainty of the measured branching fraction.

\subsubsection{$X_u$ hadronization modeling}\label{sec:xu_fragmentation_modelling}
An uncertainty on the fragmentation modeling of the $X_u$ system in nonresonant signal events is derived using the signal region $\pi^\pm$ multiplicity. It was checked beforehand in control regions that this variable is reasonably well modeled in \btoxclnu\ decays. In each of the three signal regions, the fit results are projected onto the $\pi^\pm$ multiplicity distribution and it is assumed that the remaining discrepancy between data and simulation is entirely due to the imperfect $X_u$ hadronization modeling. For each kinematic region, three scale factors are extracted in bins of the generator-level $\pi^\pm$ multiplicity of nonresonant \btoxulnu\ events (0, 1, 2 and more $\pi^\pm$) such that the simulated postfit $\pi^\pm$ multiplicity distribution matches perfectly the data distribution. The scale factors are applied to the pre-fit signal region distributions and the three fits are repeated. The difference between the measured \btoxulnu\ branching fraction and the branching fraction extracted from the rescaled distribution is taken as an uncertainty. Part of the uncertainty could originate from mismodeling of backgrounds in the $\pi^\pm$ multiplicity distribution. However, because of the relatively large signal purity in the three kinematic regions, the full difference is conservatively attributed to the \btoxulnu\ modeling.

\subsection{Signal extraction validation}\label{sec:fit_validation}
Before fitting signal region data, the signal extraction is validated using the regions where a kaon is found. A simulated \btoxulnu\ component with known normalization is injected into the CR\khigh\ experimental and simulated data and the CR\khigh\ and CR\klow\ regions are used as input to the fit. This allows the normalization of the signal injected into experimental data to be controlled while keeping the simulated data normalization fixed. The \btoxclnu\ normalization in the two regions is corrected before the fit, as is done in the actual signal fits.
\\
\\
The first test performed with this setup is to vary the normalization of the signal component injected into data in the range $0.85-1.15$ in steps of 0.05 while keeping the normalization of the signal in simulated data fixed. In this test, when the measured signal normalization is compared to the input, a slight bias is observed. The bias arises because of differences in shape and composition of the \btoxclnu\ component in the control and signal regions used in the fit. Because of these differences, the \btoxclnu\ shape correction extracted from the control region doesn't adjust perfectly the \btoxclnu\ shape in the signal region, thus forcing the fit to pull on other fit parameters, including the free-floating \btoxulnu\ normalization, to match the simulated templates to data. A correction for this bias is determined by using the response obtained in the validation setup and results in a $1-3\%$ correction for fits 1, 2 and 3 depending on the measured signal normalization.
\\
\\
Since this test relies on a single specific data configuration, an uncertainty is extracted by producing a large number of randomly sampled datasets. Each sampled dataset is created by varying the CR\khigh\ data yields within their statistical uncertainties while keeping the injected signal yields and the CR\klow\ data fixed. A set of 1,000 sampled datasets is produced for each injected signal normalization and each sampled dataset is fitted. The standard deviation of the resulting seven output signal normalization Gaussian distributions is nearly constant and the largest of the seven values for each set of kinematic selections is added as an uncertainty on the measured \btoxulnu\ branching fraction. This bias correction results in a 2.6\%, 2.3\% and 1.8\% relative uncertainty for fits 1, 2 and 3 respectively.
\\
\\
Since the correction and uncertainty are derived from the CR\khigh\ and CR\klow\ regions instead of CR\zlow\ and the signal region, an additional uncertainty related to the differences in \btoxclnu\ composition in each region is extracted. The \btoxclnu\ composition in the signal and control regions for the three sets of kinematic selections is shown in Appendix~\ref{app:xc_compo}, Table~\ref{tab:xc_compo}. The injected signal fit setup is repeated by fixing the normalization of the injected signal to unity and by rescaling the \btoxclnu\ composition of CR\khigh\ (CR\klow) such that it matches the composition of the signal region (CR\zlow). The difference in measured normalization with and without rescaling the composition is taken as an uncertainty, adding a 1.4\%, 5.7\% and 0.2\% contribution to the uncertainty of fits 1, 2 and 3, respectively. The difference in \btoxclnu\ composition between CR\khigh\ and the signal region in the phase-space region considered for fit 2 being relatively large (see Table~\ref{tab:xc_compo} in Appendix~\ref{app:xc_compo}), the composition uncertainty extracted for this fit is larger than in the other fits. Besides, in fit 3, the \elb\ distribution is not particularly sensitive to the \btoxclnu\ composition and this uncertainty is therefore relatively low. The \btoxclnu\ composition uncertainty is subleading in fits 1 and 3 but is one of the dominant sources in fit 2.

\subsection{Prefit distributions}\label{sec:prefit}
\begin{figure*}[t]
    \centering
    \includegraphics[scale=0.4]{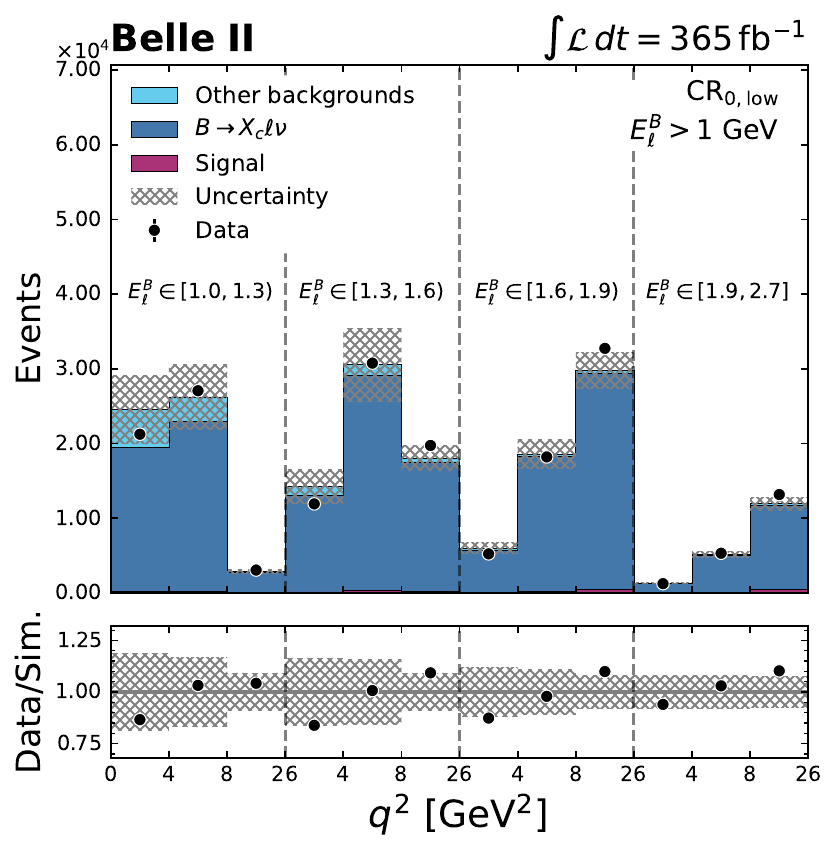}
    \includegraphics[scale=0.4]{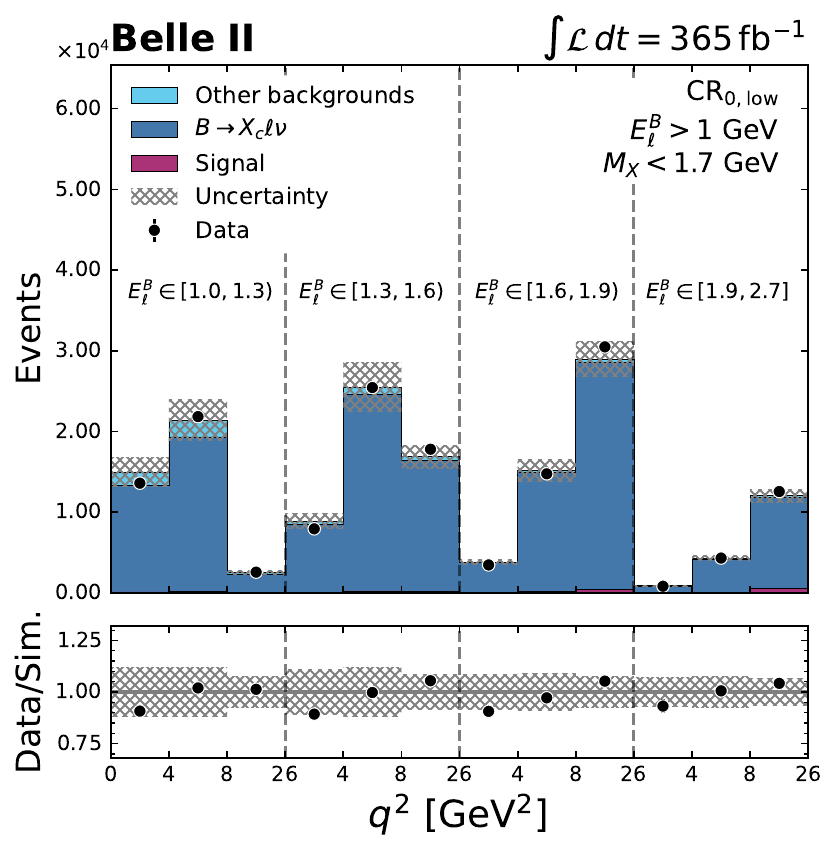}
    \includegraphics[scale=0.4]{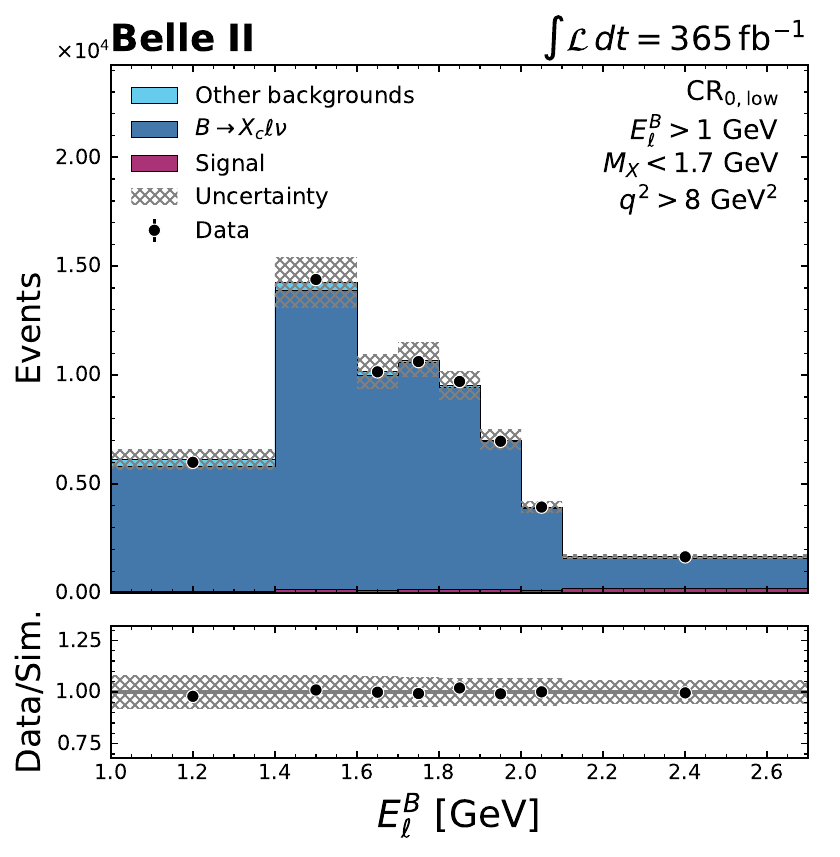}

    \includegraphics[scale=0.4]{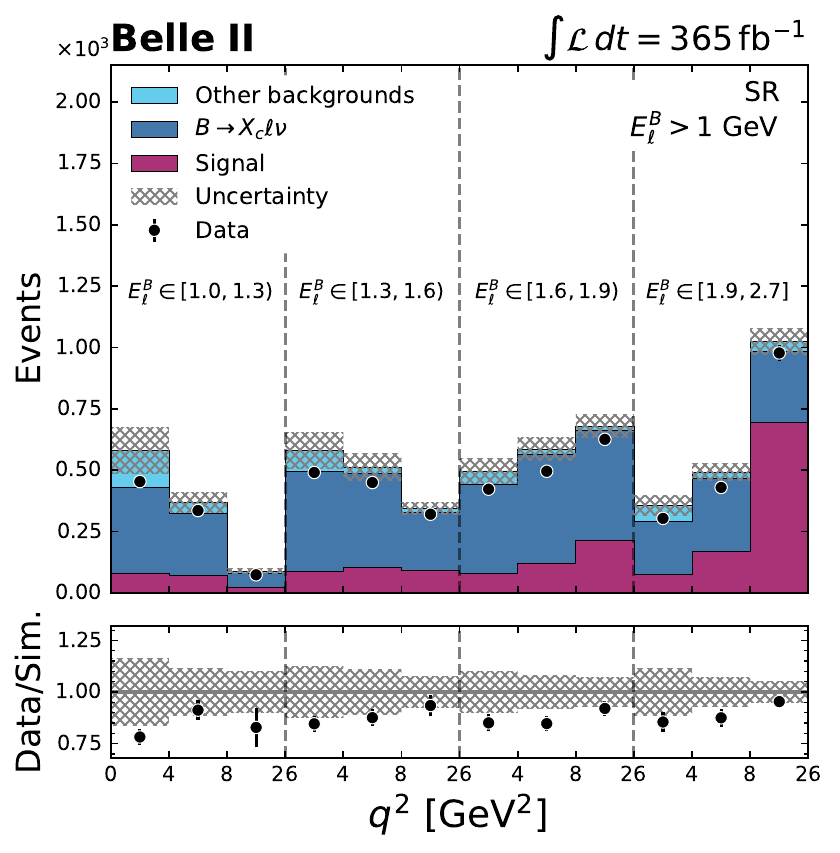}
    \includegraphics[scale=0.4]{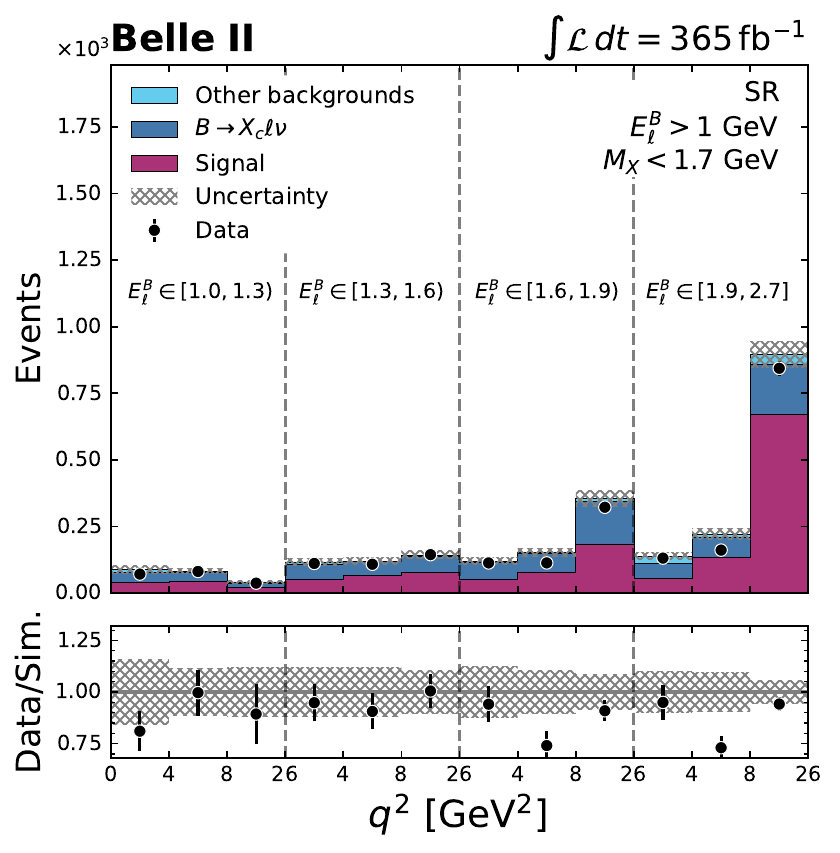}
    \includegraphics[scale=0.4]{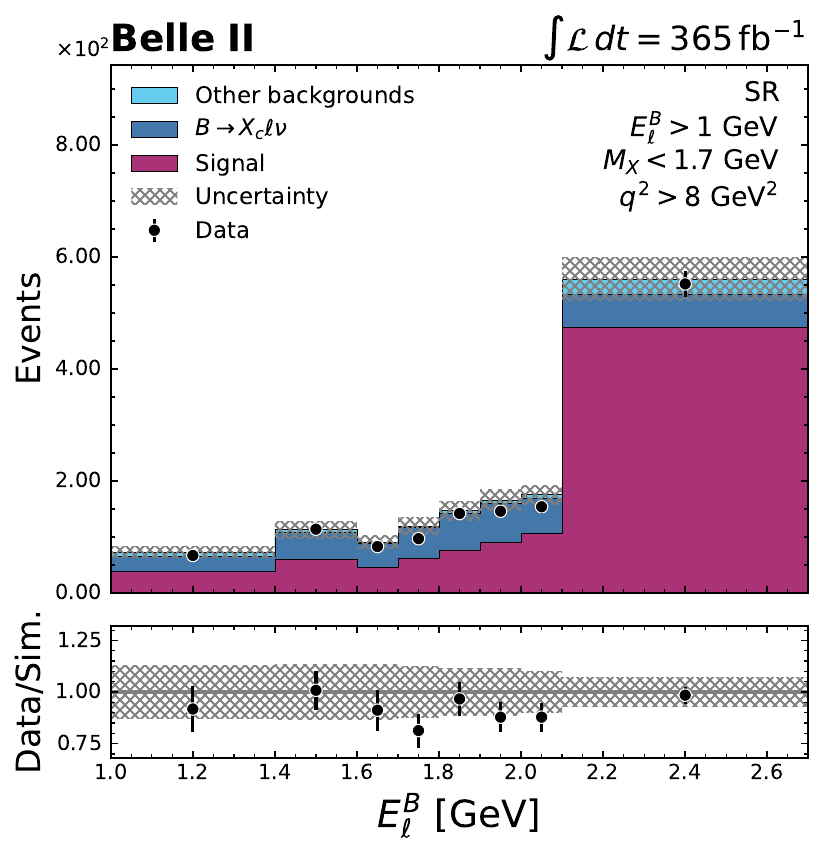}
    \caption{The \elb:\qtwo\ and \elb\ distributions in data and simulation before the fit in CR\zlow\ (top) and the signal region (bottom) considered for fit 1 (left), 2 (middle) and 3 (right). The \elb:\qtwo\ distribution is shown in bins of \elb\ (in GeV) and \qtwo\ as indicated in the figures. The \btoxclnu\ normalization correction factors have been applied (see Table~\ref{tab:xc_norms}). The signal-in and out components are merged in the \textit{Signal} template. The bottom panel shows the ratio between data and simulation yields in each bin.}
    \label{fig:fit_variables}
\end{figure*}

The three fitted variables in CR\zlow\ and the signal region are shown in Fig.~\ref{fig:fit_variables}. While the agreement between data and simulation appears to be relatively good in signal enriched regions, this is not the case in regions of the phase space where \btoxclnu\ events dominate. The shape mismodeling of this component is expected to be corrected by simultaneously fitting each distribution in the signal region and CR\zlow. However, the normalization of the charm background in the signal regions appears to be overestimated after applying the normalization correction factors ($cf.$ Table~\ref{tab:xc_norms}). To evaluate the impact of the overestimated \btoxclnu\ normalizations, the injected signal setup described in Sec.~\ref{sec:fit_validation} is used. The injected signal normalization is fixed to unity and the \btoxclnu\ template in CR\khigh\ is scaled up by an \textit{overestimation factor} in the range [1.05, 1.10, 1.15, 1.20]. After correcting for the bias described in Sec.~\ref{sec:fit_validation}, a slight bias increasing linearly with the \btoxclnu\ overestimation factor is observed. In the three signal regions considered, it is not possible to extract precisely the true \btoxclnu\ normalization without making assumptions about the \btoxulnu\ normalization, which is extracted from the fit. By checking the ratio of data to simulated data yields, it is estimated that the \btoxclnu\ normalization is overestimated by a factor roughly in the range $1.10-1.15$ in all three regions. A flat correction factor is therefore extracted assuming that the overestimation factor is equal to 1.15. The entire difference between the measured branching fraction and the corrected one is added as an uncertainty. The \btoxclnu\ normalization bias correction is tested with different correction factors assuming various overestimation factors in the range $1.00-1.30$. In all cases the assigned uncertainty is large enough to cover the true injected signal normalization after applying the bias correction.

\section{Results}\label{sec:results}
\subsection{Partial branching fractions}
\begin{figure*}[t]
    \centering
    \includegraphics[scale=0.4]{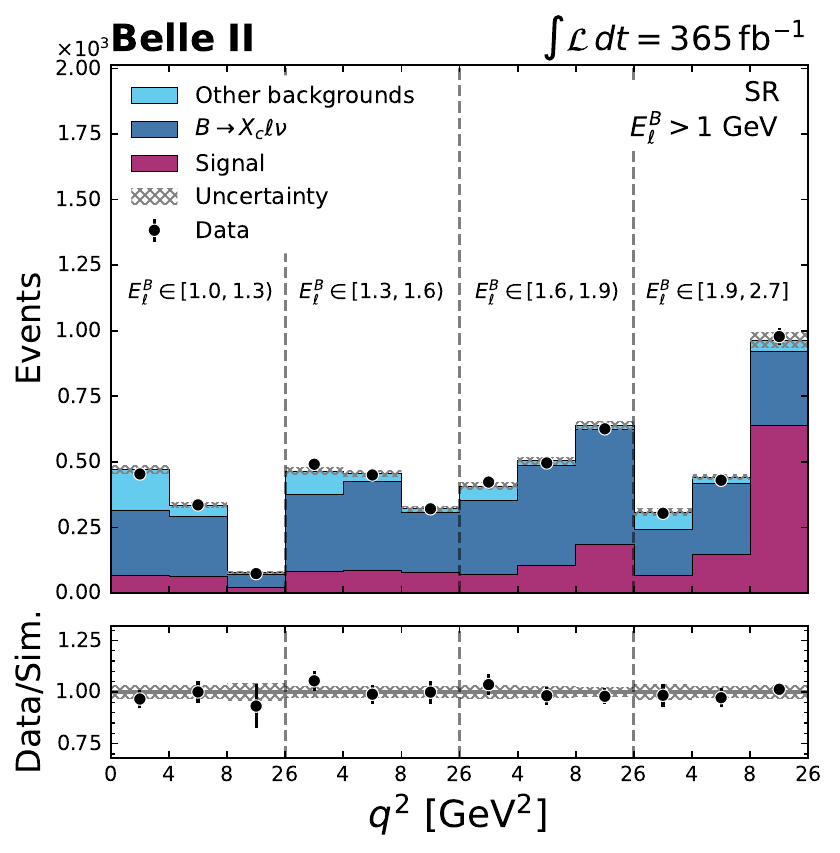}
    \includegraphics[scale=0.4]{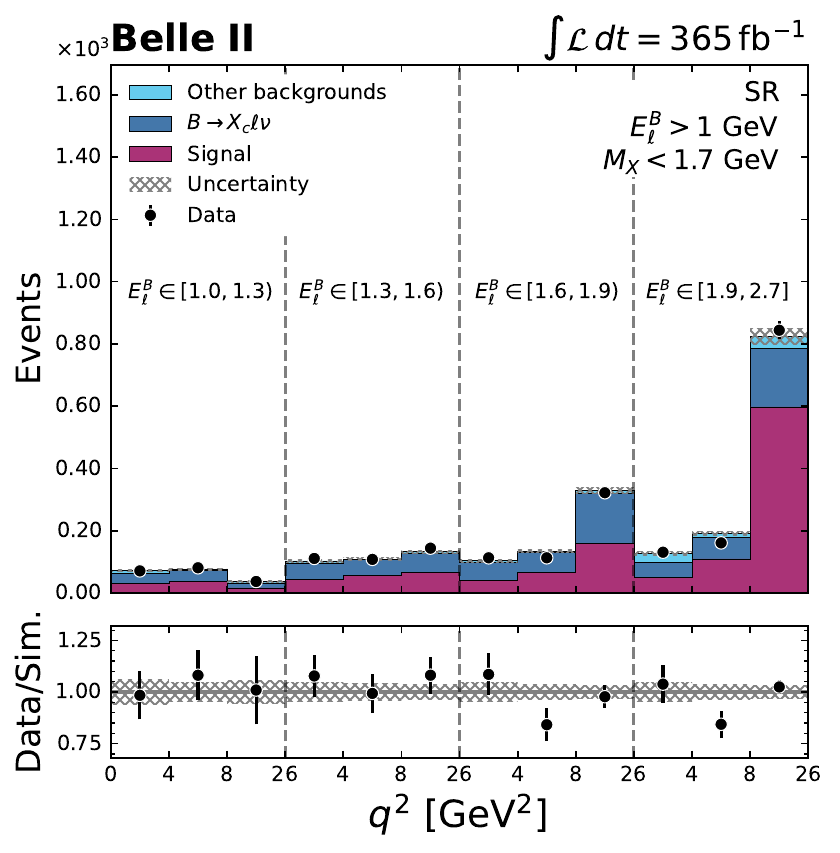}
    \includegraphics[scale=0.4]{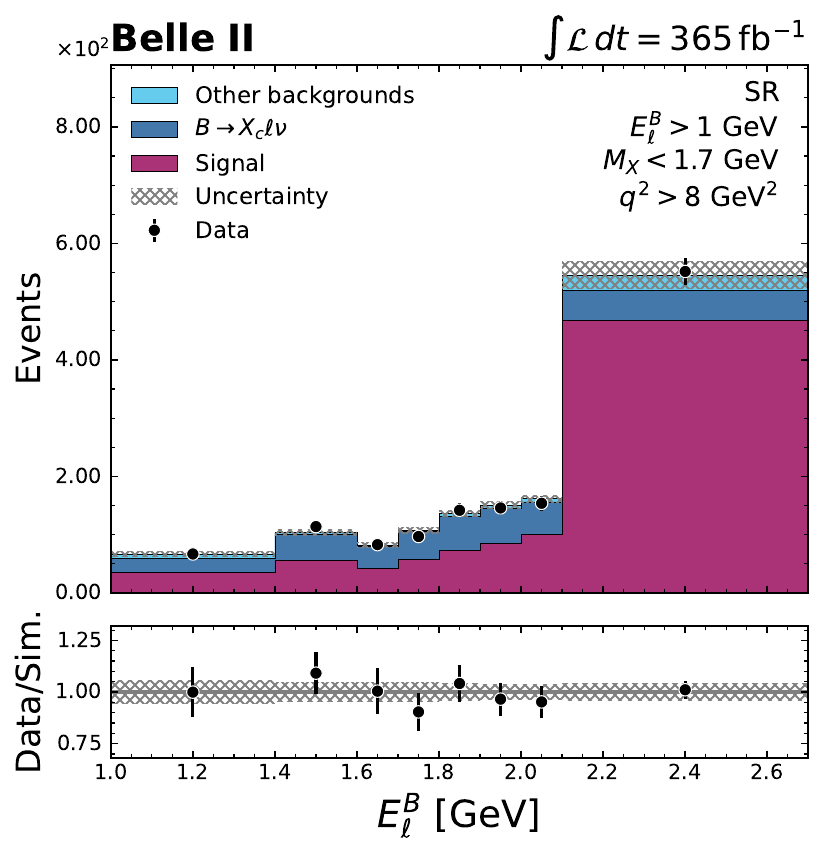}
    \caption{Postfit signal region distributions for fit 1 (left), 2 (middle) and 3 (right). The \elb:\qtwo\ variable is flattened in bins of \elb\ (in GeV) and \qtwo. The control region postfit distributions are not shown as the data/simulation agreement is almost perfect. The signal-in and out components are merged in the \textit{Signal} template. The bottom panel shows the ratio between data and simulation yields in each bin.}
    \label{fig:sr_variables_postfit}
\end{figure*}

The postfit distributions in each of the three signal regions are shown in Fig.~\ref{fig:sr_variables_postfit}. The excellent post-fit agreement is confirmed by the $p$-values obtained for each fit: 0.79, 0.20 and 0.84 for fits 1, 2 and 3, respectively. The event yields as extracted from fits 1, 2 and 3 are given in Table~\ref{tab:fit_yields}. In the kinematic regions considered for fits 1 and 2, the signal yields are similar to those reported in the latest Belle measurement~\cite{Belle:2021eni} while the background yields are approximately 30–45\% lower, despite using a dataset that is about half the size of the 711 fb$^{-1}$ dataset collected by the Belle detector. The higher signal reconstruction efficiency can be explained by the improvements in the FEI performance compared to the Full Reconstruction algorithm used in Belle~\cite{FEINDT2011432}, most notably the inclusion of more $B$ decay chains. Furthermore, in this measurement, the continuum and \btoxclnu\ backgrounds are suppressed by relying on two neural networks which are optimized separately whereas a single BDT was used for that purpose in the Belle measurement. Finally, the increased reconstruction efficiency for $\pi_s$ helps to reject a larger fraction of \btodstlnu\ events.
\begin{table}[h!]
    \centering
    \caption{\label{tab:fit_yields}Signal, background and data yields in the signal region as extracted from each fit. The error given for \btoxulnu\ and background yields is the total uncertainty as obtained from each fit. The three sets of kinematic selections used in each fit are given in Table~\ref{tab:fit_selections}.}
    \begin{tabular}{ccccc}
    \hline
    \hline
        Fit & \btoxulnu\ in & \btoxulnu\ out & Backgrounds & Data \\[0.5ex]
        Fit 1 & $1615 \pm 122$ & -- & $3782 \pm 128$ & $5383 \pm 73$ \\[0.5ex]
        Fit 2 & $1240 \pm 105$ & $33 \pm 5$ & $\phantom{1}965 \pm \phantom{1}95$ & $2236 \pm 47$ \\[0.5ex]
        Fit 3 & $\phantom{1}876 \pm \phantom{1}68$ & $42 \pm 4$ & $\phantom{1}439 \pm \phantom{1}62$ & $1355 \pm 37$ \\
    \hline
    \hline
    \end{tabular}
\end{table}
\\
\\
The inclusive \btoxulnu\ partial branching fractions obtained from fits 1, 2 and 3 are
\begin{eqnarray}
    \pBR(\btoxulnu)_1 &= (1.54 \pm 0.08 \pm 0.12) \times 10^{-3}, \label{equ:result_br1} \\
    \pBR(\btoxulnu)_2 &= (0.95 \pm 0.05 \pm 0.10) \times 10^{-3}, \label{equ:result_br3} \\
    \pBR(\btoxulnu)_3 &= (0.55 \pm 0.03 \pm 0.05) \times 10^{-3}, \label{equ:result_br5}
\end{eqnarray}
where the uncertainties are statistical and systematic, respectively. With a relative uncertainty of 9.6\%, the measurement of the branching fraction over the phase-space region defined by the selection $\elb > 1$~GeV (fit 1) is competitive with the most precise measurement performed by the BaBar collaboration (8.1\% relative uncertainty) using an inclusive tagging method~\cite{BaBar:2016rxh}. It is more precise than similar measurements performed by the Belle (11.4\%) and BaBar (14.5\%) collaborations using a hadronic tagging method~\cite{BaBar:2011xxm, Belle:2021eni}. With relative uncertainties of 11.9\% and 10.7\%, the measurements of the branching fraction over the two tighter phase-space regions are competitive with the most precise determinations from these two experiments. When extrapolated to the full phase-space region, the three measured \btoxulnu\ branching fractions are $(1.78 \pm 0.17) \times 10^{-3}$, $(1.67 \pm 0.20) \times 10^{-3}$ and $(1.76 \pm 0.19) \times 10^{-3}$ which can be compared to the latest average value listed in the PDG Review of Particle Physics~\cite{PhysRevD.110.030001}, $(1.92 \pm 0.21) \times 10^{-3}$.\footnote{The uncertainty on the value quoted by HFLAV is inflated based on theoretical errors. The branching fractions we quote are simply scaled by the phase-space acceptances given in Table~\ref{tab:fit_selections}. They lack theoretical uncertainties related to this procedure and are therefore simply quoted here as a point of comparison.} The three values are correlated. Assuming a systematic correlation of 100\% and estimating the statistical correlation from bootstrapping, it was checked that the values agree with each other within one standard deviation. The correlation matrix of the statistical uncertainties is given in Table~\ref{tab:stat_correlation}. The statistical correlation is similar between each pair of fits because there is a large overlap in signal events and because, in all three fits, the measurement is sensitive to the region with best signal purity $i.e.$ the region with high \elb\ and \qtwo.
\begin{table}[h!]
    \centering
    \caption{\label{tab:stat_correlation} Correlations between statistical uncertainties of the branching fractions measured in fits 1, 2 and 3.}
    \begin{tabular}{c|ccc}
         \hline
         \hline
               & Fit 1 & Fit 2 & Fit 3 \\[0.8ex]
               \hline
         Fit 1 & 1     & 0.86  & 0.81 \\[0.8ex]
         Fit 2 & 0.86  & 1     & 0.83 \\[0.8ex]
         Fit 3 & 0.81  & 0.83  & 1    \\[0.8ex]
         \hline
         \hline
    \end{tabular}
\end{table}
\\
\\
The total uncertainty breakdown in terms of individual sources of uncertainty is shown in Table~\ref{tab:syst_uncert_breakdown} and the numerical impact of uncertainties related to the bias correction and the sample composition on the measured partial branching fractions is summarized in Table~\ref{tab:tot_uncert_breakdown}. These individual uncertainties are intended to provide insight into the sensitivity of the measurement to inputs and assumptions. However, it is not possible to account for correlations between fit parameters when calculating the individual uncertainties and the numbers quoted are therefore approximate. Hence, the sum in quadrature of all individual sources of uncertainty does not necessarily match the total uncertainty obtained from the fit. For all three measured branching fractions, the DFN model parameter uncertainty represents one of the largest sources of uncertainty. It is larger in fit 3 than in the other two fits because the \btoxulnu\ acceptance is the smallest among the three kinematic regions making the dependence on the model stronger. One of the leading uncertainties of the branching fractions measured with fits 2 and 3 is the hadronic system fragmentation modeling uncertainty. The result obtained from fit 2 is significantly impacted by the \btoxclnu\ composition uncertainty described in Sec.~\ref{sec:fit_validation} because the relative fractions of \btoxclnu\ subcomponents are considerably different between the signal region and CR\khigh\ when applying the \elb\ and \mx\ selections (see Table~\ref{tab:xc_compo}).
\begin{table*}[t]
    \caption{\label{tab:syst_uncert_breakdown}Breakdown of the systematic uncertainties of the partial branching fraction obtained from each fit.}
    \begin{ruledtabular}
    \begin{tabular}{cccc}
         & \multicolumn{3}{c}{Relative uncertainty (\%)} \\[1.9ex]
        Uncertainty source & Fit 1 & Fit 2 & Fit 3 \\[1ex]
        \hline
        DFN parameters & 4.4 & 4.5 & 5.7 \\
        DFN $\to$ BLNP & 0.2 & 0.8 & 1.3 \\
        $\gamma_S$ & 1.7 & 2.1 & 2.1 \\[1.9ex]
        \btopilnu\ form factors & 0.3 & 0.3 & 0.3 \\
        \btorholnu\ form factors & 0.3 & 0.3 & 0.2 \\ 
        \btoomegalnu\ form factors & 0.1 & 0.1 & 0.1 \\ 
        \btoetaetaplnu\ form factors & $< 0.1$ & $< 0.1$ & $< 0.1$ \\ [1.9ex]
        $B^{\pm} \to X_u \ell \nu$ branching fractions & 0.9 & 0.6 & 0.5 \\
        $B^0 \to X_u \ell \nu$ branching fractions & 0.6 & 0.5 & 0.5 \\[1.9ex]
        $B \to D^{**}_{\rm Broad}$ form factors & 0.5 & 0.1 & 0.2 \\ 
        $B \to D^{**}_{\rm Narrow}$ form factors & 0.1 & $< 0.1$ & $< 0.1$ \\
        $B \to D/D^* \ell \nu$ form factors & $< 0.1$ & $< 0.1$ & $< 0.1$ \\[1.9ex]
        $B^{\pm} \to X_c \ell \nu$ branching fractions & 0.7 & 0.5 & 0.2 \\ 
        $B^0 \to X_c \ell \nu$ branching fractions & 0.6 & 0.2 & 0.1 \\ 
        $D$ decay branching fractions & 0.1 & 0.3 & 0.1 \\ [1.9ex]
        SR \xclnu\ normalization & 1.6 & 3.5 & 3.4 \\
        CR \xclnu\ normalization & 0.9 & 1.1 & 0.4 \\
        \textit{Other backgrounds} normalization & 0.3 & N/A & N/A \\[1.9ex]
        $X_u$ fragmentation & 0.3 & 4.4 & 3.9 \\
        $N_{\Upsilon(4S)}$ & 1.4 & 1.4 & 1.4 \\
        FEI & 1.3 & 1.3 & 1.4 \\
        Slow pion efficiency & 0.4 & 0.2 & 0.3 \\
        $\ell$ identification & 0.7 & 0.7 & 0.6 \\
        $f^{\pm/00}$ & 0.6 & 0.7 & 0.6 \\ 
        Continuum calibration & 0.2 & 0.2 & 0.2 \\ 
        Tracking & 0.3 & 0.3 & 0.3 \\ 
        \kshort\ efficiency & 0.1 & 0.1 & $< 0.1$ \\ 
        $K^{\pm}$ ID & $< 0.1$ & $< 0.1$ & $< 0.1$ \\ 
        Simulated data statistics & 1.1 & 1.1 & 0.8 \\
\end{tabular}
\end{ruledtabular}
\end{table*}
\begin{table*}[t]
    \caption{\label{tab:tot_uncert_breakdown}Breakdown of uncertainties added following the bias corrections described in Secs~\ref{sec:fit_validation} and~\ref{sec:prefit}. In addition, we show the systematic, statistical and total uncertainties of the corrected branching fractions.}
    \begin{ruledtabular}
    \begin{tabular}{cccc}
         & \multicolumn{3}{c}{Relative uncertainty (\%)} \\[1.9ex]
         & Fit 1 & Fit 2 & Fit 3 \\[1ex]
        \hline
        Fit bias & 2.6 & 2.3 & 1.8 \\
        Composition uncertainty & 1.3 & 5.7 & 0.2 \\
        \btoxclnu\ overestimation correction & 2.6 & 0.4 & 1.3 \\[1.9ex]
        Total systematic & 7.9 & 10.5 & 9.7 \\ 
        Statistical & 5.4 & 5.6 & 4.5  \\ 
        Total & 9.6 & 11.9 & 10.7 \\ 
\end{tabular}
\end{ruledtabular}
\end{table*}

\subsection{$|\vub|$ determination}
The value of $|\vub|$ can be determined from the measured partial branching fractions using Eq.~\ref{equ:vub_extraction}. We use for the value of the $B$ meson lifetime, the average of the $B^+$ and $B^0$ lifetimes, $\tau_B = 1.578 \pm 0.003$ ps~\cite{hflav_2023}. Following Ref.~\cite{hflav_2023}, three different theoretical frameworks are used to extract $|\vub|$:
\begin{itemize}
    \item the BLNP model~\cite{Lange:2005yw, BOSCH2004335} with input values in the shape-function (SF) renormalization scheme~\cite{NEUBERT200513, PhysRevD.72.074025}: $m_b^{\rm SF} = 4.600 \pm 0.022$~GeV and $\mu_\pi^{2 \, \rm SF} = 0.184^{+0.051}_{-0.062}$~GeV$^2$;
    \item the Dressed Gluon Exponentiation (DGE) model~\cite{Andersen:2005mj} with input values in the $\overline{\rm MS}$ scheme: $m_b^{\overline{\rm MS}} = 4.206 \pm 0.040$~GeV;
    \item the Gambino, Giordano, Ossola and Uraltsev (GGOU) model~\cite{Gambino:2007rp} with input values in the kinetic scheme~\cite{PhysRevD.56.4017}: $m_b^{\rm kin} = 4.573 \pm 0.012$~GeV and $\mu_\pi^{2 \, \rm kin} = 0.454 \pm 0.043$~GeV$^2$.
\end{itemize}
The values of $|\vub|$ obtained from each model and each signal extraction fit using Eq.~\ref{equ:vub_extraction} are summarized in Table~\ref{tab:decay_rate_vub}. Theoretical uncertainties, $e.g.$, those due to the assumed value of the $b$ quark mass, impact both the determination of the branching fraction and the calculation of $|\vub|$. However, the correlations between these quantities are difficult to determine due to the different parameterizations of nonperturbative effects used in the different theoretical calculations of $|\vub|$ and in the DFN model used in our simulation. Therefore, similarly to what is done in the HFLAV averaging calculation, correlations between the branching fraction and partial decay rate theoretical uncertainties are ignored when extracting $|\vub|$.
\begin{table*}
    \caption{\label{tab:decay_rate_vub}Theoretical \btoxulnu\ decay rates $\Delta\tilde{\Gamma}$ (excluding the $|\vub|^2$ term) and $|\vub|$ obtained from three different theoretical predictions in three separate phase-space regions. The values of $\Delta\tilde{\Gamma}$ are used to compute $|\vub|$ following Eq.~\ref{equ:vub_extraction}. The uncertainties on $|\vub|$ are statistical, systematic and theoretical, respectively.}
    \begin{ruledtabular}
    \begin{tabular}{ccccc}
         & Inclusive \btoxulnu\ model & Fit 1 & Fit 2 & Fit 3 \\[1.0ex]
        \hline
        & & & & \\
        & BLNP & $62.9 ^{+5.3} _{-5.9}$ & $47.5 ^{+4.5} _{-4.8}$ & $24.4 ^{+2.6} _{-2.9}$ \\[1.5ex]
        $\Delta\tilde{\Gamma}$ (ps$^{-1}$) & DGE & $59.4 ^{+2.9} _{-3.2}$ & $43.1 ^{+4.5} _{-4.2}$ & $25.0 ^{+1.8} _{-2.0}$ \\[1.5ex]
        & GGOU & $60.8 ^{+2.2} _{-2.6}$ & $47.6 ^{+3.0} _{-3.2}$ & $25.0 ^{+2.2} _{-2.6}$ \\[1.5ex]
        \hline
        & & & & \\
        & BLNP & $3.94 \pm 0.11 \pm 0.16 ^{+0.17} _{-0.18}$ & $3.55 \pm 0.10 \pm 0.19 ^{+0.17} _{-0.18}$ & $3.79 \pm 0.09 \pm 0.18 ^{+0.20} _{-0.22}$ \\[1.5ex]
        $|\vub| \times 10^3$ & DGE & $4.05 \pm 0.11 \pm 0.16 ^{+0.10} _{-0.11}$ & $3.73 \pm 0.10 \pm 0.20 ^{+0.19} _{-0.18}$ & $3.75 \pm 0.08 \pm 0.18 ^{+0.13} _{-0.15}$ \\[1.5ex]
        & GGOU & $4.01 \pm 0.11 \pm 0.16 ^{+0.07} _{-0.08}$ & $3.55 \pm 0.10 \pm 0.19 ^{+0.11} _{-0.12}$ & $3.75 \pm 0.08 \pm 0.18 ^{+0.17} _{-0.20}$ \\[1.5ex]
\end{tabular}
\end{ruledtabular}
\end{table*}

\section{Conclusions}\label{sec:conclusions}
\begin{figure}[h!]
    \centering
    \includegraphics[scale=0.4]{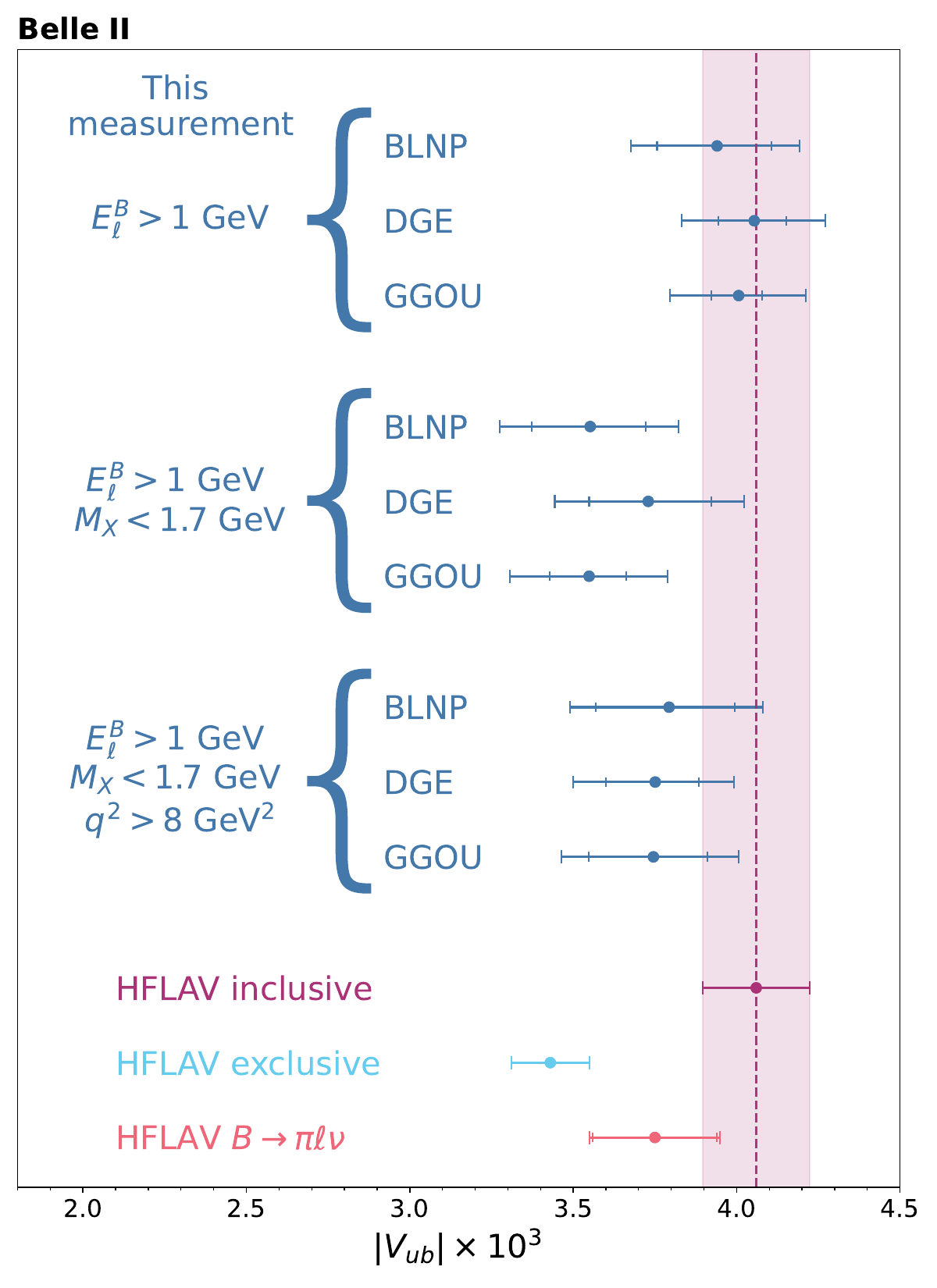}
    \caption{Comparison between the three values of $|\vub|$ obtained from fits 1, 2 and 3 (blue), the inclusive (purple band), exclusive (cyan) and \btopilnu\ (pink) averages quoted in the latest HFLAV report~\cite{hflav_2023}. The outer error bars represent the total uncertainty and, where shown, the inner error bars represent the contribution of the theoretical uncertainty.}
    \label{fig:vub_comparison}
\end{figure}

A measurement of the partial branching fraction of inclusive charmless semileptonic decays in three phase-space regions is reported. The three regions are defined by selections on the kinematic properties of \btoxulnu\ decays and cover, respectively, 87\%, 57\% and 31\% of the full phase space. These kinematic selections, in addition to the use of neural networks, help to suppress various sources of background and in particular the contribution of CKM-favored \btoxclnu\ decays. The partner $B$ meson from the $\Upsilon(4S)$ decay is reconstructed to constrain the kinematics of the signal-side decays and precisely determine several variables which play a key role in the measurement such as \qtwo\ and the variables used to suppress the \btoxclnu\ background. The reconstruction and selection strategy developed for this analysis leads to a relatively higher signal efficiency compared to previous similar measurements such as Ref.~\cite{Belle:2021eni}. The broadest phase-space region, where inclusive \btoxulnu\ theoretical predictions are most reliable, is defined by a selection $\elb > 1$~GeV. The partial branching fraction in this region is measured from a fit to the two-dimensional variable \elb:\qtwo\ and the obtained value is
\begin{equation}
    \pBR(\btoxulnu) = (1.54 \pm 0.08 \pm 0.12) \times 10^{-3},
\end{equation}
where the uncertainties are statistical and systematic, respectively. From this branching fraction, the value of $|\vub|$ obtained using a partial decay rate predicted by the GGOU framework is
\begin{equation}
    |\vub| = (4.01 \pm 0.11 \pm 0.16 ^{+0.07} _{-0.08}) \times 10^{-3},
\end{equation}
where the uncertainties are statistical, systematic and theoretical, respectively. The GGOU framework is used to allow a consistent comparison with the inclusive $|\vub|$ average reported by HFLAV, with which our result is compatible~\cite{hflav_2023}. The measured value also agrees within uncertainties with the average value obtained from measurements using \btopilnu\ decays but it exceeds the HFLAV exclusive average. The values of $|\vub|$ extracted from the three phase-space regions are compared to the latest inclusive and exclusive HFLAV averages in Fig.~\ref{fig:vub_comparison}.

\begin{acknowledgments}
The authors would like to thank Keri Vos for the useful discussions related to the theoretical description of semileptonic $B$ decays.
This work, based on data collected using the Belle II detector, which was built and commissioned prior to March 2019,
was supported by
Higher Education and Science Committee of the Republic of Armenia Grant No.~23LCG-1C011;
Australian Research Council and Research Grants
No.~DP200101792, 
No.~DP210101900, 
No.~DP210102831, 
No.~DE220100462, 
No.~LE210100098, 
and
No.~LE230100085; 
Austrian Federal Ministry of Education, Science and Research,
Austrian Science Fund (FWF) Grants
DOI:~10.55776/P34529,
DOI:~10.55776/J4731,
DOI:~10.55776/J4625,
DOI:~10.55776/M3153,
and
DOI:~10.55776/PAT1836324,
and
Horizon 2020 ERC Starting Grant No.~947006 ``InterLeptons'';
Natural Sciences and Engineering Research Council of Canada, Digital Research Alliance of Canada, and Canada Foundation for Innovation;
National Key R\&D Program of China under Contract No.~2024YFA1610503,
and
No.~2024YFA1610504
National Natural Science Foundation of China and Research Grants
No.~11575017,
No.~11761141009,
No.~11705209,
No.~11975076,
No.~12135005,
No.~12150004,
No.~12161141008,
No.~12405099,
No.~12475093,
and
No.~12175041,
and Shandong Provincial Natural Science Foundation Project~ZR2022JQ02;
the Czech Science Foundation Grant No. 22-18469S,  Regional funds of EU/MEYS: OPJAK
FORTE CZ.02.01.01/00/22\_008/0004632 
and
Charles University Grant Agency project No. 246122;
European Research Council, Seventh Framework PIEF-GA-2013-622527,
Horizon 2020 ERC-Advanced Grants No.~267104 and No.~884719,
Horizon 2020 ERC-Consolidator Grant No.~819127,
Horizon 2020 Marie Sklodowska-Curie Grant Agreement No.~700525 ``NIOBE''
and
No.~101026516,
and
Horizon 2020 Marie Sklodowska-Curie RISE project JENNIFER2 Grant Agreement No.~822070 (European grants);
L'Institut National de Physique Nucl\'{e}aire et de Physique des Particules (IN2P3) du CNRS
and
L'Agence Nationale de la Recherche (ANR) under Grant No.~ANR-23-CE31-0018 (France);
BMFTR, DFG, HGF, MPG, and AvH Foundation (Germany);
Department of Atomic Energy under Project Identification No.~RTI 4002,
Department of Science and Technology,
and
UPES SEED funding programs
No.~UPES/R\&D-SEED-INFRA/17052023/01 and
No.~UPES/R\&D-SOE/20062022/06 (India);
Israel Science Foundation Grant No.~2476/17,
U.S.-Israel Binational Science Foundation Grant No.~2016113, and
Israel Ministry of Science Grant No.~3-16543;
Istituto Nazionale di Fisica Nucleare and the Research Grants BELLE2,
and
the ICSC – Centro Nazionale di Ricerca in High Performance Computing, Big Data and Quantum Computing, funded by European Union – NextGenerationEU;
Japan Society for the Promotion of Science, Grant-in-Aid for Scientific Research Grants
No.~16H03968,
No.~16H03993,
No.~16H06492,
No.~16K05323,
No.~17H01133,
No.~17H05405,
No.~18K03621,
No.~18H03710,
No.~18H05226,
No.~19H00682, 
No.~20H05850,
No.~20H05858,
No.~22H00144,
No.~22K14056,
No.~22K21347,
No.~23H05433,
No.~26220706,
and
No.~26400255,
and
the Ministry of Education, Culture, Sports, Science, and Technology (MEXT) of Japan;  
National Research Foundation (NRF) of Korea Grants
No.~2021R1-F1A-1064008, 
No.~2022R1-A2C-1003993,
No.~2022R1-A2C-1092335,
No.~RS-2016-NR017151,
No.~RS-2018-NR031074,
No.~RS-2021-NR060129,
No.~RS-2023-00208693,
No.~RS-2024-00354342
and
No.~RS-2025-02219521,
Radiation Science Research Institute,
Foreign Large-Size Research Facility Application Supporting project,
the Global Science Experimental Data Hub Center, the Korea Institute of Science and
Technology Information (K25L2M2C3 ) 
and
KREONET/GLORIAD;
Universiti Malaya RU grant, Akademi Sains Malaysia, and Ministry of Education Malaysia;
Frontiers of Science Program Contracts
No.~FOINS-296,
No.~CB-221329,
No.~CB-236394,
No.~CB-254409,
and
No.~CB-180023, and SEP-CINVESTAV Research Grant No.~237 (Mexico);
the Polish Ministry of Science and Higher Education and the National Science Center;
the Ministry of Science and Higher Education of the Russian Federation
and
the HSE University Basic Research Program, Moscow;
University of Tabuk Research Grants
No.~S-0256-1438 and No.~S-0280-1439 (Saudi Arabia), and
Researchers Supporting Project number (RSPD2025R873), King Saud University, Riyadh,
Saudi Arabia;
Slovenian Research Agency and Research Grants
No.~J1-50010
and
No.~P1-0135;
Ikerbasque, Basque Foundation for Science,
State Agency for Research of the Spanish Ministry of Science and Innovation through Grant No. PID2022-136510NB-C33, Spain,
Agencia Estatal de Investigacion, Spain
Grant No.~RYC2020-029875-I
and
Generalitat Valenciana, Spain
Grant No.~CIDEGENT/2018/020;
The Knut and Alice Wallenberg Foundation (Sweden), Contracts No.~2021.0174, No.~2021.0299, and No.~2023.0315;
National Science and Technology Council,
and
Ministry of Education (Taiwan);
Thailand Center of Excellence in Physics;
TUBITAK ULAKBIM (Turkey);
National Research Foundation of Ukraine, Project No.~2020.02/0257,
and
Ministry of Education and Science of Ukraine;
the U.S. National Science Foundation and Research Grants
No.~PHY-1913789 
and
No.~PHY-2111604, 
and the U.S. Department of Energy and Research Awards
No.~DE-AC06-76RLO1830, 
No.~DE-SC0007983, 
No.~DE-SC0009824, 
No.~DE-SC0009973, 
No.~DE-SC0010007, 
No.~DE-SC0010073, 
No.~DE-SC0010118, 
No.~DE-SC0010504, 
No.~DE-SC0011784, 
No.~DE-SC0012704, 
No.~DE-SC0019230, 
No.~DE-SC0021274, 
No.~DE-SC0021616, 
No.~DE-SC0022350, 
No.~DE-SC0023470; 
and
the Vietnam Academy of Science and Technology (VAST) under Grants
No.~NVCC.05.02/25-25
and
No.~DL0000.05/26-27.

These acknowledgements are not to be interpreted as an endorsement of any statement made
by any of our institutes, funding agencies, governments, or their representatives.

We thank the SuperKEKB team for delivering high-luminosity collisions;
the KEK cryogenics group for the efficient operation of the detector solenoid magnet and IBBelle on site;
the KEK Computer Research Center for on-site computing support; the NII for SINET6 network support;
and the raw-data centers hosted by BNL, DESY, GridKa, IN2P3, INFN, 
and the University of Victoria.

\end{acknowledgments}

\appendix

\section{\btoxclnu\ suppression classifier input features}\label{app:xclnu_suppression}
\begin{figure*}[h!]
    \centering
    \includegraphics[scale=0.35]{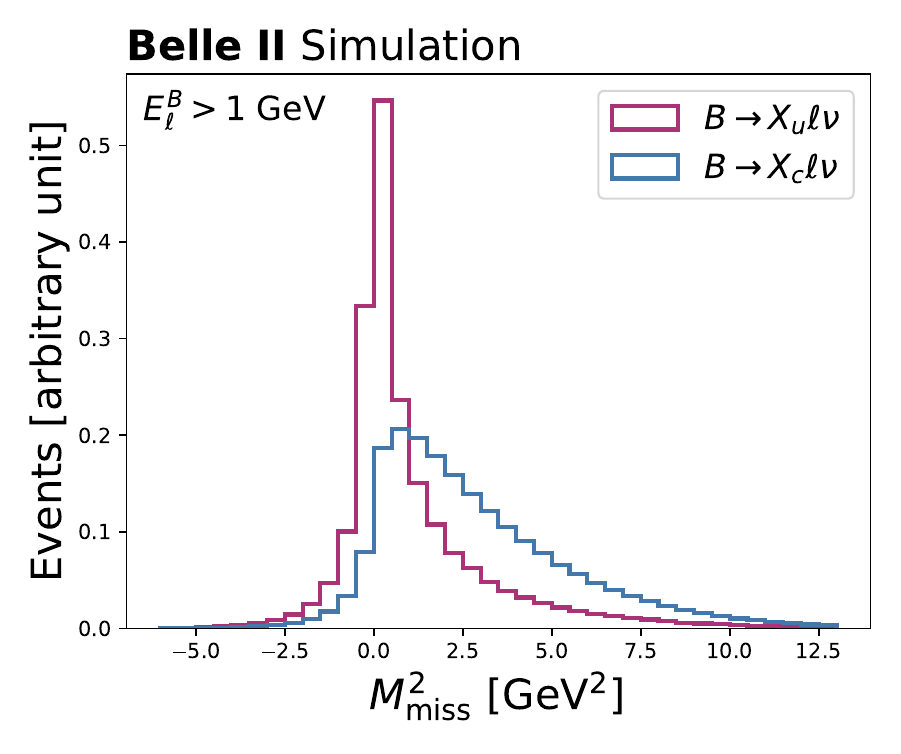}
    \includegraphics[scale=0.35]{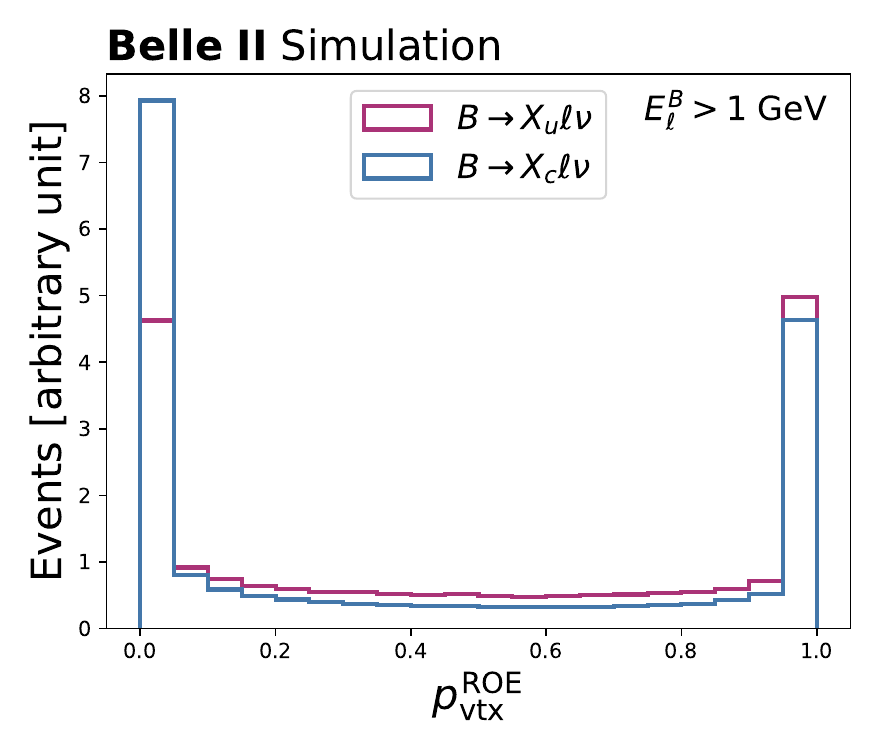}
    \includegraphics[scale=0.35]{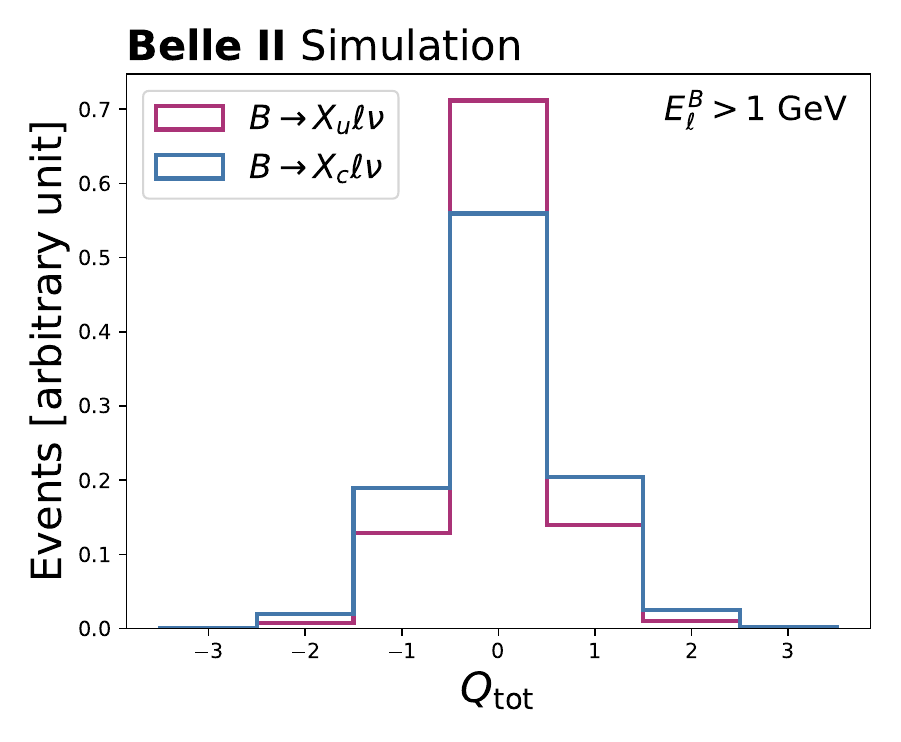}
    \includegraphics[scale=0.35]{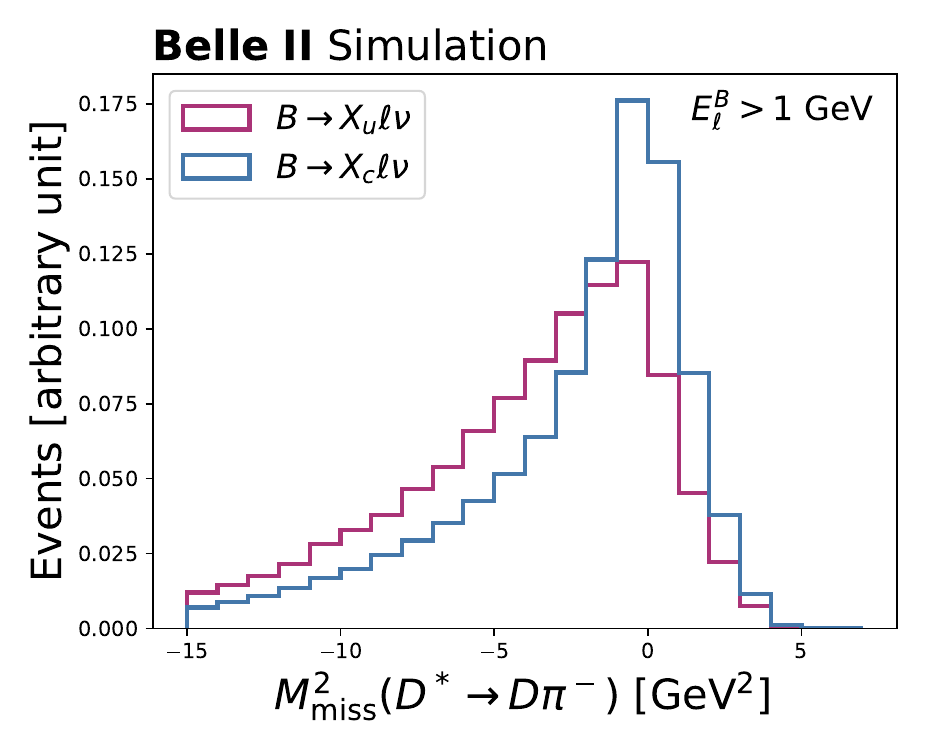}
    \includegraphics[scale=0.35]{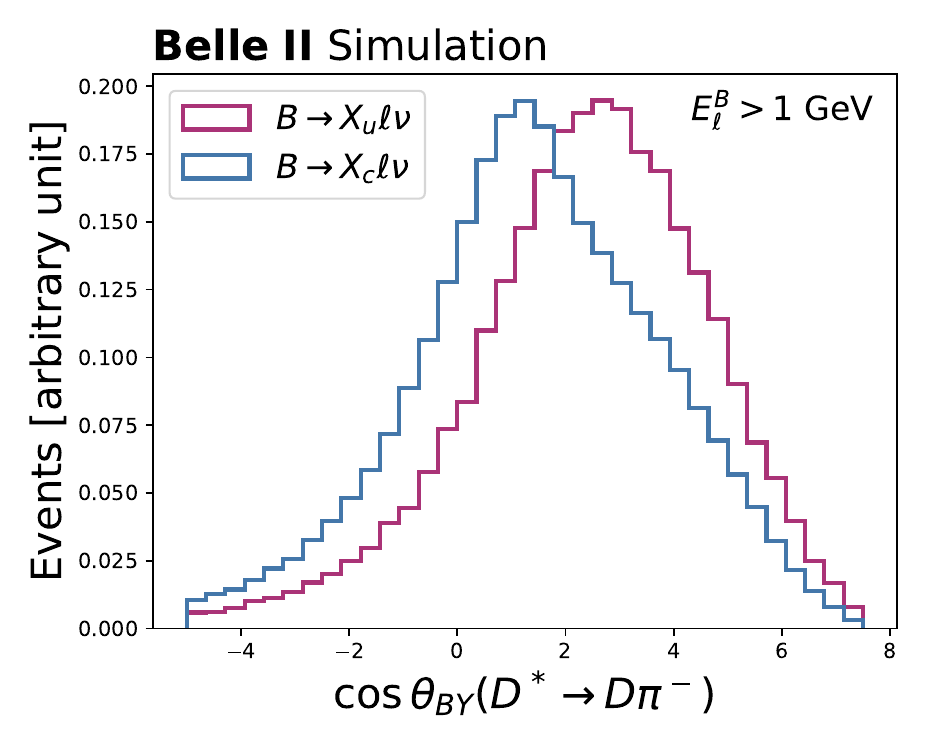}
    \includegraphics[scale=0.35]{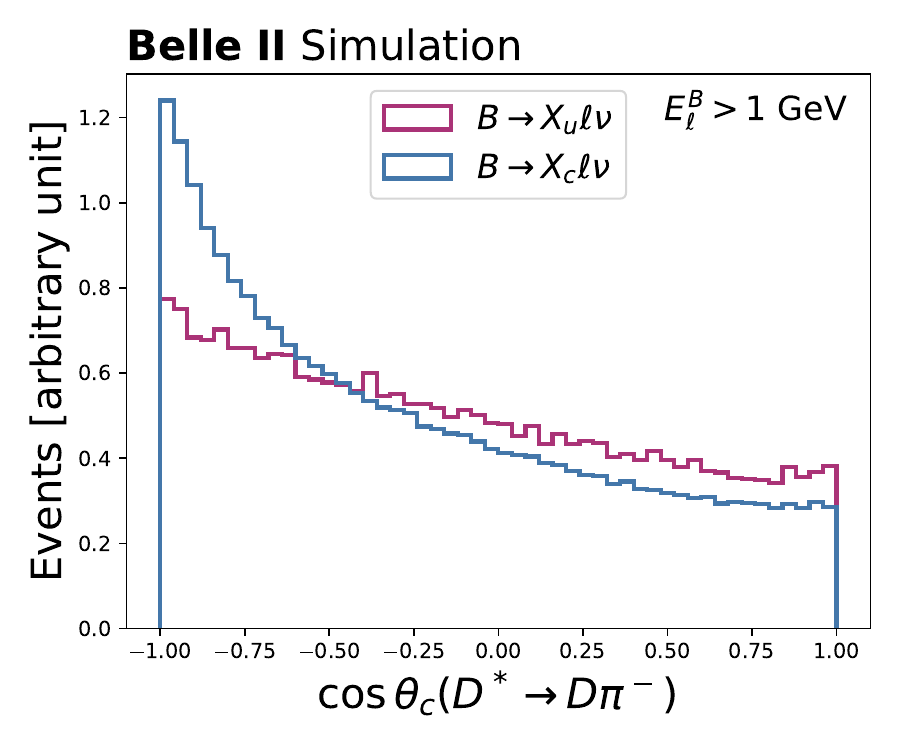}
    \includegraphics[scale=0.35]{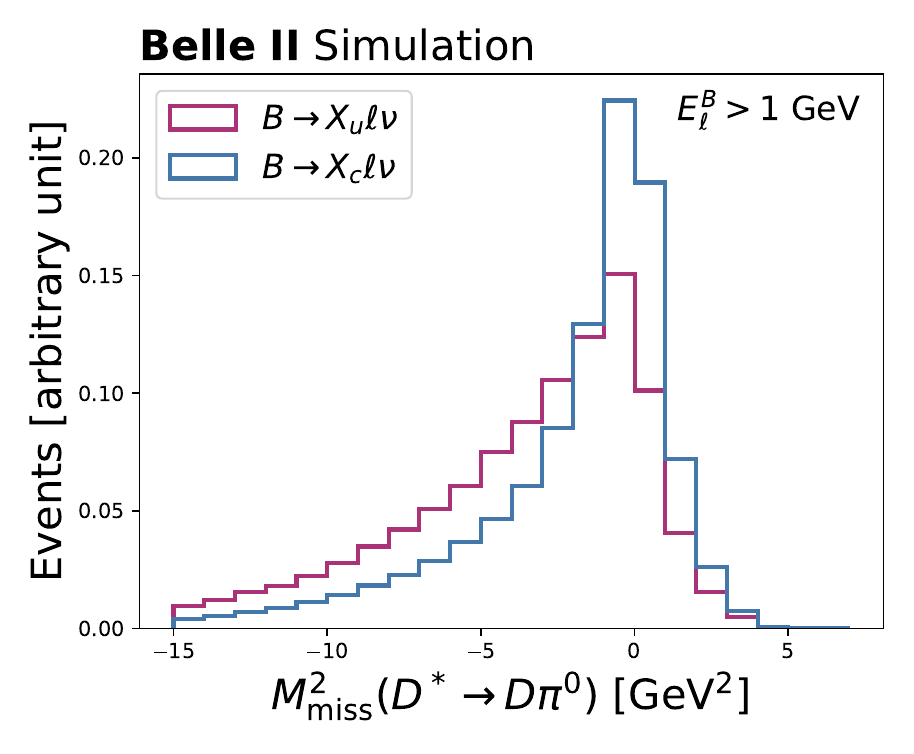}
    \includegraphics[scale=0.35]{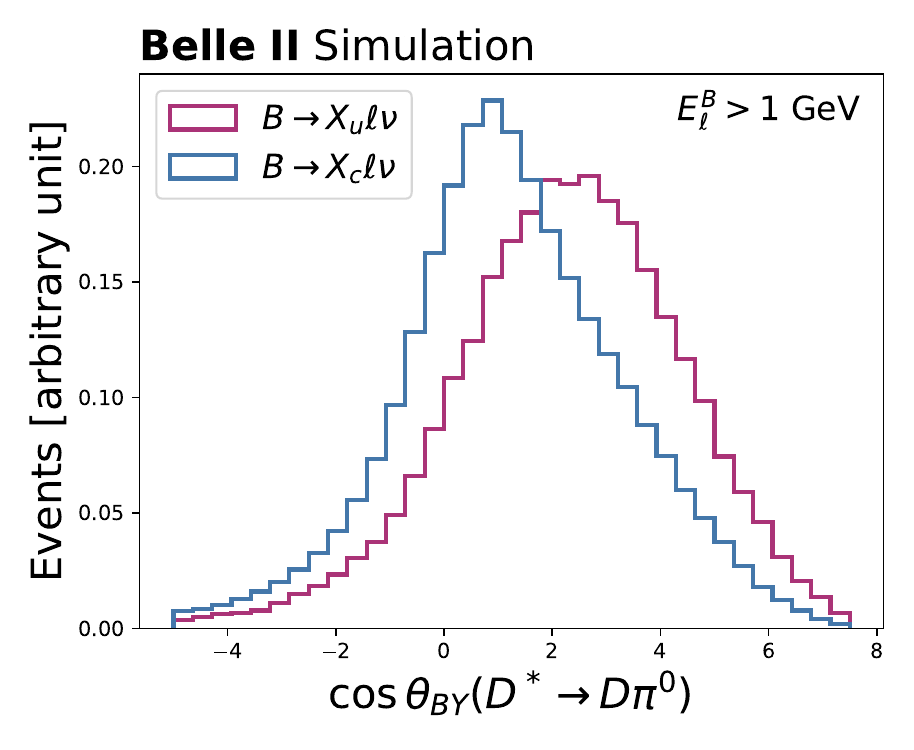}
    \includegraphics[scale=0.35]{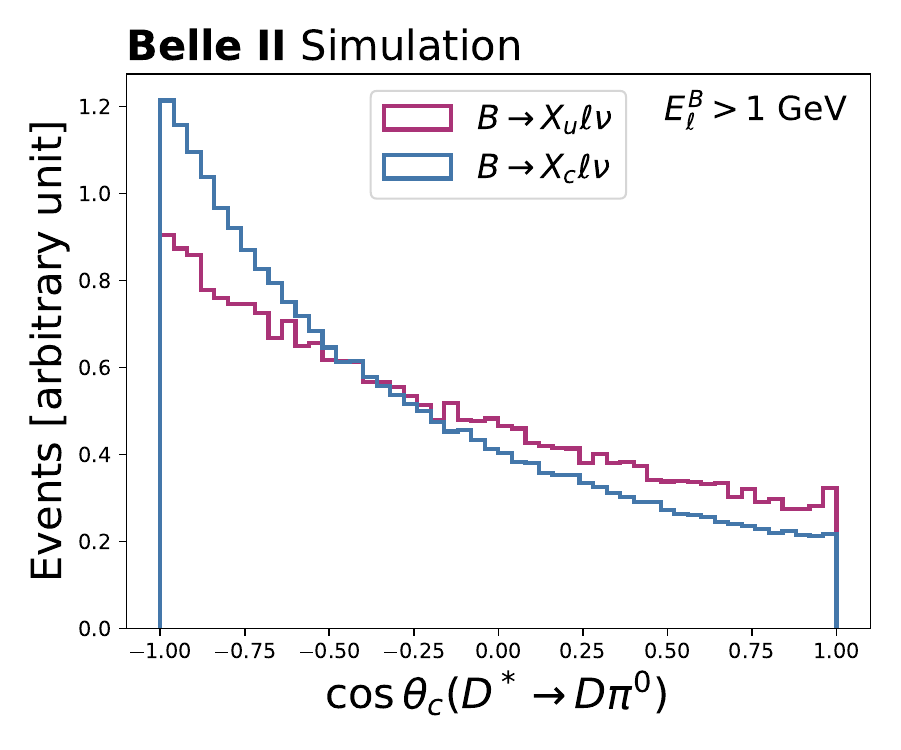}
    \caption{The shapes of the nine \btoxclnu\ suppression classifier input features are compared for \btoxulnu\ (purple) and \btoxclnu\ (blue) events in the preselected region on which the continuum suppression and kaon veto have been applied. The default values chosen for $p_{\rm vtx}^{\rm ROE}$ when the vertex fit fails and for the six $D^* \to D \pi$ variables when no low momentum $\pi$ is found in the event are not shown. These values are $-1$, $-20$ GeV$^2$, $-10$ and 1 for $p_{\rm vtx}^{\rm ROE}$, \mmsqu, $\cos\theta_{BY}$ and $\cos\theta_c$, respectively.}
    \label{fig:xclnu_mva_input}
\end{figure*}

The shapes of the nine variables used to train the \btoxclnu\ suppression classifier (see Sec.~\ref{sec:btoxclnu_suppression}) are compared for \btoxulnu\ and \btoxclnu\ events in Fig.~\ref{fig:xclnu_mva_input}. The performance of the classifier appears to be mostly driven by the event missing mass, the quality of ROE vertex fit and the missing mass in $\dst \to D\pi^0$ decays.

\section{\btoxclnu\ composition}\label{app:xc_compo}
The composition of the \btoxclnu\ component is given in Table~\ref{tab:xc_compo} for three different sets of kinematic selections and for the four control and signal regions defined in the main text. The compositions are extracted from simulated data.
\setlength{\tabcolsep}{11pt} 
\renewcommand{\arraystretch}{1.3} 
\begin{table*}[h!]
    \centering
    \caption{\btoxclnu\ composition in terms of the four types of subcomponents in each control and signal region for the three sets of kinematic selections.}
    \begin{tabular}{lrrrr}
        \hline
        \hline
         & CR\klow & CR\khigh & CR\zlow & SR \\
         \hline
         \multicolumn{5}{c}{$\elb > 1.0$~GeV} \\
         \hline
         $\btodlnu$ & 17.3\% & 40.4\%  & 19.1\%  & 37.3\%  \\
         $\btodstlnu$ & 61.2\%  & 47.3\%  &  58.7\% & 49.2\%  \\
         $\btodststlnu$ & 9.7\%  & 6.2\%  & 10.0\%  & 6.9\%  \\
         $\btodgaplnu$ &  11.8\%  & 6.1\%  & 12.2\%  & 6.6\%  \\
         \hline
         \multicolumn{5}{c}{$\elb > 1.0$~GeV, $\mx < 1.7$~GeV} \\
         \hline
         $\btodlnu$ & 22.0\% & 64.3\%  & 21.5\%  & 47.7\%  \\
         $\btodstlnu$ & 65.3\%  & 33.6\%  &  61.9\% & 47.3\%  \\
         $\btodststlnu$ & 6.4\%  & 1.5\%  & 8.0\%  & 2.7\%  \\
         $\btodgaplnu$ &  6.3\%  & 0.6\%  & 8.6\%  & 2.3\%  \\
         \hline
         \multicolumn{5}{c}{$\elb > 1.0$~GeV, $\mx < 1.7$~GeV, $\qtwo > 8$~GeV$^2$} \\
         \hline
         $\btodlnu$ & 14.6\% & 35.9\%  & 16.5\%  & 32.1\%  \\
         $\btodstlnu$ & 79.9\%  & 61.4\%  &  75.2\% & 63.2\%  \\
         $\btodststlnu$ & 3.5\%  & 2.0\%  & 5.1\%  & 3.2\%  \\
         $\btodgaplnu$ &  1.9\%  & 0.8\% & 3.3\% & 1.5\%  \\
         \hline
         \hline
    \end{tabular}
    \label{tab:xc_compo}
\end{table*}

\clearpage
\bibliography{apssamp}

@article{LANGE2001152,
title = {The {EvtGen} particle decay simulation package},
journal = {Nucl. Instrum. Meth. A},
volume = {462},
number = {1},
pages = {152-155},
year = {2001},
note = {{Beauty 2000}, Proceedings of the 7th Int. Conf. on B-Physics at Hadron Machines},
issn = {0168-9002},
doi = {https://doi.org/10.1016/S0168-9002(01)00089-4},
url = {https://www.sciencedirect.com/science/article/pii/S0168900201000894},
author = {David J. Lange}
}

@article{Andersen:2005mj,
  author = {Andersen, Jeppe R. and Gardi, Einan},
  year = {2006},
  journal = {J. High Energy Phys.},
  volume = {01},
  number = {2006},
  pages = {097},
  issn = {1029-8479},
  doi = {10.1088/1126-6708/2006/01/097},
  url = {http://arxiv.org/abs/hep-ph/0509360},
  urldate = {2022-08-12},
  archiveprefix = {arxiv},
  title = {Inclusive Spectra in Charmless Semileptonic {$B$} Decays by Dressed Gluon Exponentiation}
}

@article{BaBar:2011xxm,
  author = {Lees, J.P. and others},
  year = {2012},
  journal = {Phys. Rev. D},
  volume = {86},
  number = {3},
  primaryclass = {hep-ex},
  pages = {032004},
  issn = {1550-7998, 1550-2368},
  doi = {10.1103/PhysRevD.86.032004},
  url = {http://arxiv.org/abs/1112.0702},
  urldate = {2022-02-21},
  archiveprefix = {arxiv},
  collaboration = {BaBar Collaboration},
  title = {Study of {$\bar{B}\to X_u \ell \bar{\nu}$} Decays in {$B\bar{B}$} Events Tagged by a Fully Reconstructed {{B-meson}} Decay and Determination of {$|V_{ub}|$}}
}

@article{BaBar:2016rxh,
  author = {Lees, J.P. and others},
  year = {2017},
  month = apr,
  journal = {Phys. Rev. D},
  volume = {95},
  number = {7},
  primaryclass = {hep-ex},
  pages = {072001},
  issn = {2470-0010, 2470-0029},
  doi = {10.1103/PhysRevD.95.072001},
  url = {http://arxiv.org/abs/1611.05624},
  urldate = {2022-02-21},
  archiveprefix = {arxiv},
  collaboration = {BaBar Collaboration},
  title = {Measurement of the Inclusive Electron Spectrum from {$B$} Meson Decays and Determination of {$|V_{ub}|$}}
}

@article{Belle:2021eni,
  author = {Cao, L. and others},
  year = {2021},
  month = jul,
  journal = {Phys. Rev. D},
  volume = {104},
  number = {1},
  primaryclass = {hep-ex},
  pages = {012008},
  issn = {2470-0010, 2470-0029},
  doi = {10.1103/PhysRevD.104.012008},
  url = {http://arxiv.org/abs/2102.00020},
  urldate = {2022-02-15},
  archiveprefix = {arxiv},
  collaboration = {Belle Collaboration},
  title = {Measurements of {{Partial Branching Fractions}} of {{Inclusive}} {$B \to X_u \, \ell^+\, \nu_{\ell}$} {{Decays}} with {{Hadronic Tagging}}}
}

@article{Bernlochner:2017jxt,
  author = {Bernlochner, Florian U. and Ligeti, Zoltan and Robinson, Dean J.},
  year = {2018},
  month = apr,
  journal = {Phys. Rev. D},
  volume = {97},
  number = {7},
  primaryclass = {hep-ph},
  pages = {075011},
  issn = {2470-0010, 2470-0029},
  doi = {10.1103/PhysRevD.97.075011},
  url = {http://arxiv.org/abs/1711.03110},
  urldate = {2023-09-25},
  archiveprefix = {arxiv},
  title = {Model Independent Analysis of Semileptonic {$B$} Decays to {$D^{**}$} for Arbitrary New Physics}
}

@article{Bernlochner:2020tfi,
  title = {Das Ist Der {{HAMMER}}: {{Consistent}} New Physics Interpretations of Semileptonic Decays},
  shorttitle = {Das Ist Der {{HAMMER}}},
  author = {Bernlochner, Florian U. and Duell, Stephan and Ligeti, Zoltan and Papucci, Michele and Robinson, Dean J.},
  year = {2020},
  month = sep,
  journal = {Eur. Phys. J. C},
  volume = {80},
  number = {9},
  primaryclass = {hep-ph},
  pages = {883},
  issn = {1434-6044, 1434-6052},
  doi = {10.1140/epjc/s10052-020-8304-0},
  url = {http://arxiv.org/abs/2002.00020},
  urldate = {2023-09-25},
  archiveprefix = {arxiv}
}

@article{Bernlochner:2021rel,
  shorttitle = {B \textrightarrow{} {$\rho$} l {$\nu$} \textasciimacron{} and {$\omega$} l {$\nu$} \textasciimacron{} in and beyond the {{Standard Model}}},
  author = {Bernlochner, Florian U. and Prim, Markus T. and Robinson, Dean J.},
  year = {2021},
  month = aug,
  journal = {Phys. Rev. D},
  volume = {104},
  number = {3},
  primaryclass = {hep-ph},
  pages = {034032},
  issn = {2470-0010, 2470-0029},
  doi = {10.1103/PhysRevD.104.034032},
  url = {https://link.aps.org/doi/10.1103/PhysRevD.104.034032},
  urldate = {2022-04-02},
  archiveprefix = {arxiv},
  langid = {english},
  title = {{$B \to \rho \ell \bar{\nu}$} and {$\omega \ell \bar{\nu}$} in and beyond the {{Standard Model}}: {{Improved}} Predictions and {$|V_{ub}|$}}
}

@article{Bernlochner:2022ywh,
  shorttitle = {Constrained Second-Order Power Corrections in {{HQET}}},
  author = {Bernlochner, Florian U. and Ligeti, Zoltan and Papucci, Michele and Prim, Markus T. and Robinson, Dean J. and Xiong, Chenglu},
  year = {2022},
  month = nov,
  journal = {Phys. Rev. D},
  volume = {106},
  number = {9},
  primaryclass = {hep-ph},
  pages = {096015},
  issn = {2470-0010, 2470-0029},
  doi = {10.1103/PhysRevD.106.096015},
  url = {http://arxiv.org/abs/2206.11281},
  urldate = {2023-09-25},
  archiveprefix = {arxiv},
  title = {Constrained Second-Order Power Corrections in {{HQET}}: {$R(D^{(*)})$}, {$|V_{cb}|$}, and New Physics}
}

@article{SJOSTRAND2008852,
title = {{A brief introduction to PYTHIA 8.1}},
journal = {Computer Physics Communications},
volume = {178},
number = {11},
pages = {852-867},
year = {2008},
issn = {0010-4655},
doi = {https://doi.org/10.1016/j.cpc.2008.01.036},
url = {https://www.sciencedirect.com/science/article/pii/S0010465508000441},
author = {Torbjörn Sjöstrand and Stephen Mrenna and Peter Skands},
}

@article{Bourrely:2008za,
  author = {Bourrely, Claude and Caprini, Irinel and Lellouch, Laurent},
  year = {2009},
  journal = {Phys. Rev. D},
  volume = {79},
  number = {1},
  primaryclass = {hep-ph},
  pages = {013008},
  publisher = {{American Physical Society}},
  doi = {10.1103/PhysRevD.82.099902},
  url = {https://link.aps.org/doi/10.1103/PhysRevD.79.013008},
  urldate = {2022-08-12},
  archiveprefix = {arxiv},
  lccn = {as},
  title = {Model-Independent Description of {$B \to\pi \ell \nu$} Decays and a Determination of {$|V_{ub}|$}}
}

@article{Cabibbo:1963yz,
  title = {Unitary {{Symmetry}} and {{Leptonic Decays}}},
  author = {Cabibbo, Nicola},
  year = {1963},
  journal = {Phys. Rev. Lett.},
  volume = {10},
  number = {12},
  pages = {531--533},
  publisher = {{American Physical Society}},
  doi = {10.1103/PhysRevLett.10.531},
  url = {https://link.aps.org/doi/10.1103/PhysRevLett.10.531},
  urldate = {2022-02-21}
}

@article{CLEO:1995rok,
  title = {Search for Exclusive Charmless Hadronic {$B$} Decays},
  author = {Asner, D.M. and others},
  year = {1996},
  journal = {Phys. Rev. D},
  volume = {53},
  pages = {1039--1050},
  doi = {10.1103/PhysRevD.53.1039},
  url = {https://arxiv.org/abs/hep-ex/9508004},
  archiveprefix = {arxiv},
  collaboration = {CLEO Collaboration}
}

@misc{loshchilov2019,
      title={Decoupled Weight Decay Regularization}, 
      author={Ilya Loshchilov and Frank Hutter},
      eprint={1711.05101},
      archivePrefix={arXiv},
      url={https://arxiv.org/abs/1711.05101}, 
}

@article{DeFazio:1999ptt,
  author = {De Fazio, Fulvia and Neubert, Matthias},
  year = {1999},
  journal = {J. High Energy Phys.},
  volume = {06},
  number = {1999},
  pages = {017},
  issn = {1029-8479},
  doi = {10.1088/1126-6708/1999/06/017},
  url = {http://arxiv.org/abs/hep-ph/9905351},
  urldate = {2022-03-30},
  archiveprefix = {arxiv},
  langid = {english},
  title = {{$B \to X_u \ell \bar{\nu}_{\ell}$} Decay Distributions to Order {$\alpha_s$}}
}

@article{Duplancic:2015zna,
  author = {Duplancic, Goran and Melic, Blazenka},
  year = {2015},
  month = nov,
  journal = {J. High Energy Phys.},
  volume = {11},
  number = {2015},
  primaryclass = {hep-ph},
  pages = {138},
  issn = {1029-8479},
  doi = {10.1007/JHEP11(2015)138},
  url = {http://arxiv.org/abs/1508.05287},
  urldate = {2022-04-02},
  archiveprefix = {arxiv},
  title = {Form Factors of {$B, B_{s} \to \eta^{(\prime)}$} and {$D, D_{s} \to \eta^{(\prime)}$} Transitions from {{QCD}} Light-Cone Sum Rules}
}

@article{EFFORT2022,
  title = {{{eFFORT}}},
  year = {2022},
  month = may,
  journal = {https://github.com/b2-hive/eFFORT},
  url = {https://github.com/b2-hive/eFFORT},
  urldate = {2022-08-12},
  copyright = {MIT},
  howpublished = {b2-hive}
}

@article{FlavourLatticeAveragingGroupFLAG:2021npn,
  title = {{{FLAG Review}} 2021},
  author = {Aoki, Y. and others},
  year = {2022},
  month = oct,
  journal = {Eur. Phys. J. C},
  volume = {82},
  number = {10},
  primaryclass = {hep-lat},
  pages = {869},
  doi = {10.1140/epjc/s10052-022-10536-1},
  url = {http://arxiv.org/abs/2111.09849},
  urldate = {2022-08-11},
  archiveprefix = {arxiv},
  collaboration = {Flavour Lattice Averaging Group}
}

@article{Fox:1978vu,
  title = {Observables for the {{Analysis}} of {{Event Shapes}} in $e^+e^-$ {{Annihilation}} and {{Other Processes}}},
  author = {Fox, Geoffrey C. and Wolfram, Stephen},
  year = {1978},
  journal = {Phys. Rev. Lett.},
  volume = {41},
  number = {23},
  pages = {1581},
  publisher = {{American Physical Society}},
  doi = {10.1103/PhysRevLett.41.1581},
  url = {https://link.aps.org/doi/10.1103/PhysRevLett.41.1581}
}

@article{Gambino:2007rp,
doi = {10.1088/1126-6708/2007/10/058},
url = {https://doi.org/10.1088/1126-6708/2007/10/058},
year = {2007},
month = {oct},
publisher = {},
volume = {10},
number = {2007},
pages = {058},
author = {Paolo Gambino and Paolo Giordano and Giovanni Ossola and Nikolai Uraltsev},
title = {{Inclusive semileptonic $B$ decays and the  determination of $|V_{ub}|$}},
journal = {J. High Energy Phys.}
}

@article{JADE:1983KaonProd,
  author = {Bartel, W. and others},
  year = {1983},
  month = sep,
  journal = {Z. Phys. C},
  volume = {20},
  number = {3},
  pages = {187--206},
  issn = {1431-5858},
  doi = {10.1007/BF01574851},
  url = {https://doi.org/10.1007/BF01574851},
  urldate = {2023-09-25},
  collaboration = {JADE Collaboration},
  langid = {english},
  title = {Charged Particle and Neutral Kaon Production in {$e^+ e^-$} Annihilation at {{PETRA}}}
}

@article{Kagan:1998ym,
  author = {Kagan, Alexander L. and Neubert, Matthias},
  year = {1999},
  journal = {Eur. Phys. J. C},
  volume = {7},
  number = {1},
  pages = {5--27},
  issn = {1434-6044, 1434-6052},
  doi = {10.1007/s100529800959},
  url = {http://arxiv.org/abs/hep-ph/9805303},
  urldate = {2022-04-02},
  archiveprefix = {arxiv},
  title = {{{QCD}} Anatomy of {$B\to X_s \gamma$} Decays}
}

@article{Keck2019,
author={Keck, T. and others},
title={The Full Event Interpretation},
journal={Computing and Software for Big Science},
year={2019},
month={Feb},
day={25},
volume={3},
number={1},
pages={6},
issn={2510-2044},
doi={10.1007/s41781-019-0021-8},
url={https://doi.org/10.1007/s41781-019-0021-8}
}

@article{Kobayashi:1973fv,
  title = {{{CP Violation}} in the {{Renormalizable Theory}} of {{Weak Interaction}}},
  author = {Kobayashi, Makoto and Maskawa, Toshihide},
  year = {1973},
  journal = {Prog. Theor. Phys.},
  volume = {49},
  number = {2},
  pages = {652--657},
  issn = {0033-068X},
  doi = {10.1143/PTP.49.652},
  url = {https://doi.org/10.1143/PTP.49.652},
  urldate = {2022-02-21}
}

@article{Kuhr:2018lps,
  title = {The {{Belle II Core Software}}},
  author = {Kuhr, T. and Pulvermacher, C. and Ritter, M. and Hauth, T. and Braun, N.},
  year = {2018},
  month = nov,
  journal = {Comput. Softw. Big Sci.},
  volume = {3},
  number = {1},
  primaryclass = {physics.comp-ph},
  pages = {1},
  issn = {2510-2036, 2510-2044},
  doi = {10.1007/s41781-018-0017-9},
  url = {http://arxiv.org/abs/1809.04299},
  urldate = {2022-04-06},
  archiveprefix = {arxiv},
  collaboration = {Belle-II Framework Software Group},
  langid = {english}
}

@article{Lange:2005yw,
  author = {Lange, Bjorn O. and Neubert, Matthias and Paz, Gil},
  year = {2005},
  journal = {Phys. Rev. D},
  volume = {72},
  number = {7},
  pages = {073006},
  issn = {1550-7998, 1550-2368},
  doi = {10.1103/PhysRevD.72.073006},
  url = {http://arxiv.org/abs/hep-ph/0504071},
  urldate = {2022-04-03},
  archiveprefix = {arxiv},
  title = {Theory of Charmless Inclusive {$B$} Decays and the Extraction of {$V_{ub}$}}
}

@article{leeEvidencePiPi2003,
  author = {Lee, S.H. and others},
  year = {2003},
  journal = {Phys. Rev. Lett.},
  volume = {91},
  number = {26},
  pages = {261801},
  publisher = {{American Physical Society}},
  doi = {10.1103/PhysRevLett.91.261801},
  url = {https://link.aps.org/doi/10.1103/PhysRevLett.91.261801},
  archiveprefix = {arxiv},
  collaboration = {Belle Collaboration},
  title = {Evidence for {$B^0 \to \pi^0 \pi^0$}}
}

@article{Martschei:2012pr,
  title = {{Advanced Event Reweighting Using Multivariate Analysis}},
  author = {Martschei, D. and Feindt, M. and Honc, S. and {Wagner-Kuhr}, J.},
  year = {2012},
  journal = {J. Phys. Conf. Ser.},
  volume = {368},
  number = {1},
  pages = {012028},
  issn = {1742-6596},
  doi = {10.1088/1742-6596/368/1/012028},
  url = {https://dx.doi.org/10.1088/1742-6596/368/1/012028},
  urldate = {2023-09-27},
  langid = {english}
}

@article{Milesi:2020esq,
  title = {Lepton Identification in {{Belle II}} Using Observables from the Electromagnetic Calorimeter and Precision Trackers},
  author = {Milesi, Marco and Tan, Justin and Urquijo, Phillip},
  year = {2020},
  journal = {EPJ Web Conf.},
  volume = {245},
  pages = {06023},
  publisher = {{EDP Sciences}},
  issn = {2100-014X},
  doi = {10.1051/epjconf/202024506023},
  url = {https://www.epj-conferences.org/articles/epjconf/pdf/2020/21/epjconf_chep2020_06023.pdf},
  urldate = {2023-03-02},
  copyright = {\textcopyright{} The Authors, published by EDP Sciences, 2020},
  langid = {english},
}

@misc{ramachandranSearchingActivationFunctions2017,
  title = {Searching for {{Activation Functions}}},
  author = {Ramachandran, Prajit and Zoph, Barret and Le, Quoc V.},
  year = {2017},
  month = oct,
  publisher = {{arXiv}},
  eprint={1710.05941},
  url = {http://arxiv.org/abs/1710.05941},
  urldate = {2023-05-10},
  archiveprefix = {arXiv}
}

@article{Ramirez:1989yk,
  author = {Ramirez, Carlos and Donoghue, John F. and Burdman, Gustavo},
  year = {1990},
  journal = {Phys. Rev. D},
  volume = {41},
  number = {5},
  pages = {1496},
  publisher = {{American Physical Society}},
  doi = {10.1103/PhysRevD.41.1496},
  url = {https://link.aps.org/doi/10.1103/PhysRevD.41.1496},
  urldate = {2022-08-12},
  title = {Semileptonic {$b \to u$} Decay}
}

@article{TASSO:1984nda,
  author = {Althoff, M. and others},
  year = {1985},
  journal = {Z. Phys. C},
  volume = {27},
  number = {1},
  pages = {27},
  issn = {1431-5858},
  doi = {10.1007/BF01642477},
  url = {https://doi.org/10.1007/BF01642477},
  urldate = {2023-09-25},
  collaboration = {TASSO},
  langid = {english},
  title = {A {{Detailed Study}} of {{Strange Particle Production}} in {$e^+ e^-$} {{Annihilation}} at {{High-energy}}}
}

@article{Cheema:2024iek,
    author = "Cheema, Priyanka",
    title = "{Suppressing Beam Background and Fake Photons at Belle II using Machine Learning}",
    doi = "10.1051/epjconf/202429509035",
    journal = "EPJ Web Conf.",
    volume = "295",
    pages = "09035",
    year = "2024"
}

@article{JADACH2023108556,
title = {{Multi-photon Monte Carlo event generator KKMCee for lepton and quark pair production in lepton colliders}},
journal = {Computer Physics Communications},
volume = {283},
pages = {108556},
year = {2023},
issn = {0010-4655},
doi = {https://doi.org/10.1016/j.cpc.2022.108556},
url = {https://www.sciencedirect.com/science/article/pii/S0010465522002752},
author = {S. Jadach and B.F.L. Ward and Z. Was and S.A. Yost and A. Siodmok}
}

@article{PhysRevD.73.073008,
  title = {Fit to moments of inclusive {$B \to X_c \ell \bar{\nu}$} and {$B\ensuremath{\rightarrow}{X}_{s}\ensuremath{\gamma}$} decay distributions using heavy quark expansions in the kinetic scheme},
  author = {Buchm\"uller, O. L. and Fl\"acher, H. U.},
  journal = {Phys. Rev. D},
  volume = {73},
  issue = {7},
  pages = {073008},
  numpages = {11},
  year = {2006},
  month = {Apr},
  publisher = {American Physical Society},
  doi = {10.1103/PhysRevD.73.073008},
  url = {https://link.aps.org/doi/10.1103/PhysRevD.73.073008}
}

@article{PhysRevD.85.094033,
  title = {A proposal to solve some puzzles in semileptonic {$B$} decays},
  author = {Bernlochner, Florian U. and Ligeti, Zoltan and Turczyk, Sascha},
  journal = {Phys. Rev. D},
  volume = {85},
  issue = {9},
  pages = {094033},
  numpages = {5},
  year = {2012},
  month = {May},
  publisher = {American Physical Society},
  doi = {10.1103/PhysRevD.85.094033},
  url = {https://link.aps.org/doi/10.1103/PhysRevD.85.094033}
}

@misc{hflav_2023,
      title={Averages of $b$-hadron, $c$-hadron, and $\tau$-lepton properties as of 2023}, 
      author={Sw. Banerjee and others},
      collaboration={Heavy Flavor Averaging Group},
      year={2024},
      eprint={2411.18639},
      archivePrefix={arXiv},
      primaryClass={hep-ex},
      url={https://arxiv.org/abs/2411.18639}, 
}

@misc{optuna,
      title={{Optuna: A Next-generation Hyperparameter Optimization Framework}}, 
      author={Takuya Akiba and Shotaro Sano and Toshihiko Yanase and Takeru Ohta and Masanori Koyama},
      year={2019},
      eprint={1907.10902},
      archivePrefix={arXiv},
      url={https://arxiv.org/abs/1907.10902},
}

@article{Heinrich2021,
  doi = {10.21105/joss.02823},
  url = {https://doi.org/10.21105/joss.02823},
  year = {2021},
  publisher = {The Open Journal},
  volume = {6},
  number = {58},
  pages = {2823},
  author = {Lukas Heinrich and Matthew Feickert and Giordon Stark and Kyle Cranmer},
  title = {pyhf: pure-{Python} implementation of {HistFactory} statistical models},
  journal = {Journal of Open Source Software}
}

@article{BOSCH2004335,
title = {Factorization and shape-function effects in inclusive {$B$}-meson decays},
journal = {Nuclear Physics B},
volume = {699},
number = {1},
pages = {335-386},
year = {2004},
issn = {0550-3213},
doi = {https://doi.org/10.1016/j.nuclphysb.2004.07.041},
url = {https://www.sciencedirect.com/science/article/pii/S0550321304005723},
author = {S.W. Bosch and B.O. Lange and M. Neubert and G. Paz}
}

@article{PhysRevD.110.030001,
  title = {Review of Particle Physics},
  author = {Navas, S. and others},
  collaboration = {Particle Data Group},
  journal = {Phys. Rev. D},
  volume = {110},
  issue = {3},
  pages = {030001},
  numpages = {5},
  year = {2024},
  month = {Aug},
  publisher = {American Physical Society},
  doi = {10.1103/PhysRevD.110.030001},
  url = {https://link.aps.org/doi/10.1103/PhysRevD.110.030001},
  note = {updated results and plots available at: https://pdglive.lbl.gov/Viewer.action},
}

@article{PhysRevD.52.2783,
  title = {Semileptonic meson decays in the quark model: An update},
  author = {Scora, Daryl and Isgur, Nathan},
  journal = {Phys. Rev. D},
  volume = {52},
  issue = {5},
  pages = {2783--2812},
  numpages = {0},
  year = {1995},
  month = {Sep},
  publisher = {American Physical Society},
  doi = {10.1103/PhysRevD.52.2783},
  url = {https://link.aps.org/doi/10.1103/PhysRevD.52.2783}
}

@article{Bharucha2016,
author={Bharucha, Aoife and Straub, David M. and Zwicky, Roman},
title={{$B \to V\ell^+\ell^-$} in the {Standard} {Model} from light-cone sum rules},
journal={J. High Energy Phys.},
year={2016},
month={Aug},
day={16},
volume={8},
number={2016},
pages={98},
issn={1029-8479},
doi={10.1007/JHEP08(2016)098},
url={https://doi.org/10.1007/JHEP08(2016)098}
}

@article{MANNEL1994396,
title = {Operator product expansion for inclusive semileptonic decays in heavy quark effective field theory},
journal = {Nuclear Physics B},
volume = {413},
number = {1},
pages = {396-410},
year = {1994},
issn = {0550-3213},
doi = {https://doi.org/10.1016/0550-3213(94)90625-4},
url = {https://www.sciencedirect.com/science/article/pii/0550321394906254},
author = {Thomas Mannel}
}

@article{CHAY1990399,
title = {{Lepton energy distributions in heavy meson decays from QCD}},
journal = {Physics Letters B},
volume = {247},
number = {2},
pages = {399-405},
year = {1990},
issn = {0370-2693},
doi = {https://doi.org/10.1016/0370-2693(90)90916-T},
url = {https://www.sciencedirect.com/science/article/pii/037026939090916T},
author = {Junegone Chay and Howard Georgi and Benjamin Grinstein}
}

@article{HadronIDBelleII,
author={Adachi, I. and others},
title={{Charged-hadron identification at Belle II}},
journal={Eur. Phys. J C},
year={2025},
month={Nov},
collaboration={Belle II Collaboration},
day={01},
volume={85},
number={11},
pages={1237},
issn={1434-6052},
doi={10.1140/epjc/s10052-025-14627-7},
url={https://doi.org/10.1140/epjc/s10052-025-14627-7}
}

@article{AKAI2018188,
title = {{SuperKEKB} {Collider}},
journal = {Nucl. Instrum. Meth. A},
volume = {907},
pages = {188-199},
year = {2018},
issn = {0168-9002},
doi = {https://doi.org/10.1016/j.nima.2018.08.017},
url = {https://www.sciencedirect.com/science/article/pii/S0168900218309616},
author = {Kazunori Akai and Kazuro Furukawa and Haruyo Koiso}
}

@article{FEINDT2011432,
title = {A hierarchical NeuroBayes-based algorithm for full reconstruction of {$B$} mesons at {$B$} factories},
journal = {Nucl. Instrum. Meth. A},
volume = {654},
number = {1},
pages = {432-440},
year = {2011},
issn = {0168-9002},
doi = {https://doi.org/10.1016/j.nima.2011.06.008},
url = {https://www.sciencedirect.com/science/article/pii/S0168900211011193},
author = {M. Feindt and F. Keller and M. Kreps and T. Kuhr and S. Neubauer and D. Zander and A. Zupanc}
}

@article{PhysRevD.109.112006,
  title = {Evidence for {${B}^{+}\ensuremath{\rightarrow}{K}^{+}\ensuremath{\nu}\overline{\ensuremath{\nu}}$} decays},
  author = {Adachi, I. and others},
  collaboration = {Belle II Collaboration},
  journal = {Phys. Rev. D},
  volume = {109},
  issue = {11},
  pages = {112006},
  numpages = {29},
  year = {2024},
  month = {Jun},
  publisher = {American Physical Society},
  doi = {10.1103/PhysRevD.109.112006},
  url = {https://link.aps.org/doi/10.1103/PhysRevD.109.112006}
}

@article{AGOSTINELLI2003250,
title = {Geant4—a simulation toolkit},
journal = {Nucl. Instrum. Meth. A},
volume = {506},
number = {3},
pages = {250-303},
year = {2003},
issn = {0168-9002},
doi = {https://doi.org/10.1016/S0168-9002(03)01368-8},
url = {https://www.sciencedirect.com/science/article/pii/S0168900203013688},
author = {S. Agostinelli and others}
}

@article{NEUBERT200513,
title = {Two-loop relations for heavy-quark parameters in the shape-function scheme},
journal = {Physics Letters B},
volume = {612},
number = {1},
pages = {13-20},
year = {2005},
issn = {0370-2693},
doi = {https://doi.org/10.1016/j.physletb.2005.02.055},
url = {https://www.sciencedirect.com/science/article/pii/S0370269305003072},
author = {Matthias Neubert}
}

@article{PhysRevD.72.074025,
  title = {{Advanced predictions for moments of the $\overline{B}\ensuremath{\rightarrow}{X}_{s}\ensuremath{\gamma}$ photon spectrum}},
  author = {Neubert, Matthias},
  journal = {Phys. Rev. D},
  volume = {72},
  issue = {7},
  pages = {074025},
  numpages = {17},
  year = {2005},
  month = {Oct},
  publisher = {American Physical Society},
  doi = {10.1103/PhysRevD.72.074025},
  url = {https://link.aps.org/doi/10.1103/PhysRevD.72.074025}
}

@article{PhysRevD.107.092003,
  title = {{First observation of $B\ensuremath{\rightarrow}{\overline{D}}_{1}(\ensuremath{\rightarrow}\overline{D}{\ensuremath{\pi}}^{+}{\ensuremath{\pi}}^{\ensuremath{-}}){\ensuremath{\ell}}^{+}{\ensuremath{\nu}}_{\ensuremath{\ell}}$ and measurement of the $B\ensuremath{\rightarrow}{\overline{D}}^{(*)}\ensuremath{\pi}{\ensuremath{\ell}}^{+}{\ensuremath{\nu}}_{\ensuremath{\ell}}$ and $B\ensuremath{\rightarrow}{\overline{D}}^{(*)}{\ensuremath{\pi}}^{+}{\ensuremath{\pi}}^{\ensuremath{-}}{\ensuremath{\ell}}^{+}{\ensuremath{\nu}}_{\ensuremath{\ell}}$ branching fractions with hadronic tagging at Belle}},
  author = {Meier, F. and others},
  collaboration = {Belle Collaboration},
  journal = {Phys. Rev. D},
  volume = {107},
  issue = {9},
  pages = {092003},
  numpages = {23},
  year = {2023},
  month = {May},
  publisher = {American Physical Society},
  doi = {10.1103/PhysRevD.107.092003},
  url = {https://link.aps.org/doi/10.1103/PhysRevD.107.092003}
}

@article{PhysRevLett.101.261802,
  title = {{Measurement of the Branching Fractions of $\overline{B}\ensuremath{\rightarrow}{D}^{**}{\ensuremath{\ell}}^{\ensuremath{-}}{\overline{\ensuremath{\nu}}}_{\ensuremath{\ell}}$ Decays in Events Tagged by a Fully Reconstructed $B$ Meson}},
  author = {Aubert, B. and others},
  collaboration = {BABAR Collaboration},
  journal = {Phys. Rev. Lett.},
  volume = {101},
  issue = {26},
  pages = {261802},
  numpages = {7},
  year = {2008},
  month = {Dec},
  publisher = {American Physical Society},
  doi = {10.1103/PhysRevLett.101.261802},
  url = {https://link.aps.org/doi/10.1103/PhysRevLett.101.261802}
}

@article{PhysRevD.57.2691,
  title = {Illustrative example of how quark-hadron duality might work},
  author = {Blok, B. and Shifman, M. and Zhang, Da-Xin},
  journal = {Phys. Rev. D},
  volume = {57},
  issue = {5},
  pages = {2691--2700},
  numpages = {0},
  year = {1998},
  month = {Mar},
  publisher = {American Physical Society},
  doi = {10.1103/PhysRevD.57.2691},
  url = {https://link.aps.org/doi/10.1103/PhysRevD.57.2691}
}

@article{PhysRevD.56.4017,
  title = {High power $n$ of ${m}_{b}$ in $b$-flavored widths and $n=5\ensuremath{\rightarrow}\mathbf{\ensuremath{\infty}}$ limit},
  author = {Bigi, I. and Shifman, M. and Uraltsev, N. and Vainshtein, A.},
  journal = {Phys. Rev. D},
  volume = {56},
  issue = {7},
  pages = {4017--4030},
  numpages = {0},
  year = {1997},
  month = {Oct},
  publisher = {American Physical Society},
  doi = {10.1103/PhysRevD.56.4017},
  url = {https://link.aps.org/doi/10.1103/PhysRevD.56.4017}
}

@article{Aaij2021,
author={Aaij, R. and others},
title={Measurement of the branching fraction of the {$B^0 \to D_s^+ \pi^-$} decay},
journal={Eur. Phys. J. C},
year={2021},
month={Apr},
day={12},
volume={81},
number={4},
pages={314},
issn={1434-6052},
doi={10.1140/epjc/s10052-020-08790-2},
url={https://doi.org/10.1140/epjc/s10052-020-08790-2}
}

@misc{Snowmass2022,
      title={{Snowmass White Paper: Belle II physics reach and plans for the next decade and beyond}}, 
      author={Latika Aggarwal and others},
      year={2022},
      eprint={2207.06307},
      archivePrefix={arXiv},
      primaryClass={hep-ex},
      url={https://arxiv.org/abs/2207.06307}, 
}


\end{document}